\documentclass[letterpaper,english,reprint,prx,longbibliography,aps]{revtex4-1}
\usepackage[T1]{fontenc}
\usepackage[latin9]{inputenc}
\usepackage{babel}
\usepackage{verbatim}
\usepackage{mathtools}
\usepackage{bm}
\usepackage{amsmath}
\usepackage{amssymb}
\usepackage{graphicx}
\usepackage{color}
\usepackage[unicode=true,
 bookmarks=true,bookmarksnumbered=false,bookmarksopen=false,
 breaklinks=false,pdfborder={0 0 1},backref=false,colorlinks=false]
 {hyperref}

\usepackage{ulem}

\makeatletter

\pdfpageheight\paperheight
\pdfpagewidth\paperwidth

\usepackage{babel}

\newcommand{\subsecthead}[1]{{\textit {#1}}}

\newcommand{\traj}{\varphi}
\newcommand{\trajdot}{\dot{\varphi}}

\newcommand{\trajddotvec}{\ddot{\trajvec}}
\newcommand{\trajTwo}{\phi}

\newcommand{\LOM}{\mathcal{L}_{\mathrm{OM}}}
\newcommand{\LOMDL}{\tilde{\mathcal{L}}_{\mathrm{OM}}}
\newcommand{\Ltwo}{{\mathcal{L}}^{(2)}}

\newcommand{\tDL}{\tilde{t}}
\newcommand{\tzeroDL}{\tilde{t}_i}
\newcommand{\tzero}{t_i}
\newcommand{\tfinalDL}{\tilde{t}_f}
\newcommand{\tfinal}{{t}_f}
\newcommand{\tinitial}{\tzero}
\newcommand{\tinitialDL}{\tzeroDL}

\newcommand{\BR}{B_R^{\trajvec}}

\newcommand{\BDL}{\tilde{B}}

\newcommand{\tDLdummyone}{\tilde{t}\,'}
\newcommand{\tDLdummytwo}{\tilde{t}\,''}
\newcommand{\xDL}{\tilde{x}}

\newcommand{\xzeroDLvec}{\tilde{\bm{x}}_i}
\newcommand{\xveczeroDL}{\xzeroDLvec}

\newcommand{\ParityDL}{\tilde{\mathcal{P}}}

\newcommand{\ParityZero}{p_n}

\newcommand{\vecX}{\bm{X}} 
\newcommand{\Xvec}{\vecX}
\newcommand{\Xveczero}{\vecX_{\tzero}}
\newcommand{\vecx}{\bm{x}} 
\newcommand{\xvec}{\vecx} 

\newcommand{\vecF}{\bm{F}}
\newcommand{\Fvec}{\vecF}
\newcommand{\vectraj}{\bm{\traj}}
\newcommand{\trajvec}{\vectraj}
\newcommand{\vectrajdot}{\dot{\vectraj}}
\newcommand{\trajdotvec}{\vectrajdot}
\newcommand{\grad}{\bm{\nabla}}
\newcommand{\gradvec}{\grad}

\newcommand{\vectrajTwo}{\bm{\trajTwo}}
\newcommand{\vectrajdotTwo}{\dot{\bm{\trajTwo}}}

\newcommand{\vecxDL}{\tilde{\bm{x}}}
\newcommand{\xvecDL}{\vecxDL}
\newcommand{\xDLvec}{\vecxDL}
\newcommand{\vecXDL}{\tilde{\bm{X}}}
\newcommand{\XvecDL}{\vecXDL}

\newcommand{\vecXzeroDL}{\tilde{\bm{X}}_{{\tilde{t}_i}}}
\newcommand{\XzerovecDL}{\vecXzeroDL}

\newcommand{\vecFDL}{\tilde{\bm{F}}}
\newcommand{\FvecDL}{\vecFDL}
\newcommand{\FDLvec}{\vecFDL}
\newcommand{\vectrajDL}{\tilde{\bm{\traj}}}
\newcommand{\trajvecDL}{\vectrajDL}

\newcommand{\vectrajdotDL}{\dot{\tilde{\vectraj}}}
\newcommand{\trajdotvecDL}{\vectrajdotDL}
\newcommand{\trajdotDLvec}{\vectrajdotDL}
\newcommand{\gradvecDL}{\tilde{\bm{\nabla}}}
\newcommand{\gradDL}{\gradvecDL}

\newcommand{\laplacevec}{{\bm{\nabla}}^2}

\newcommand{\laplaceDLvec}{\tilde{\bm{\nabla}}^2}

\newcommand{\FDLeffvec}{\vecFDL_{\mathrm{app}}}

\newcommand{\alphavec}{\bm{\alpha}}

\newcommand{\noisevec}{\bm{\xi}}

\newcommand{\Integral}{\tilde{\mathcal{I}}}

\newcommand{\rDL}{\tilde{P}}

\newcommand{\FDLeff}{\tilde{F}_{\mathrm{app}}}
\newcommand{\trajDL}{\tilde{\varphi}}
\newcommand{\trajdotDL}{\dot{\tilde{\varphi}}}
\newcommand{\reqDL}{\tilde{\rho}_{\mathrm{ss}}}
\newcommand{\jveceqDL}{\tilde{\bf{j}}_{\mathrm{ss}}}

\newcommand{\trajset}{\mathcal{X}_R^{\trajvec}}

\newcommand{\Pabsorbing}{P^{\,\bm{\varphi}}_R}
\newcommand{\Pabsorbingdot}{\dot{P}^{\,\bm{\varphi}}_R}
\newcommand{\PabsorbingTwo}{P^{\,\bm{\phi}}_R}

\newcommand{\PabsorbingDL}{\tilde{P}^{\,\bm{\varphi}}_\epsilon}
\newcommand{\PabsorbingDLdot}{\dot{\tilde{P}}^{\,\trajvec}_\epsilon}
\newcommand{\PabsorbingdotDL}{\PabsorbingDLdot}

\newcommand{\PabsorbingNormalizedDL}{\tilde{P}^{\,n,\varphi}_\epsilon}
\newcommand{\PabsorbingNormalizedvecDL}{\tilde{P}^{\,n,\trajvec}_\epsilon}

\newcommand{\setbar}{\bigl\vert}

\newcommand{\aexit}{\alpha_{R}^{\trajvec}}
\newcommand{\aexitDL}{\tilde{\alpha}}
\newcommand{\aexitfreeDL}{\tilde{\alpha}_{\mathrm{free}}}
\newcommand{\aexitzeroDL}{\tilde{\alpha}^{(0)}}
\newcommand{\aexittwoDL}{\tilde{\alpha}^{(2)}}

\newcommand{\PDLnoeps}{\tilde{P}}

\newcommand{\PDLzero}{\tilde{P}_i}
\newcommand{\Pzero}{{P}_i}
\newcommand{\Pinitial}{\Pzero}

\newcommand{\PnCoeffADL}{\tilde{N}_{\mathrm{s}}}
\newcommand{\PnCoeffBDL}{\tilde{N}_{\mathrm{c}}}

\newcommand{\pert}{\epsilon}

\newcommand{\td}{\tau_D}

\newcommand{\trel}{\tau_{\mathrm{rel}}}
\newcommand{\trelDL}{\tilde{\tau}_{\mathrm{rel}}}

\newcommand{\LFPDLapp}{\tilde{\mathcal{F}}_{\mathrm{app}}}

\newcommand{\evn}{\tilde{\lambda}_{n}}
\newcommand{\evm}{\tilde{\lambda}_{m}}

\newcommand{\eig}{\mathrm{eig}}
\newcommand{\evone}{\tilde{\lambda}_{1}}

\newcommand{\efn}{\tilde{\rho}_n}
\newcommand{\efninhom}{\tilde{\rho}_{n,\mathrm{inhom}}}
\newcommand{\efnhom}{\tilde{\rho}_{n,\mathrm{hom}}}
\newcommand{\efm}{\tilde{\rho}_m}
\newcommand{\efndot}{\dot{\tilde{\rho}}_n}
\newcommand{\efmdot}{\dot{\tilde{\rho}}_m}
\newcommand{\efone}{\tilde{\rho}_1}
\newcommand{\efonedot}{\dot{\tilde{\rho}}_1}

\newcommand{\bcoeffm}{\tilde{b}_{m}}
\newcommand{\bcoeffn}{\tilde{b}_{n}}
\newcommand{\bcoeffk}{\tilde{b}_{k}}
\newcommand{\bcoeffl}{\tilde{b}_{l}}

\newcommand{\bcoeffndot}{\dot{\tilde{b}}_{n}}

\newcommand{\bcoeffone}{\tilde{b}_{1}}

\newcommand{\acoeffm}{\tilde{a}_{m}}
\newcommand{\acoeffn}{\tilde{a}_{n}}

\newcommand{\acoeffndot}{\dot{\tilde{a}}_{n}}

\newcommand{\acoeffone}{\tilde{a}_{1}}

\newcommand{\Qs}{\tilde{Q}_{n,\mathrm{s}}}
\newcommand{\Qc}{\tilde{Q}_{n,\mathrm{c}}}

\newcommand{\Xns}{\tilde{A}_{n}}
\newcommand{\Xnc}{\tilde{B}_{n}}

\newcommand{\Ln}{\tilde{\Lambda}_n}
\newcommand{\Lone}{\tilde{\Lambda}_1}
\newcommand{\Lm}{\tilde{\Lambda}_m}
\newcommand{\Lmn}{\Delta \tilde{\Lambda}_{mn}}
\newcommand{\Lmk}{\Delta \tilde{\Lambda}_{mk}}
\newcommand{\Lmone}{\Delta \tilde{\Lambda}_{m1}}
\newcommand{\Lnone}{\Delta \tilde{\Lambda}_{n1}}
\newcommand{\Lnm}{\Delta \tilde{\Lambda}_{nm}}
\newcommand{\Cmat}{\tilde{C}}
\newcommand{\Cmatdot}{\dot{\tilde{C}}}

\newcommand{\Padn}{\tilde{\mathcal{P}}^{\mathrm{ad}}_{n}}
\newcommand{\Padm}{\tilde{\mathcal{P}}^{\mathrm{ad}}_{m}}

\newcommand{\Padone}{\tilde{\mathcal{P}}^{\mathrm{ad}}_{1}}
\newcommand{\Padmn}{\tilde{\mathcal{P}}^{\mathrm{ad}}_{mn}}
\newcommand{\Padnm}{\tilde{\mathcal{P}}^{\mathrm{ad}}_{nm}}
\newcommand{\Padkm}{\tilde{\mathcal{P}}^{\mathrm{ad}}_{km}}
\newcommand{\Padmk}{\tilde{\mathcal{P}}^{\mathrm{ad}}_{mk}}

\newcommand{\Padlk}{\tilde{\mathcal{P}}^{\mathrm{ad}}_{lk}}
\newcommand{\Padonen}{\tilde{\mathcal{P}}^{\mathrm{ad}}_{1n}}

\newcommand{\MMatrixOne}{\mathcal{M}^{(1)}}
\newcommand{\MMatrixTwo}{\mathcal{M}^{(2)}}

\newcommand{\EDL}{\tilde{E}}
\newcommand{\EDLvec}{\tilde{\bm{E}}}

\makeatother

\begin{document}
\title{Stochastic action for tubes: Connecting path probabilities to measurement}
\author{Julian Kappler}
\email{jk762@cam.ac.uk}
\affiliation{DAMTP, Centre for Mathematical Sciences, University of Cambridge, Wilberforce Road, Cambridge CB3 0WA, UK}
\author{Ronojoy Adhikari}
\affiliation{DAMTP, Centre for Mathematical Sciences, University of Cambridge, Wilberforce Road, Cambridge CB3 0WA, UK}
\date{\today}
\begin{abstract}
The trajectories of diffusion processes are continuous but non-differentiable,
and each occurs with vanishing probability. This introduces a gap between
theory, where path probabilities are used in many contexts, and experiment, 
where
only events with non-zero probability are measurable.
Here we bridge this gap
by considering the probability of diffusive trajectories to remain
within a tube of small but finite radius around a smooth path.
This probability can be measured in experiment,
via the rate at which trajectories exit  the tube
for the first time,
thereby establishing a link between path probabilities
 and physical observables. 
Considering $N$-dimensional overdamped
Langevin dynamics, we show that the tube probability can be
obtained theoretically from the solution of the Fokker-Planck equation.
Expressing
the resulting exit rate as a functional of the path and ordering it as a power
series in the tube radius, we identify the zeroth-order term as the
Onsager-Machlup stochastic action,
thereby elevating it from a mathematical construct to a physical observable.
 The higher-order terms reveal,
for the first time, the form of the finite-radius contributions which
account for fluctuations around the path. To demonstrate the experimental relevance
of this  action functional for tubes, we numerically sample trajectories
of Brownian motion in a double-well potential, compute their exit
rate, and show an excellent agreement with our analytical
results. Our work shows that smooth tubes are surrogates for non-differentiable
diffusive trajectories, and provide a direct way of comparing theoretical
results on single trajectories, such as
path-wise definitions of irreversibility,
 to measurement. 
\end{abstract}
\maketitle

\section{Introduction}

Stochastic effects are ubiquitous in physical systems, and are widely
modeled by diffusion processes \cite{risken_fokker-planck_1996,oksendal_stochastic_2007,kampen_stochastic_2007,gardiner_stochastic_2009}.
Physical examples include the motion of individual colloidal particles
\cite{perrin_mouvement_1909,haw_colloidal_2002,nelson_biological_2014,bera_fast_2017,caciagli_optical_2017},
the dynamics of polymers and proteins \cite{gebhardt_full_2010,konig_single-molecule_2015,amitai_polymer_2017,kappler_cyclization_2019},
or of active particles such as driven colloidal systems, cells, or
bacteria \cite{aranson_active_2013,bruckner_stochastic_2019}. Diffusion
processes are also employed beyond the physical sciences, for example
in quantitative finance \cite{friz_large_2015} or the dynamics of
ecosystems \cite{nolting_balls_2015}.

A fundamental concern in stochastic dynamics is to meaningfully
quantify the probability 
of 
a given trajectory. 
These probabilities fully characterize a given stochastic
dynamics and
are indispensable in applications.
For example, path-wise definitions of irreversibility as 
ratios of probabilities of forward- and time-reversed trajectories,
are central to the field of 
 stochastic thermodynamics \cite{seifert_entropy_2005,seifert_stochastic_2012}.
As a second example, reaction pathways between states, obtained from
the most probable path connecting them, are essential to the study of  rare events such as chemical reactions
or conformational changes in biomolecules \cite{e_string_2002,ren_transition_2005,e_transition_2005}.

For any diffusive dynamics,
as for example the
overdamped Langevin equation \cite{oksendal_stochastic_2007,kampen_stochastic_2007,gardiner_stochastic_2009},
which is the most widely used model for stochastic dynamics, the probability
of any single trajectory is zero. Consequently, over the last decades,
much work has been going into quantifying relative probabilities of
 Langevin paths \citep{onsager_fluctuations_1953,graham_path_1977,langouche_functional_1979,dekker_path_1980,weber_master_2017,cugliandolo_building_2018,stratonovich_probability_1971,horsthemke_onsager-machlup_1975,durr_onsager-machlup_1978,ito_probabilistic_1978,williams_probability_1981,fujita_onsager-machlup_1982,ikeda_stochastic_1989}.
However, because it is not possible to directly access experimentally
the ratio of two vanishingly small quantities, hitherto these theoretical
results could not be put to the experimental test. More generally,
the fact that a given individual stochastic trajectory occurs with
probability zero is the reason that no theoretical result pertaining
to individual stochastic trajectories can be checked directly in experiment.

\begin{figure}[ht]
\centering \includegraphics[width=1\columnwidth]{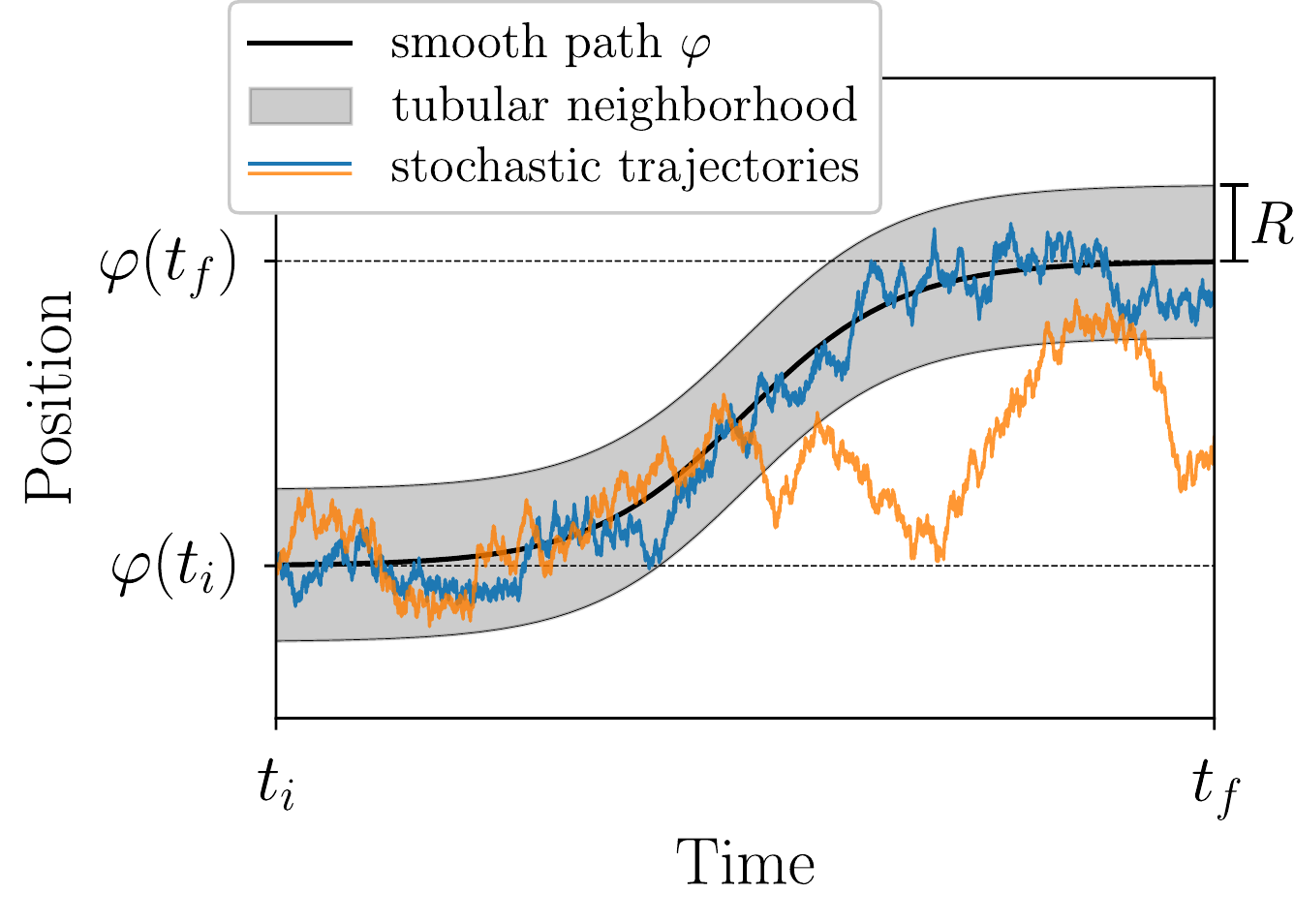} \caption{ \label{fig:illustration} 
For a one-dimensional system, a smooth path $\traj$ is shown as black
solid line, around which a tube of radius $R$ is indicated as grey
shaded area. Initial and final position of $\traj$ are shown as horizontal
dotted lines. While the blue trajectory is a realization of the Langevin
Eq.~\eqref{eq:Langevin_eq_intro} which stays inside the tube at
all times, the orange trajectory leaves the tube before the final
time $\tfinal$, and therefore contributes to the exit rate from the
tube. }
\end{figure}

We here overcome this limitation, by shifting the focus from individual
stochastic trajectories to the finite-radius tubular ensemble, comprised
of all stochastic trajectories that remain within a small but finite
threshold distance $R$ from a smooth reference path $\trajvec(t)$,
see Fig.~\ref{fig:illustration} for an illustration. 
 The name tubular
ensemble is motivated by the fact that this neighborhood around the
reference path is a tube in spacetime. 
The probability
to observe any of those stochastic trajectories, which is called the
sojourn probability, is nonzero, 
and can be  measured directly in experiment or simulation, simply by
counting which ratio of observed trajectories remains within the threshold
distance from the reference path until the final observation time.
Thus, considering this ensemble
yields a systematic approach to regularizing and connecting to experiment
the theoretical discussion of path probabilities, which are recovered as tubes
shrink to zero radius.
Importantly, our work elevates  stochastic actions,
 a widely used theoretical concept to quantify ratios of path probabilities,
to physical observables.
This allows, for the first time, the testing of theoretical 
results involving path probabilities directly in experiment.

The relevance of the tubular ensemble, however,  goes beyond serving as a 
bridge between theory and experiment. In physical applications, one is typically
not interested in a single path, but rather in a pathway, that is a family of trajectories
that remain within a threshold distance of a reference path. This is precisely the
family of trajectories that the tubular ensemble describes.

Our work establishes the tubular ensemble as a generalization of the
very concept of an individual stochastic trajectory, which allows
to connect to experiment or simulation any question related to individual
paths in systems subject to stochastic dynamics. For the overdamped
Langevin equation, we provide a conceptually simple derivation of
the sojourn probability. In the limit $R\rightarrow0$ we recover
the Onsager-Machlup (OM) stochastic action Lagrangian, which is known
to characterize relative path likelihoods \citep{onsager_fluctuations_1953,graham_path_1977,langouche_functional_1979,dekker_path_1980,weber_master_2017,cugliandolo_building_2018,stratonovich_probability_1971,horsthemke_onsager-machlup_1975,durr_onsager-machlup_1978,ito_probabilistic_1978,williams_probability_1981,fujita_onsager-machlup_1982,ikeda_stochastic_1989};
in particular, we show explicitly that this Lagrangian appears as
a contribution to the exit rate with which stochastic trajectories
first leave the tubular neighborhood around $\trajvec$. By calculating
the first radius-dependent corrections to the OM Lagrangian, we go
beyond single-trajectory asymptotics.

The remainder of this paper is organized as follows. In Sect.~\ref{sec:GeneralTheory}
we discuss our general theory for $N$-dimensional Langevin dynamics.
In Sect.~\ref{sec:one_dimensional} we illustrate our general results
by considering explicitly the special case of barrier crossing in
a one-dimensional system, $N=1$. We in particular show how our theoretical
predictions can be compared directly to observables from simulated
Langevin time series. We close in Sect.~\ref{sec:conclusions} by
summarizing our results, and discussing their further implications.

\section{Theory}

\label{sec:GeneralTheory}

We consider the overdamped Langevin equation, which for an $N$-dimensional
coordinate $\Xvec_{t}\equiv\Xvec(t)\equiv(X_{1}(t),...,X_{N}(t))$,
is given by 
\begin{equation}
\dot{\vecX}_{t}=D\beta\vecF(\vecX_{t},t)+\sqrt{2D}\noisevec_{t},\label{eq:Langevin_eq_intro}
\end{equation}
where $D=k_{\mathrm{B}}T/\gamma$ is the diffusivity, $\beta^{-1}=k_{\mathrm{B}}T$
is the inverse thermal energy with $k_{\mathrm{B}}$ the Boltzmann
constant and $T$ the temperature, $\gamma$ is the friction coefficient,
$\vecF(\vecx,t)$ is a deterministic, possibly time-dependent, force, and $\noisevec$ is Gaussian
white noise with vanishing mean and unit covariance matrix. We assume
that $D$ is position-independent, extension of our results to position-dependent
diffusivity is discussed in the conclusions.

\subsection{The tubular ensemble}

One approach to relative path likelihoods of overdamped Langevin dynamics
is to derive a formal path-integral representation of the propagator
associated with Eq.~\eqref{eq:Langevin_eq_intro}, and then to use
the resulting symbolic expression as a basis for relative path probabilities
\citep{onsager_fluctuations_1953,tisza_fluctuations_1957,graham_path_1977,langouche_functional_1979,dekker_path_1980,lau_state-dependent_2007,weber_master_2017,cugliandolo_building_2018}.
However, this approach suffers from ambiguities arising from the time-discretization
of the short-time propagator \citep{wissel_manifolds_1979,adib_stochastic_2008}.
In essence, the formal expression one obtains depends on which of
infinitely many time-discretization schemes one uses \citep{wissel_manifolds_1979};
while for most purposes these discretizations are equivalent, the
theoretically derived most probable path, which is sometimes thought of representing
the typical behavior of the dynamics,
 depends on the choice of
discretization \citep{adib_stochastic_2008}.

A different route towards quantifying relative path probabilities
is to consider the tubular ensemble, which consists of those realizations
$\Xvec_{t}$ of the Langevin Eq.~\eqref{eq:Langevin_eq_intro} that
stay inside a ball of radius $R$ with center a smooth reference path
$\trajvec(t)$, $t\in[\tzero,\tfinal]$ \citep{stratonovich_probability_1971,horsthemke_onsager-machlup_1975,durr_onsager-machlup_1978,ito_probabilistic_1978,williams_probability_1981,fujita_onsager-machlup_1982,zeitouni_onsager-machlup_1989,ikeda_stochastic_1989},
up to time $t\leq\tfinal$, 
\begin{equation}
\trajset(t)\equiv\left\{ \,\vecX~\setbar~||\vecX_{s}-\trajvec(s)||<R~~\forall~s\in[\tinitial,t]\,\right\} ,\label{eq:def_trajset}
\end{equation}
where $||\vecX||\equiv\sqrt{X_{1}^{2}+...+X_{N}^{2}}$; see Fig.~\ref{fig:illustration}
for an illustration of $\trajset$. We use the name tubular ensemble
for $\trajset$ because a ball with time-dependent center is a tube
in spacetime $(\xvec,t)$, c.f.~Fig.~\ref{fig:illustration}.

The corresponding sojourn probability 
\begin{equation}
\Pabsorbing(t)\equiv P\left(\vecX\in\trajset(t);\vecX_{\tinitial}\sim\Pinitial\right)\label{eq:sojourn_prob_intro}
\end{equation}
is the probability for a stochastic trajectory $\Xvec$ to remain
closer than a distance $R$ to $\trajvec$ until time $t$; for finite
$R$ this probability of course depends on the distribution of initial
positions $\Xvec_{\tinitial}\sim\Pinitial$ inside the tube. Because
the probability of any individual trajectory is zero for Langevin
dynamics, the sojourn probability vanishes as $R\rightarrow0$. The
relative probability for two reference paths $\trajvec$, $\bm{\trajTwo}$
can still be quantified by \citep{stratonovich_probability_1971,horsthemke_onsager-machlup_1975,durr_onsager-machlup_1978,ito_probabilistic_1978,williams_probability_1981,fujita_onsager-machlup_1982,ikeda_stochastic_1989}
\begin{equation}
\frac{e^{-S[\trajvec]}}{e^{-S[\bm{\trajTwo}]}}\equiv\lim_{R\rightarrow0}\frac{\Pabsorbing(\tfinal)}{\PabsorbingTwo(\tfinal)},\label{eq:action_difference_intro}
\end{equation}
where the stochastic action $S[\trajvec]$, which is a functional
of the smooth path $\trajvec$, is found to be 
\begin{equation}
S[\trajvec]=\int_{\tinitial}^{\tfinal}\mathrm{d}t~\LOM(\vectraj(t),\vectrajdot(t),t),\label{eq:OM_action_def}
\end{equation}
with the Onsager-Machlup (OM) Lagrangian 
\begin{equation}
\LOM(\vectraj,\vectrajdot)=\frac{1}{4D}\left[\vectrajdot-D\beta\vecF(\vectraj)\right]^{2}+\frac{1}{2}D\beta\bm{\nabla}\cdot\vecF(\vectraj).\label{eq:OM_intro}
\end{equation}
The literature concerned with deriving Eq.~\eqref{eq:OM_intro} via
the ensemble Eq.~\eqref{eq:def_trajset} is rather technical \citep{stratonovich_probability_1971,horsthemke_onsager-machlup_1975,durr_onsager-machlup_1978,ito_probabilistic_1978,williams_probability_1981,fujita_onsager-machlup_1982},
and is focused on the tubular ensemble in the singular single-trajectory
limit $R\rightarrow0$.

The key difference between the previous literature and our derivation,
is that, instead of working directly with the Langevin Eq.~\eqref{eq:Langevin_eq_intro},
we consider the equivalent description of the stochastic process inside
the tube via the Fokker-Planck equation (FPE) \citep{kampen_stochastic_2007,gardiner_stochastic_2009}
\begin{equation}
\partial_{t}\Pabsorbing(\vecx,t)=-\bm{\nabla}\cdot\left[D\beta\vecF(\vecx,t)\Pabsorbing(\vecx,t)\right]+\laplacevec\left[D\Pabsorbing(\vecx,t)\right],\label{eq:1D_FP_eq}
\end{equation}
with a time-dependent spatial domain given at time $t$ by 
\begin{equation}
\vecx\in\BR(t)\equiv\left\{ \,\vecx~\setbar~||\vecx-\vectraj(t)||<R\right\} \label{eq:def_B}
\end{equation}
as illustrated by the grey shaded area in Fig.~\ref{fig:illustration},
and subject to absorbing boundary conditions at the tube boundary,
\begin{equation}
\label{eq:absorbing_bc_main}
\Pabsorbing(\xvec,t)=0  \qquad \forall \vecx\in\BR(t),
\end{equation}
 so that $\Pabsorbing(\xvec,t)$
describes the distribution of those particles that have never left
the tube until time $t$. Once Eq.~\eqref{eq:1D_FP_eq} is solved
for given initial condition $\Xvec_{\tinitial}\sim\Pzero$, the sojourn
probability up to time $t$ is simply the survival probability 
\begin{equation}
\Pabsorbing(t)=\int_{\BR(t)}\mathrm{d}^{N}\vecx~\Pabsorbing(\vecx,t),\label{eq:sojourn_prob_FP_sol_intro}
\end{equation}
where here and in the following we suppress the dependence on the
initial condition $\Pzero$ unless it is relevant for the discussion.
From Eq.~\eqref{eq:sojourn_prob_FP_sol_intro} in turn we obtain
the instantaneous exit rate $\aexit(t)$ at which stochastic trajectories
leave the tube for the first time, defined by 
\begin{equation}
\Pabsorbing(t)=\exp\left[-\int_{\tinitial}^{t}\mathrm{d}s~\aexit(s)\right].\label{eq:aexit_def_intro}
\end{equation}
As we show in the following subsections, this yields 
\begin{align}
\aexit(t) & =\frac{D\evone^{(0)}}{R^{2}}+\LOM(\trajvec(t),\trajdotvec(t),t)\label{eq:axit_result_intro} \\
 & \qquad+R^{2}\Ltwo(\trajvec(t),\trajdotvec(t),\trajddotvec(t),t) 
+\mathcal{O}(R^{4}),
\nonumber 
\end{align}
where 
\begin{equation}
\alpha_{\mathrm{free}}\equiv\frac{D\evone^{(0)}}{R^{2}}
\end{equation}
is the free-diffusion steady-state exit rate out of a ball of radius
$R$, with $\evone^{(0)}$ the negative of the eigenvalue with the smallest absolute value
 of the Laplace
operator on the unit ball $B_{1}$ with absorbing boundary conditions,
$\LOM$ is the OM Lagrangian defined in Eq.~\eqref{eq:OM_intro},
and $\Ltwo$
is a  quadratic correction to
the exit rate, which we calculate in this work. 
According to Eq.~\eqref{eq:axit_result_intro}, for small radius $R$ the
exit rate is dominated by free diffusion.
The OM Lagrangian is the first correction to freely diffusive exit
from the tube,
and with $\Ltwo$
we include finite-radius effects beyond OM theory.
Our derivation directly relates $\LOM$ to an experimentally
measurable exit rate from a fictitious tube around a smooth
reference path $\trajvec$; 
despite the appearance of the term $\alpha_{\mathrm{free}}$
in the mathematical literature on the subject \citep{williams_probability_1981,fujita_onsager-machlup_1982},
this connection between stochastic action and a physical exit rate
has not been made explicit before.

In the following subsections we discuss our general theory, outlined
just above, for $N$-dimensional Langevin dynamics. In Sect.~\ref{sec:perturbative_solution_vec}
we derive a perturbative expression for the propagator of the FPE, 
Eq.~\eqref{eq:1D_FP_eq}, with absorbing boundary conditions. 
Based on this propagator, we in Sect.~\ref{sec:exit_rate_vec}
calculate the instantaneous exit rate, defined in Eq.~\eqref{eq:aexit_def_intro},
as a power series in the tube radius $R$, which finally leads to
Eq.~\eqref{eq:axit_result_intro}.

\subsection{Perturbative solution of FPE in tube interior}

\label{sec:perturbative_solution_vec}

\textit{FPE in dimensionless streaming coordinates.} To eliminate the
time-dependence of the spatial domain Eq.~\eqref{eq:def_B}, we introduce
the dimensionless streaming variables 
\begin{align}
\tDL(t) & \equiv\frac{t}{\td}, & \vecxDL(\xvec,t) & \equiv\frac{\vecx-\vectraj(t)}{R},\label{eq:def_tDL}
\end{align}
where $\td\equiv{L^{2}}/{D}$ is the time scale on which a particle
diffuses over the typical length scale $L$ of the external force
$\Fvec$. The domain for $\vecxDL$ is then independent of time and
given by the unit ball, 
\begin{equation}
\vecxDL\in\BDL\equiv\left\{ \,\vecxDL~\setbar~||\vecxDL||<1\,\right\} .\label{eq:def_BDL}
\end{equation}
We furthermore define a dimensionless probability density $\rDL$,
dimensionless force $\FvecDL$, and a dimensionless path $\trajvecDL$
as 
\begin{align}
\PabsorbingDL(\xvecDL,\tDL) & \equiv R^{N}\Pabsorbing\left(\xvec,t\right),\label{eq:def_ProbAbsorbingDL}\\
\vecFDL(\xvecDL,\tDL) & \equiv L\beta\vecF\left(\vecx,t\right),\label{eq:def_FDL}\\
\vectrajDL(\tDL) & \equiv\vectraj(t)/L,\label{eq:def_trajDL}
\end{align}
where $(\vecx,t)$ and $(\vecxDL,\tDL)$ are related as defined in
Eq.~\eqref{eq:def_tDL}. 
Here and below, dimensionless quantities  are always indicated by a tilde.
In dimensionless form the FPE,
Eq.~\eqref{eq:1D_FP_eq}, becomes 
\begin{align}
\pert^{2}\partial_{\tDL}\PabsorbingDL & =\LFPDLapp\PabsorbingDL,\label{eq:1D_FP_eq_DL}
\end{align}
with the dimensionless tube radius 
\begin{equation}
\pert\equiv\frac{R}{L},\label{eq:def_pert}
\end{equation}
and the dimensionless apparent Fokker-Planck (FP) operator $\LFPDLapp$,
given by 
\begin{equation}
\LFPDLapp\PabsorbingDL\equiv-\epsilon\gradvecDL\cdot\left[\left(\FvecDL-\vectrajdotDL\right)\PabsorbingDL\right]+\laplaceDLvec\PabsorbingDL,\label{eq:1D_FPO_DL}
\end{equation}
where $\gradvecDL$ denotes the gradient with respect to $\xDL$ with components
$\tilde{\nabla}_{j}\equiv \partial/\partial \xDL_j$, and where $\trajdotvecDL\equiv\partial_{\tDL}\trajvecDL$.
A dot over a function in dimensionless (dimensionful) form always signifies a derivative with respect 
to  dimensionless (dimensionful) time. For example, $\trajdotvec=L/\td\,\trajdotvecDL$.
Dots are used interchangeably with the symbols $\partial_{t}$, $\partial_{\tDL}$.
As can be seen directly from Eq.~\eqref{eq:1D_FPO_DL},
with respect to the coordinate system $(\xvecDL,\tDL)$, the velocity
of the path $\trajvec$ acts as a fictitious spatially constant force
inside the tube, so that we obtain an apparent total force 
\begin{equation}
\FDLeffvec=\FDLvec-\trajdotDLvec,\label{eq:effective_total_force}
\end{equation}
which is why we call $\LFPDLapp$ the apparent dimensionless FP
operator. In dimensionless streaming coordinates, the time-depedendent absorbing boundary condition, Eq.~\eqref{eq:absorbing_bc_main}, becomes
 \begin{equation}
 \PabsorbingDL(\xDLvec,\tDL)=0\qquad
\forall~||\xDLvec||=1,
\end{equation}
which is independent
of time.
This is  the principal advantage of transforming to streaming coordinates. 

\textit{FPE in terms of the instantaneous eigenbasis.}
We expand the probability distribution $\PabsorbingDL$ in Eq.~\eqref{eq:1D_FP_eq_DL}
in terms of the instantaneous FP eigenstates $\efn(\xDLvec,\tDL)$
as 
\begin{equation}
\PabsorbingDL(\xDLvec,\tDL)=\sum_{m=1}^{\infty}\acoeffm(\tDL)\efm(\xDLvec,\tDL).\label{eq:1D_rho_expansion}
\end{equation}
At time $\tDL$ the eigenvalues $-\evn(\tDL)$ and eigenfunctions
$\efn(\xDLvec,\tDL)$ of the apparent dimensionless FP
operator $\LFPDLapp(\tDL)$ fulfill the eigenvalue equation 
\begin{equation}
\LFPDLapp(\tDL)\efn(\xDLvec,\tDL)=-\evn(\tDL)\efn(\xDLvec,\tDL)\label{eq:1D_FP_eq_DL_linearized_ev}
\end{equation}
and the absorbing boundary conditions $\efn(\xDLvec,\tDL)=0$ for
$||\xDLvec||=1$. We assume the eigenvalues to be ordered, i.e.~$\evn\leq\evm$
 for $n<m$, and due to the absorbing boundary condition
we have $\evone>0$. We assume that at any time $\tDL$ there exists
a steady-state solution $\reqDL(\xDLvec,\tDL)$ of Eq.~\eqref{eq:1D_FP_eq_DL}
with reflecting boundary conditions at $\BDL$; we do not require
$\reqDL$ to be normalized. Using this instantaneous steady-state
we introduce the instantaneous inner product 
\begin{equation}
\langle f,g\rangle\equiv\int_{\BDL}\mathrm{d}^{N}\xDLvec~f(\xDLvec)g(\xDLvec)/\reqDL(\xDLvec,\tDL).\label{eq:def_inner_product}
\end{equation}
With respect to this inner product, the FP operator $\LFPDLapp$
is self-adjoint so that the absorbing-boundary eigenfunctions $\efn$
can be chosen orthogonal at each time $\tDL$ \cite{gardiner_stochastic_2009}.
If at any time $\tDL$ the force $\vecF(\xvec,t)$ inside the domain
$\BR(t)$ originates from a potential $U(\vecx,t)$, such that $\Fvec=-\gradvec U$,
then the instantaneous steady-state solution is given by 
\begin{equation}
\reqDL(\xDLvec,\tDL)=\exp\left[-\epsilon\,\tilde{U}(\vecxDL,\tDL)-\epsilon\,\vecxDL\cdot\trajdotDLvec\right],\label{eq:reqDL}
\end{equation}
where $\tilde{U}(\vecxDL,\tDL)\equiv\beta U(x,t)$, and the dot indicates
the standard Euclidean inner product on $\mathbb{R}^{N}$. We emphasize
that Eq.~\eqref{eq:reqDL} does not require a global potential for
$\Fvec$, but only a local potential inside the ball $\BR(t)$. If
such a local potential does not exist, the instantaneous non-equilibrium
steady state $\reqDL$ has to be determined by other means \citep{bouchet_perturbative_2016}.

Expanding the probability distribution $\PabsorbingDL$ in Eq.~\eqref{eq:1D_FP_eq_DL}
in terms of the instantaneous FP eigenstates as given by
Eq.~\eqref{eq:1D_rho_expansion}, and projecting the equation onto
$\efn$ using the inner product Eq.~\eqref{eq:def_inner_product},
yields 
\begin{equation}
-{\acoeffndot}=\frac{\evn}{\pert^{2}}\acoeffn+\sum_{m=1}^{\infty}\frac{\langle\efn,\efmdot\rangle}{\langle\efn,\efn\rangle}\acoeffm,\label{eq:1D_FP_in_eigenbasis}
\end{equation}
where $n\in\mathbb{N}$ and a dot here denotes a derivative with respect
to $\tDL$. Because the apparent FP operator is time-dependent,
both the eigenvalues $\evn$ and the inner products $\langle\efn,\efmdot\rangle$,
$\langle\efn,\efn\rangle$, are functions of $\tDL$. The FPE,
Eq.~\eqref{eq:1D_FP_eq}, with absorbing boundary conditions is equivalent
to Eq.~\eqref{eq:1D_FP_in_eigenbasis}; once the latter is solved,
the dimensionless probability density inside the tube is obtained
from Eq.~\eqref{eq:1D_rho_expansion}, which can be recast in physical
units using Eq.~\eqref{eq:def_ProbAbsorbingDL}.

Since $\LFPDLapp$ depends on $\epsilon$, so do the quantities $\evn$,
$\langle\efn,\efmdot\rangle$, $\langle\efn,\efn\rangle$, which appear
in Eq.~\eqref{eq:1D_FP_in_eigenbasis}. From Eq.~\eqref{eq:1D_FPO_DL}
it is apparent that the ratio of the drift to the diffusion is of order $\epsilon$ 
and, therefore, to lowest order the spectrum is that of a free diffusion inside
a unit ball. The eigenvalues, eigenfunctions, and steady-state distributions 
are independent of $\tDL$ at this order, and therefore, any time-dependence
of the eigenfunctions must be at least of order $\epsilon$. This implies that the
ratio of the off-diagonal to diagonal  terms in Eq.~\eqref{eq:1D_FP_in_eigenbasis}
is at least of order $\epsilon^3$.
Thus, mode-coupling effects are sub-dominant and the uncoupled dynamics provides
a good first approximation for small $\epsilon$.  
In the context of time-dependent perturbation theory in quantum
mechanics, this is known as the adiabatic approximation \citep{ballentine_quantum_2010}.

\textit{Perturbative calculation of the instantaneous FP
spectrum.} 
In App.~\ref{app:perturbative_calculations_N}, we
discuss in detail the calculation of both the instantaneous eigenvalues
and eigenfunctions as perturbation series in $\epsilon$, 
\begin{align}
\evn & =\sum_{k=0}^{\infty}\epsilon^{k}\evn^{(k)}, & \efn & =\sum_{k=0}^{\infty}\epsilon^{k}\efn^{(k)},\label{eq:evn_power_series}
\end{align}
and calculate explicit expressions for the eigenvalues $\evn$ to
order $\epsilon^{3}$, and for the eigenfunctions $\efn$ to order
$\epsilon$. For $n=1$, and if the force $\Fvec$ inside the tube
is given by a potential also for $n>1$, we furthermore calculate
explicitly the contribution $\efn^{(2)}$ at order $\epsilon^{2}$.

\textit{Perturbative solution of the FPE.} In App.~\ref{app:perturbative_solution_FP}
we in detail derive a solution to Eq.~\eqref{eq:1D_FP_in_eigenbasis},
given by
\begin{align}
\acoeffone(\tDL) & \approx\exp\left[-\dfrac{1}{\epsilon^{2}}\int_{\tzeroDL}^{\tDL}\mathrm{d}\tDLdummyone~\Lone(\tDLdummyone)\right]\label{eq:acoeffone_final_result}\\
 & \quad\times\left[\acoeffone(\tzeroDL)-\epsilon^{2}\sum_{m=2}^{\infty}\left.\frac{\langle\efone,\efmdot\rangle}{\langle\efone,\efone\rangle}\right|_{\tzeroDL}\frac{\acoeffm(\tzeroDL)}{\Lmone(\tzeroDL)}+\mathcal{O}(\epsilon^{k})\right],\nonumber \\
\acoeffn(\tDL) & \approx-\epsilon^{2}\left.\frac{\langle\efn,\efonedot\rangle}{\langle\efn,\efn\rangle}\right|_{\tDL}\frac{\acoeffone(\tDL)}{\Lnone(\tDL)}+\mathcal{O}(\epsilon^{k}),\label{eq:acoeffn_final_result_2}
\end{align}
where $n>1$ in Eq.~\eqref{eq:acoeffn_final_result_2}, for a one-dimensional
system $N=1$ we have $k=6$ and for $N\geq2$ we have $k=5$, and
where we define 
\begin{align}
\Ln & \equiv\evn+\epsilon^{2}\frac{\langle\efn,\efndot\rangle}{\langle\efn,\efn\rangle},\label{eq:def_Lambda_n_main_text}\\
\Lmn & \equiv\Lm-\Ln.\label{eq:def_Lmn_main_text}
\end{align}
The solution Eqs.~\eqref{eq:acoeffone_final_result}, \eqref{eq:acoeffn_final_result_2},
is valid after an initial transient time, i.e.~for 
\begin{equation}
\tDL-\tzeroDL\gtrsim\trelDL\equiv\frac{\epsilon^{2}}{\Delta\Lambda_{21}},\label{eq:trel_def}
\end{equation}
and neglects terms that are exponentially small as compared to Eqs.~\eqref{eq:acoeffone_final_result},
\eqref{eq:acoeffn_final_result_2}.

The form of Eqs.~\eqref{eq:acoeffone_final_result},
\eqref{eq:acoeffn_final_result_2} 
allows for an intuitive interpretation. 
Initially all eigenmodes are excited, with their respective
amplitude $\acoeffn(\tzeroDL)$ determined by the initial condition.
The dynamics of each mode is dominated by the adiabatic exponential
decay, and 
after an initial relaxation time the mode $n=1$ (which decays slowest)
 dominates the probability distribution Eq.~\eqref{eq:1D_rho_expansion};
this is represented by the first term in the bracket in Eq.~\eqref{eq:acoeffone_final_result}.
The leading-order effect of the mode coupling is twofold. 
First, during their initial decay the modes $n>1$ can
transfer some of their initial amplitude $\acoeffn(\tzeroDL)$ to
the $n=1$ mode, as described by the second term in the bracket in 
Eq.~\eqref{eq:acoeffone_final_result}. 
Second, after their initial decay the $n>1$ modes
can be excited instantaneously by the lowest mode $n=1$,
as described by Eq.~\eqref{eq:acoeffn_final_result_2}.

For a particle initially localized at $\xveczeroDL$, we have a delta-peak
initial condition, $\PabsorbingDL(\xvecDL,\tzeroDL)\equiv\PDLzero(\xvecDL)=\delta(\xvecDL-\xveczeroDL)$,
so that the initial amplitude of the $n$-th mode is given by 
\begin{equation}
\acoeffn(\tzeroDL)=\left.\frac{\langle\PabsorbingDL,\efn\rangle}{\langle\efn,\efn\rangle}\right|_{\tzeroDL}=\frac{\efn(\xveczeroDL,\tzeroDL)}{\reqDL(\xveczeroDL,\tzeroDL){\langle\efn,\efn\rangle|_{\tzeroDL}}}.
\end{equation}
Substituting the resulting coefficients Eq.~\eqref{eq:acoeffone_final_result},
\eqref{eq:acoeffn_final_result_2}, into the eigenmode expansion Eq.~\eqref{eq:1D_rho_expansion}
of the propagator then yields 
\begin{widetext}
\begin{align}
\label{eq:approximate_propagator_steady}
	\PabsorbingDL(\,\xDLvec,\tDL~\setbar~\xzeroDLvec,\tzeroDL\,) &=
	 \exp\left[ - \dfrac{1}{\epsilon^2}\int_{\tzeroDL}^{\tDL}\mathrm{d}\tDLdummyone~\Lone(\tDLdummyone) \right] 
	 \,
	 \frac{1}{\reqDL(\xzeroDLvec,\tzeroDL) {\langle \efone,\efone \rangle|_{\tzeroDL}} } \\
\nonumber
	 &\quad \times
 \left[ \efone(\xDLvec,\tDL) -\epsilon^2 \sum_{m=2}^{\infty}
   \dfrac{1}{\Lmone(\tDL)} \left.\frac{\langle \efm,\efonedot\rangle}{\langle \efm,\efm\rangle}\right|_{\tDL}\, \efm(\xDLvec,\tDL) \right] 
   %
 \left[ \efone(\xzeroDLvec,\tzeroDL) -\epsilon^2 \sum_{m=2}^{\infty}
   \dfrac{1}{\Lmone(\tzeroDL)}
   \left.\frac{\langle \efone,\efmdot\rangle}{\langle \efm,\efm\rangle}\right|_{\tzeroDL}\, \efm(\xzeroDLvec,\tzeroDL) \right] \\
\nonumber
   	 &\qquad 
   + \mathcal{O}(\epsilon^k),
\end{align}
\end{widetext} 
where $k=6$ for a one-dimensional system, $N=1$, and $k=5$ for
$N\geq2$. Equation \eqref{eq:approximate_propagator_steady} is an
approximate solution to the FPE, Eq.~\eqref{eq:1D_FP_eq_DL},
valid after an initial decay time $\trelDL$ defined in Eq.~\eqref{eq:trel_def}.
With the definitions Eqs.~\eqref{eq:def_Lambda_n_main_text}, \eqref{eq:def_Lmn_main_text},
the propagator Eq.~\eqref{eq:approximate_propagator_steady} is fully
expressed in terms of the instantaneous eigenvalues and eigenvectors
of the FP operator. Note that Eq.~\eqref{eq:approximate_propagator_steady}
is factorized into a part that only depends on $(\xvecDL,\tDL)$,
and a part that only depends on $(\xveczeroDL,\tzeroDL)$; Thus, while
the total probability to have remained inside the tube until time
$\tDL$ is affected by the initial condition, after the initial relaxation
time $\trelDL$ the spatial probability distribution inside the tube
is independent of the initial condition.

Using Eq.~\eqref{eq:approximate_propagator_steady}, we can express
the solution for an arbitrary initial distribution $\PDLzero$ inside
the tube as 
\begin{equation}
\PabsorbingDL(\xvecDL,\tDL~\setbar~\XvecDL_{\tinitialDL}\sim\PDLzero)=\int_{\BDL}\mathrm{d}^{N}\xveczeroDL~\PabsorbingDL(\xvecDL,\tDL~\setbar~\xveczeroDL,\tzeroDL)\PDLzero(\xveczeroDL),\label{eq:approximate_FP_solution_steady}
\end{equation}
from which the survival probability,
Eq.~\eqref{eq:sojourn_prob_FP_sol_intro},
 follows in dimensionless form as
\begin{equation}
\PabsorbingDL(\tDL~\setbar~\XvecDL_{\tinitialDL}\sim\PDLzero)=\int_{\BDL}\mathrm{d}^{N}\xvecDL\,
\PabsorbingDL(\xvecDL,\tDL~\setbar~\XvecDL_{\tinitialDL}\sim\PDLzero).\label{eq:survival_probability_formula}
\end{equation}
Complementary to the survival probability is the normalized probability
density $\PabsorbingNormalizedDL$ inside the tube at any time $\tDL$,
defined as 
\begin{align}
\PabsorbingNormalizedvecDL(\xvecDL,\tDL) & \equiv\frac{\PabsorbingDL(\,\xvecDL,\tDL~\setbar~\XvecDL_{\tinitialDL}\sim\PDLzero\,)}{\int_{\BDL}\mathrm{d}^{N}\xvecDL'~\PabsorbingDL(\,\xvecDL',\tDL~\setbar~\XvecDL_{\tinitialDL}\sim\PDLzero\,)},\label{eq:approximate_normalized_density}
\end{align}
which describes the distribution inside the tube of those particles
that have stayed until the current time $\tDL$.
Using Eqs.~\eqref{eq:approximate_propagator_steady}, \eqref{eq:approximate_FP_solution_steady},
the distribution Eq.~\eqref{eq:approximate_normalized_density} can be 
shown to be independent of $\PDLzero$.

\subsection{Exit rate from tube}

\label{sec:exit_rate_vec}

For a particle starting at time $\tzero$ according to a distribution
$\Xveczero\sim\Pzero$ inside the tube, the instantaneous exit rate
is given by 
\begin{equation}
\aexit(t)=-\frac{\Pabsorbingdot(t)}{\Pabsorbing(t)},\label{eq:exit_rate_with_dimensions}
\end{equation}
where $\Pabsorbing(t)\equiv\Pabsorbing(\,t\mid\Xveczero\sim\Pzero\,)$
is the survival probability defined in Eq.~\eqref{eq:sojourn_prob_FP_sol_intro}.
Using Eq.~\eqref{eq:def_tDL},
\eqref{eq:def_ProbAbsorbingDL}, the dimensionless instantaneous exit
rate Eq.~\eqref{eq:exit_rate_with_dimensions} is defined as 
\begin{equation}
\aexitDL_{\epsilon}^{\trajvec}(\tDL)\equiv\td\,\aexit(t)=-\frac{\PabsorbingdotDL(\tDL)}{\PabsorbingDL(\tDL)}\label{eq:exit_rate_DL}
\end{equation}
where the dot denotes a derivative with respect to $\tDL$, and $\PabsorbingDL(\tDL)\equiv\PabsorbingDL(\,\tDL\mid\XzerovecDL\sim\PDLzero\,)$
is the survival probability in dimensionless form, with $\PDLzero(\xvecDL)=R^{N}\Pzero(\xvec)$.
Using the steady-state FP solution Eqs.~(\ref{eq:approximate_propagator_steady}-\ref{eq:survival_probability_formula}),
the exit rate Eq.~\eqref{eq:exit_rate_DL} is evaluated to yield
\begin{align}
\aexitDL_{\epsilon}^{\trajvec}(\tDL) & =
\frac{\evone}{\epsilon^2}+\frac{\langle\efone,\efonedot\rangle}{\langle\efone,\efone\rangle}
-
\frac{ \dot{\Integral}_1}{\Integral_1}
+ \mathcal{O}(\epsilon^4),
\label{eq:exit_rate_intermediate_expression}
\end{align}
with
\begin{equation}
\Integral_{n}(\tDL)\equiv\int_{\BDL}\mathrm{d}^{N}\xDLvec~\efn(\xvecDL,\tDL),\label{eq:definition_integral_n}
\end{equation}
and where we used that $\langle \efm,\efonedot\rangle \Integral_m$
is of order $\epsilon^2$, c.f.~App.~\ref{app:power_series_properties_parity}.

Equation \eqref{eq:exit_rate_intermediate_expression}, which is valid
after the initial transient decay time $\trelDL$ defined in Eq.~\eqref{eq:trel_def},
is independent of the initial distribution $\PDLzero$; this is because
in Eq.~\eqref{eq:approximate_propagator_steady} the initial condition
only contributes an overall prefactor independent of $(\xvecDL,\tDL)$,
which does not affect the relative change of particles inside the
tube quantified by Eq.~\eqref{eq:exit_rate_DL}. With Eq.~\eqref{eq:exit_rate_intermediate_expression}
the instantaneous exit rate is expressed solely in terms of the instantaneous
FP spectrum inside the tube. Expanding the quantities that
appear in Eq.~\eqref{eq:exit_rate_intermediate_expression} in powers
of $\epsilon$, and using the symmetry properties of these quantities,
 c.f.~App.~\ref{app:perturbative_calculations_N},
a power series expansion of the exit rate is obtained as 
\begin{align}
\aexitDL_{\epsilon}^{\trajvec}=\frac{\evone^{(0)}}{\epsilon^{2}}+\aexitDL^{(0)}+\epsilon^{2}\aexitDL^{(2)}+\mathcal{O}(\epsilon^{4}),\label{eq:exit_rate_Ndim_expansion_DL}
\end{align}
where
\begin{align}
\aexitfreeDL & =\frac{\evone^{(0)}}{\epsilon^{2}},\label{eq:exit_rate_Ndim_free}\\
\aexitDL^{(0)} & =\evone^{(2)}=\td\LOM\label{eq:exit_rate_Ndim_0}\\
\aexitDL^{(2)} & =\evone^{(4)}+\frac{\langle\efone,\efonedot\rangle^{(2)}}{\langle\efone,\efone\rangle^{(0)}}-\frac{\dot{\Integral}_{1}^{(2)}}{\Integral_{1}^{(0)}},\label{eq:exit_rate_Ndim_2}
\end{align}
where at Eq.~\eqref{eq:exit_rate_Ndim_0} we use the perturbative
result for $\evn^{(2)}$, c.f.~App.~\ref{app:perturbative_calculations_N},
the definition of the OM Lagrangian $\LOM$ is given in Eq.~\eqref{eq:OM_intro},
and where
\begin{equation}
\Integral_{n}^{(k)}(\tDL)\equiv \int_{\BDL}\mathrm{d}^{N}\xDLvec~\efn^{(k)}(\xDLvec,\tDL).\label{eq:int_rhon_series}
\end{equation}
Note that we suppress
the dependence on $\trajvec$ in the notation of the $\aexitDL^{(k)}$,
and that $\aexitfreeDL$ is independent of $\trajvec$.

Using Eq.~\eqref{eq:exit_rate_DL} the exit rate in physical units
can be obtained from Eqs.~(\ref{eq:exit_rate_Ndim_expansion_DL}-\ref{eq:exit_rate_Ndim_2});
note that according to Eq.~\eqref{eq:exit_rate_DL}, a scaling $\epsilon^{k}$
in $\aexitDL_{\epsilon}^{\trajvec}$ (dimensionless form) translates
to a scaling $R^{k}$ in $\aexit$ (physical units). The order-$\epsilon^2$
term in Eq.~\eqref{eq:axit_result_intro} is thus given by $\Ltwo=\aexitDL^{(2)}/(\td L^{2})$; 
according to Eq.~\eqref{eq:1D_FPO_DL}
the instantaneous FP spectrum depends on $(\trajvec,\dot{\trajvec})$;
because of the additional time derivative in Eq.~\eqref{eq:exit_rate_Ndim_2},
 the term $\Ltwo$ additionally
depends on $\ddot{\trajvec}$.

Equations (\ref{eq:exit_rate_Ndim_expansion_DL}-\ref{eq:exit_rate_Ndim_2}),
which express the exit rate $\aexitDL_{\epsilon}^{\trajvec}$ fully in terms of the perturbative
spectrum of the FP operator inside the tube, are one of
the main results of this paper. The equations show that for small
tube radius $\epsilon\ll1$, the exit from the tube is dominated by
the steady-state free-diffusion exit rate given by Eq.~\eqref{eq:exit_rate_Ndim_free};
this is consistent with the fact that the Langevin Eq.~\eqref{eq:Langevin_eq_intro}
is on short times dominated by the noise term $\noisevec$ (as opposed
to the deterministic force $\Fvec$). From Eq.~\eqref{eq:exit_rate_Ndim_expansion_DL}
we see that the free-diffusion exit rate in fact diverges as $1/\epsilon^{2}$,
which gives a physical picture as to why the probability for observing
the single path $\trajvec$ is zero.

According to Eqs.~\eqref{eq:exit_rate_Ndim_expansion_DL}, \eqref{eq:exit_rate_Ndim_0},
the first correction to the free-diffusion exit rate, which occurs
at order $\epsilon^{0}$, is given by the OM Lagrangian $\LOM$; this
establishes a direct link between $\LOM$ and the physical observable
$\aexit$. The next correction Eq.~\eqref{eq:exit_rate_Ndim_2},
which is quadratic in the tube radius, is still in the adiabatic limit,
meaning that only the $n=1$ eigenvalue and eigenfunction appear in
Eq.~\eqref{eq:exit_rate_Ndim_2}.

\section{one-dimensional systems and numerical experiments}

\label{sec:one_dimensional}

In the present section we consider the special case of a one-dimensional
system, $N=1$, for which it is straightforward to calculate explicit
expressions for the results derived in Sect.~\ref{sec:GeneralTheory}.
To illustrate and verify our perturbative analytical results, we compare
to numerical simulations throughout; in Sect.~\ref{sec:one_dim_intro}
we introduce the corresponding example system, a double-well system
with a barrier-crossing transition path $\traj$. While in Sect.~\ref{sec:example_solution}
we discuss the normalized probability density inside the tube, we
in Sect.~\ref{sec:ExitRate} consider the exit rate.

\subsection{Model}

\label{sec:one_dim_intro}

\begin{figure}[ht]
\centering \includegraphics[width=1\columnwidth]{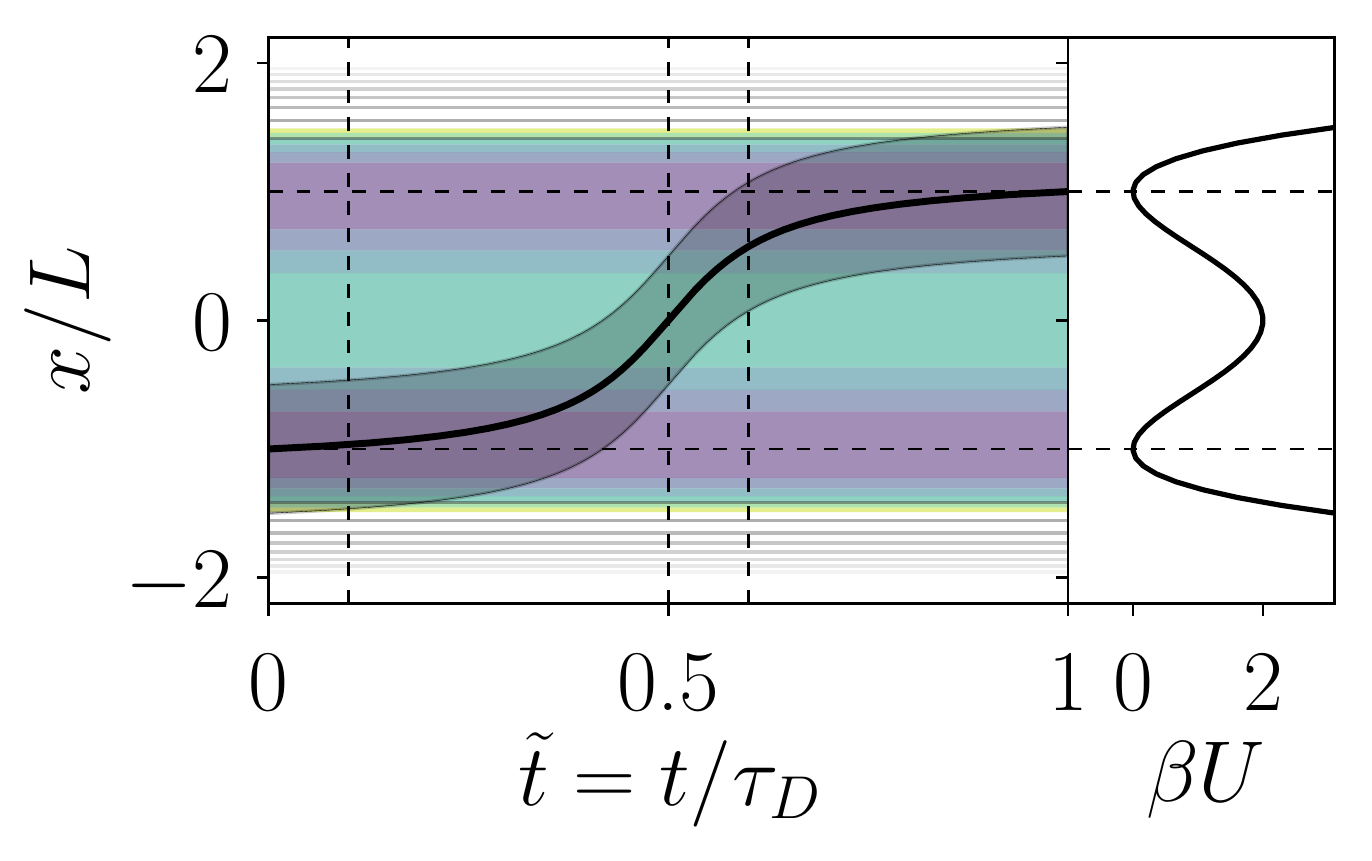} \caption{ \label{fig:potential_and_trajectory} 
{Potential and path
considered in the numerical examples in Sect.~\ref{sec:one_dimensional}.}
The plot on the right-hand side shows the quartic double well potential
Eq.~\eqref{eq:U_quartic} for barrier height $\beta U_{0}=2$. In
the plot on the left-hand side, the potential is shown as colormap
in the background, with the two minima of the potential represented
by horizontal dashed lines. The reference path $\traj$, defined in
Eq.~\eqref{eq:def_traj}, is shown as solid black line. Around the
reference path, a tube of radius $\epsilon=R/L=0.5$ is depicted by a shaded
grey region. The vertical dashed lines denote the times $\tDL=0.1$,
$0.5$, $0.6$, which are considered in Fig.~\ref{fig:densities_and_rates}. }
\end{figure}

For a length scale $L$ and a time scale $T$, we consider the Langevin
Eq.~\eqref{eq:Langevin_eq_intro} with a diffusion coefficient $D=L^{2}/T$,
so that $\td=T$. We consider a force $F$ that is given as the gradient
of a potential, $F(x)=-(\partial_{x}U)(x)$, and for the potential
$U(x)$ use a quartic double well, 
\begin{equation}
U(x)=U_{0}\left[\left(\frac{x}{L}\right)^{2}-1\right]^{2},\label{eq:U_quartic}
\end{equation}
with $\beta U_{0}=2$, as illustrated on the right-hand side of Fig.~\ref{fig:potential_and_trajectory}.
For the smooth reference path $\varphi$ we choose a barrier crossing
path, parametrized as 
\begin{equation}
\traj(t)=\frac{L}{\arctan(\kappa/2)}\arctan\left[\kappa\cdot\left(\frac{t-\tfinal/2}{\td}\right)\right],\label{eq:def_traj}
\end{equation}
where for $\kappa$, which controls the maximal barrier crossing speed,
we use $\kappa=10$; we furthermore choose $\tinitial=0$, $\tfinal=\td$.
The prefactor in Eq.~\eqref{eq:def_traj} ensures that the path starts
at $x=-L$ and ends at $x=L$. The reference path Eq.~\eqref{eq:def_traj}
is illustrated in Fig.~\ref{fig:potential_and_trajectory}. 

\subsection{Perturbative solution of FPE in tube interior}

\label{sec:example_solution}

As we show in detail in App.~\ref{app:perturbative_spectrum}, for
$N=1$ the eigenvalue/eigenfunction Eq.~\eqref{eq:1D_FP_eq_DL_linearized_ev}
can be solved recursively for increasing $k$, and the solution at
order $k$ is of the form 
\begin{align}
\efn^{(k)}(\xDL,\tDL) & =\Qs^{(k)}(\xDL,\tDL)\,\sin\left[n\frac{\pi}{2}(\xDL+1)\right]\label{eq:form_of_perturbation_terms_main_text}\\
 & \qquad+\Qc^{(k)}(\xDL,\tDL)\,\cos\left[n\frac{\pi}{2}(\xDL+1)\right],\nonumber 
\end{align}
where $\Qs^{(k)}(\xDL,\tDL)$, $\Qc^{(k)}(\xDL,\tDL)$ are polynomials
in $\xDL$ of order $\leq k$, and depend on $\tDL$ via $\EDL_{l}(\tDL)$,
$1 \leq l\leq k$, which are given by 
\begin{align}
\EDL_{k}(\tDL) & \equiv-\frac{L^{k}\beta}{k!}\left.\frac{\partial^{k-1}F}{\partial x^{k-1}}\right|_{(\traj(t),t)}+\delta_{k,1}\trajdotDL(\tDL).\label{eq:EDL_def_onedim}
\end{align}
In App.~\ref{app:perturbative_spectrum}, we give explicit expressions
for $\evn^{(k)}$, $\Qs^{(k)}$, $\Qc^{(k)}$,
up to $k=5$.

Using the perturbative spectrum Eq.~\eqref{eq:form_of_perturbation_terms_main_text},
the propagator Eq.~\eqref{eq:approximate_propagator_steady} can
be calculated as a power series in $\epsilon$.
From the propagator, in turn, the perturbation series for the normalized
probability density $\PabsorbingNormalizedDL$ inside the tube is
obtained using Eq.~\eqref{eq:approximate_normalized_density}. It
is found that $\PabsorbingNormalizedDL$ is of the form 
\begin{align}
\PabsorbingNormalizedDL(\xDL,\tDL) & =\sum_{k=0}^{5}\epsilon^{k}\left\{ \PnCoeffADL^{(k)}(\xDL,\tDL)\,\sin\left[\frac{\pi}{2}(\xDL+1)\right]\right.\label{eq:PnormalizedDL_power_series}\\
 & \qquad\qquad\left.+\PnCoeffBDL^{(k)}(\xDL,\tDL)\,\cos\left[\frac{\pi}{2}(\xDL+1)\right]\right\} +\mathcal{O}(\epsilon^{6}),\nonumber 
\end{align}
where the coefficients $\PnCoeffADL^{(k)}(\xDL,\tDL)$, $\PnCoeffBDL^{(k)}(\xDL,\tDL)$,
are polynomials in $\xDL$ of order $\leq k$, and depend on $\tDL$
via $\EDL_{l}(\tDL)$, $1 \leq l\leq k$, as defined in Eq.~\eqref{eq:EDL_def_onedim}.
The explicit expressions for $\PnCoeffADL^{(k)}$, $\PnCoeffBDL^{(k)}$
for $k\leq5$ are given in App.~\ref{app:normalized_density_in_tube}.

\begin{figure*}[ht]
\centering \includegraphics[width=0.97\textwidth]{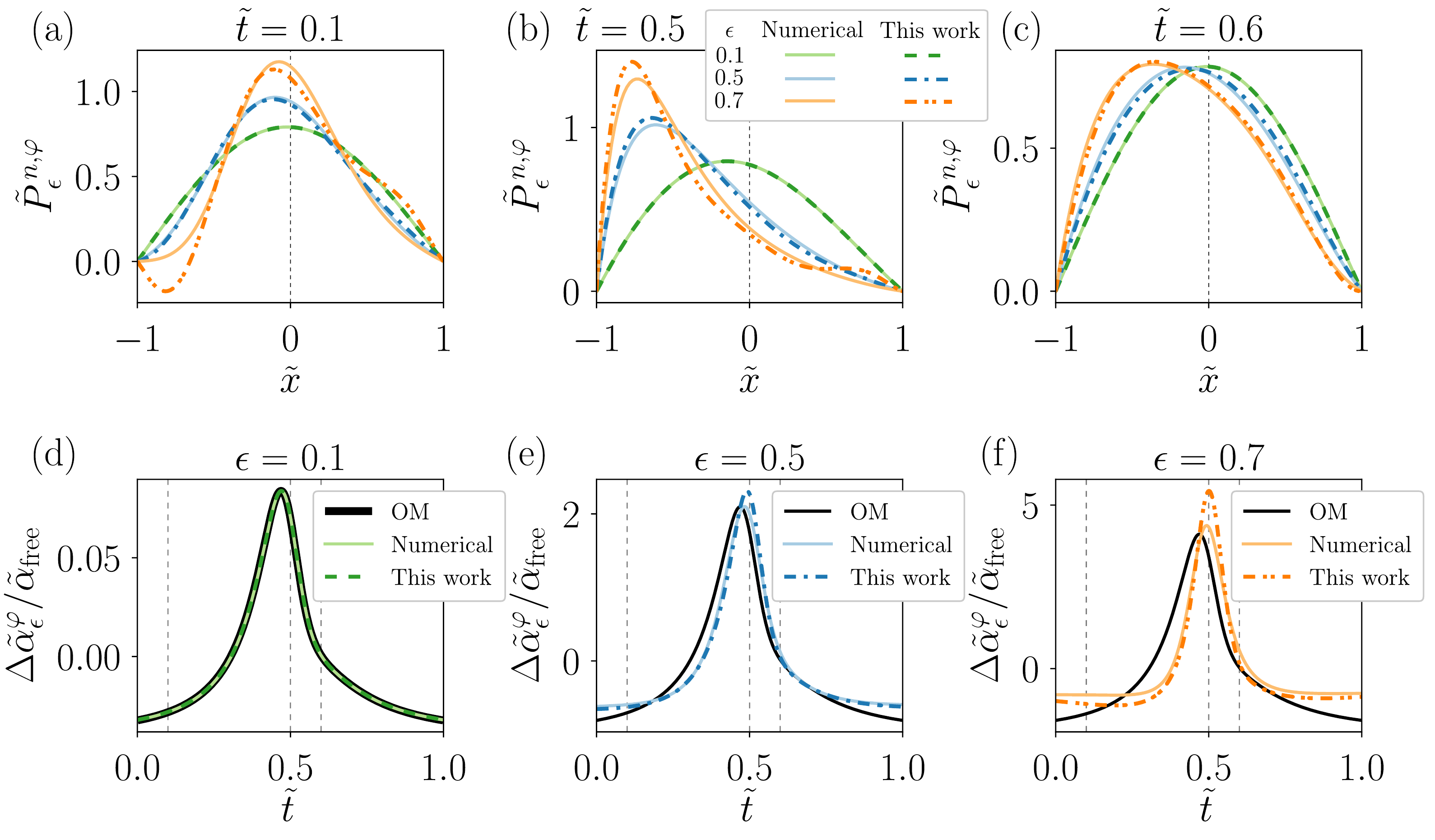} 
\caption{\label{fig:densities_and_rates} 
{Subplots {(a), (b), (c)}
show the normalized probability density} $\PabsorbingNormalizedDL(\xDL,\tDL)$,
defined in Eq.~\eqref{eq:PnormalizedDL_power_series}, as function
of position $\xDL$ for time (a) $\tDL=0.1$, (b) $\tDL=0.5$, and
(c) $\tDL=0.6$, and for tube radius $\epsilon=0.1$ (green), $\epsilon=0.3$
(blue), and $\epsilon=0.5$ (orange). The legend given in subplot
(b) is valid for subplots (a), (b), (c). Solid colored lines denote
results from numerical simulation of the FPE, Eq.~\eqref{eq:1D_FP_eq_DL},
see App.~\ref{app:numerics} for details on the numerical algorithm.
Colored broken lines denote the perturbative result Eq.~\eqref{eq:PnormalizedDL_power_series},
calculated to order $\epsilon^{5}$. Vertical dashed lines indicate
the tube center $\xDL=0$. {Subplots (d), (e), (f) show the
exit rate} $\aexitDL_{\epsilon}^{\traj}$, defined in 
Eq.~\eqref{eq:exit_rate_Ndim_expansion_DL},
as a function of time $\tDL$, for tube radii (d) $\epsilon=0.1$,
(e) $\epsilon=0.5$, and (f) $\epsilon=0.7$. From all rates the free-diffusion
exit rate is subtracted and the result is divided by the free-diffusion
exit rate, as defined in Eq.~\eqref{eq:def_relative_rate}. Colored
solid lines denote exit rates calculated from numerical simulation
of the FPE, Eq.~\eqref{eq:1D_FP_eq_DL}. Colored
broken lines show perturbative exit rates according to 
Eqs.~\eqref{eq:exit_rate_Ndim_expansion_DL}, 
(\ref{eq:aexit_free_DL}-\ref{eq:aexit_two_DL}).
Black solid lines denote the OM Lagrangian Eq.~\eqref{eq:aexit_zero_DL},
from which the free-diffusion exit rate has already been subtracted
so that in fact $\aexitzeroDL/\aexitfreeDL$ is plotted. Vertical
dashed lines indicate the times $\tDL=0.1$, $0.5$, $0.6$ used for
subplots (a), (b), (c). }
\end{figure*}

In Fig.~\ref{fig:densities_and_rates} (a), (b), (c), we compare
the perturbative analytical probability density Eq.~\eqref{eq:PnormalizedDL_power_series}
to order $\epsilon^{5}$ with results from direct numerical solution
of the FPE, Eq.~\eqref{eq:1D_FP_eq_DL}. Figure
\ref{fig:densities_and_rates} shows the probability density inside
the tube at times (a) $\tDL=0.1$, (b) $\tDL=0.5$, and (c) $\tDL=0.6$,
as indicated by vertical dashed lines in Fig.~\ref{fig:potential_and_trajectory}.
At each time we show results for radii $\epsilon=0.1$ (green), $\epsilon=0.5$
(blue), and $\epsilon=0.7$ (orange). 
Note that the intermediate tube
radius $\epsilon\equiv R/L=0.5$ is in fact so large that during the
ascent of the path $\traj$ towards the barrier top, there is a time
at which the interval $B_{R}^{\traj}(t)=[\traj(t)-R,\traj(t)+R]$
spans from the minimum $x=-L$ to the barrier top $x=0$.

For all times displayed, we observe that while for $\epsilon = 0.1$, $0.5$,
numerical and perturbative results agree very well with each other, for
the largest radius considered, 
 $\epsilon = 0.7$, clear deviations between the two are discernible.
At the time $\tDL=0.1$ considered in Fig.~\ref{fig:densities_and_rates}
(a), the path $\traj$ is close to the minimum at $x=-L$ and has
a small velocity, c.f.~Fig.~\ref{fig:potential_and_trajectory}.
While for the smallest radius $\epsilon=0.1$ the probability density
is almost symmetric around $\xDL=0$, indicating that the dynamics
inside the tube is dominated by free diffusion, for $\epsilon=0.5$,
$0.7$ the influence of the potential leads to a slight shift of the
most probable position towards small negative values of $\xDL$.
The perturbative probability density for $\epsilon = 0.7$ takes on negative
values close to $\xDL=-1$, which is clearly unphysical and signifies
a breakdown of the perturbative results of order $\epsilon^{5}$.
 In
Fig.~\ref{fig:densities_and_rates} (b) we show probability densities
at time $\tDL=0.5$, which according to Fig.~\ref{fig:potential_and_trajectory}
is when the path $\traj$ traverses the barrier top. 
Despite the fact that at the maximum the potential $U$
is a symmetric function of $\xDL$, all probability densities shown
in Fig.~\ref{fig:densities_and_rates} (b) are tilted towards negative
values of $\xDL$. This is because the velocity of the path $\traj$
leads to a ficticious force, as seen explicitly in Eq.~\eqref{eq:effective_total_force};
due to this fictitious force the symmetry of the potential $U$ is
broken at the barrier top, which leads to the tilted probability densities
observed in the figure. This effect is less pronounced at small $\epsilon$,
where the dynamics inside the tube is dominated by free diffusion,
as compared to the apparent deterministic force due to $U$ and $\trajdot$.
In Fig.~\ref{fig:densities_and_rates} (c) we consider the time $\tDL=0.6$,
at which the path $\traj$ descents from the barrier top towards the
minimum at $x=L$, c.f.~Fig.~\ref{fig:potential_and_trajectory}.
Here we observe that even though
the force resulting from the potential $U$ pushes towards the positive
$\xDL$-direction, due to the velocity of the path $\traj$ the apparent
force Eq.~\eqref{eq:effective_total_force} leads to a probability
density that is still slightly tilted towards the negative $\xDL$-direction,
i.e.~uphill in the potential energy landscape.

In the Supplementary Material (SM) \cite{supplement} we provide videos
that show the full time evolution of the normalized probability density
for radius $\epsilon=0.1$, $0.3$, $0.5$, $0.7$. For $\epsilon=0.1$,
$0.3$, numerical and analytical results show perfect agreement throughout.
Consistent with Fig.~\ref{fig:densities_and_rates}, for $\epsilon=0.5$
small deviations between numerical and analytical density are observed,
and become most pronounced as the path $\traj$ ascends the barrier
($\tDL\approx0.45$); however, given the size of the tube the agreement
between numerical and analytical probability density is remarkably
good overall. For $\epsilon=0.7$ the breakdown of our perturbative
results can be observed; the analytical probability takes on negative
values and at times deviates considerably from the numerical data.

Overall, from Fig.~\ref{fig:densities_and_rates} (a), (b), (c),
and also the supplementary videos, we conclude that for small to intermediate
tube radius, our analytic result Eq.~\eqref{eq:PnormalizedDL_power_series}
very well approximates the actual FP dynamics inside the
tube.

\subsection{Exit rate from tube}

\label{sec:ExitRate}

Using the explicit expressions for the spectrum given in App.~\ref{app:one_dimensional_system},
the expansion of the exit rate Eq.~\eqref{eq:exit_rate_Ndim_expansion_DL}
in powers of $\epsilon$ is given by 
\begin{align}
\aexitfreeDL & =\frac{\pi^{2}}{4\epsilon^{2}},\label{eq:aexit_free_DL}\\[1ex]
\aexitzeroDL & \equiv\LOMDL=\frac{\EDL_{1}^{2}}{4}-\EDL_{2},\label{eq:aexit_zero_DL}\\
\aexittwoDL & =-\frac{\EDL_{1}\dot{\EDL}_{1}}{4}\left(1-\frac{8}{\pi^{2}}\right)+\frac{\dot{\EDL}_{2}}{3}\left(1-\frac{9}{\pi^{2}}\right)\label{eq:aexit_two_DL}\\
 & \qquad+\left(\frac{\EDL_{1}\EDL_{3}}{2}+\frac{\EDL_{2}^{2}}{3}-2\EDL_{4}\right)\left(1-\frac{6}{\pi^{2}}\right),\nonumber 
\end{align}
where the $\EDL_{l}\equiv\EDL_{l}(\tDL)$ are defined in Eq.~\eqref{eq:EDL_def_onedim},
and a dot denotes a derivative with respect to $\tDL$. 
We again consider barrier crossing
in the double-well system, as defined in Eqs.~\eqref{eq:U_quartic},
\eqref{eq:def_traj}, and illustrated in Fig.~\ref{fig:potential_and_trajectory}.
In Fig.~\ref{fig:densities_and_rates} (d), (e), (f), we compare
numerically calculated exit rates to perturbative results obtained
from Eqs.~\eqref{eq:exit_rate_Ndim_expansion_DL}, (\ref{eq:aexit_free_DL}-\ref{eq:aexit_two_DL}).
In the plots the exit rate is shifted and rescaled according to 
\begin{equation}
\frac{\Delta\aexitDL_{\epsilon}^{\traj}}{\aexitfreeDL}\equiv\frac{\aexitDL_{\epsilon}^{\traj}-\aexitfreeDL}{\aexitfreeDL},\label{eq:def_relative_rate}
\end{equation}
so that i) the sign of a curve indicates whether the exit rate is
enhanced or diminished as compared to the free-diffusion limit $\aexitfreeDL$,
and ii) the magnitude yields the relative importance of the terms
Eqs.~\eqref{eq:aexit_zero_DL}, \eqref{eq:aexit_two_DL} as compared
to $\aexitfreeDL$. Numerical data is shown as solid colored lines,
perturbative analytical results are given as broken colored lines.
To gauge the importance of the quadratic correction Eq.~\eqref{eq:aexit_two_DL}
relative to the OM Lagrangian Eq.~\eqref{eq:aexit_zero_DL},
we furthermore include plots of the latter as solid black lines in
Fig.~\ref{fig:densities_and_rates} (d), (e), (f). As detailed in
App.~\ref{app:numerics}, we use as initial distribution $\PDLzero$
for our simulations the instantaneous steady state of the FPE, 
so that there is no transient initial decay in our numerical
data; a brief discussion of the transient effects of the initial condition
on the exit rate is given in App.~\ref{app:equilibration}.

In Fig.~\ref{fig:densities_and_rates} (d) we consider the radius
$\epsilon=0.1$. As can be seen, the numerical and analytical results
agree perfectly with each other, and also with the OM Lagrangian Eq.~\eqref{eq:aexit_zero_DL}.
This means that the quadratic correction Eq.~\eqref{eq:aexit_two_DL}
is not yet relevant at this radius. 
Relative deviations from the free-diffusion
exit rate $\aexitfreeDL$ are less than 10\% throughout, so that the exit
rate is dominated by free diffusion. 
Figure \ref{fig:densities_and_rates}
(e) shows data for the intermediate radius $\epsilon=0.5$. Numerical
and perturbative analytical results agree very well with each other,
with minor deviations at $\tDL\approx0.5$. However, clear deviations
between numerical data and the OM Lagrangian Eq.~\eqref{eq:aexit_zero_DL}
are visible, meaning  that the quadratic correction Eq.~\eqref{eq:aexit_two_DL} to the exit
rate is  now relevant.
The deviations between our perturbative/numerical results and OM theory
are twofold. First, when the path is close to the minima, 
the OM action underestimates the true (numerical)
exit rate. During these times, the numerical exit rate is rather insensitive
to the exact position of the tube center within the well, 
 because the rate limiting
step to exit the tube is to climb the potential barrier, which is
expected to be rather insensitive to the exact position of the tube center
 in the well. 
The second difference between our perturbative/numerical
results and OM theory is that during barrier crossing, the numerical
exit rate is delayed as compared to the OM Lagrangian.
From the magnitude of the rescaled exit rate Eq.~\eqref{eq:def_relative_rate},
we conclude that for $\epsilon=0.5$, the free-diffusion exit rate is
 of the same order as the corrections Eq.~\eqref{eq:aexit_zero_DL}, 
 \eqref{eq:aexit_two_DL}. 
 Figure \ref{fig:densities_and_rates}
(f) shows data for the largest radius $\epsilon=0.7$. Overall the
perturbative result Eq.~\eqref{eq:exit_rate_Ndim_expansion_DL}
still shows reasonable agreement with the numerical exit rate, which
is surprising since the corresponding probability density at times
deviates strongly from the numerical results, c.f.~Fig.~\ref{fig:densities_and_rates}
(a), and the supplementary videos. However, clear deviations between
numerical and analytical exit rate can be discerned, most prominently
during barrier crossing at $\tDL\approx0.5$. 
Numerical exit rate and OM Lagrangian Eq.~\eqref{eq:aexit_zero_DL} 
 disagree considerably, showing
the importance of the quadratic correction Eq.~\eqref{eq:aexit_two_DL}.
During
barrier crossing, the contributions to the exit rate from Eqs.~\eqref{eq:aexit_zero_DL},
\eqref{eq:aexit_two_DL} are about 5 times larger than the free-diffusion
exit rate $\aexitfreeDL$.

In summary, Fig.~\ref{fig:densities_and_rates} (d), (e), (f) shows
that our perturbative results Eqs.~\eqref{eq:exit_rate_Ndim_expansion_DL}, 
(\ref{eq:aexit_free_DL}-\ref{eq:aexit_two_DL})
describe the exit rate quantitatively up to a tube radius well comparable
to the typical length scale of the potential $U$, and in particular
beyond the regime where the OM Lagrangian is applicable.

\begin{figure}[t]
\centering \includegraphics[width=\columnwidth]{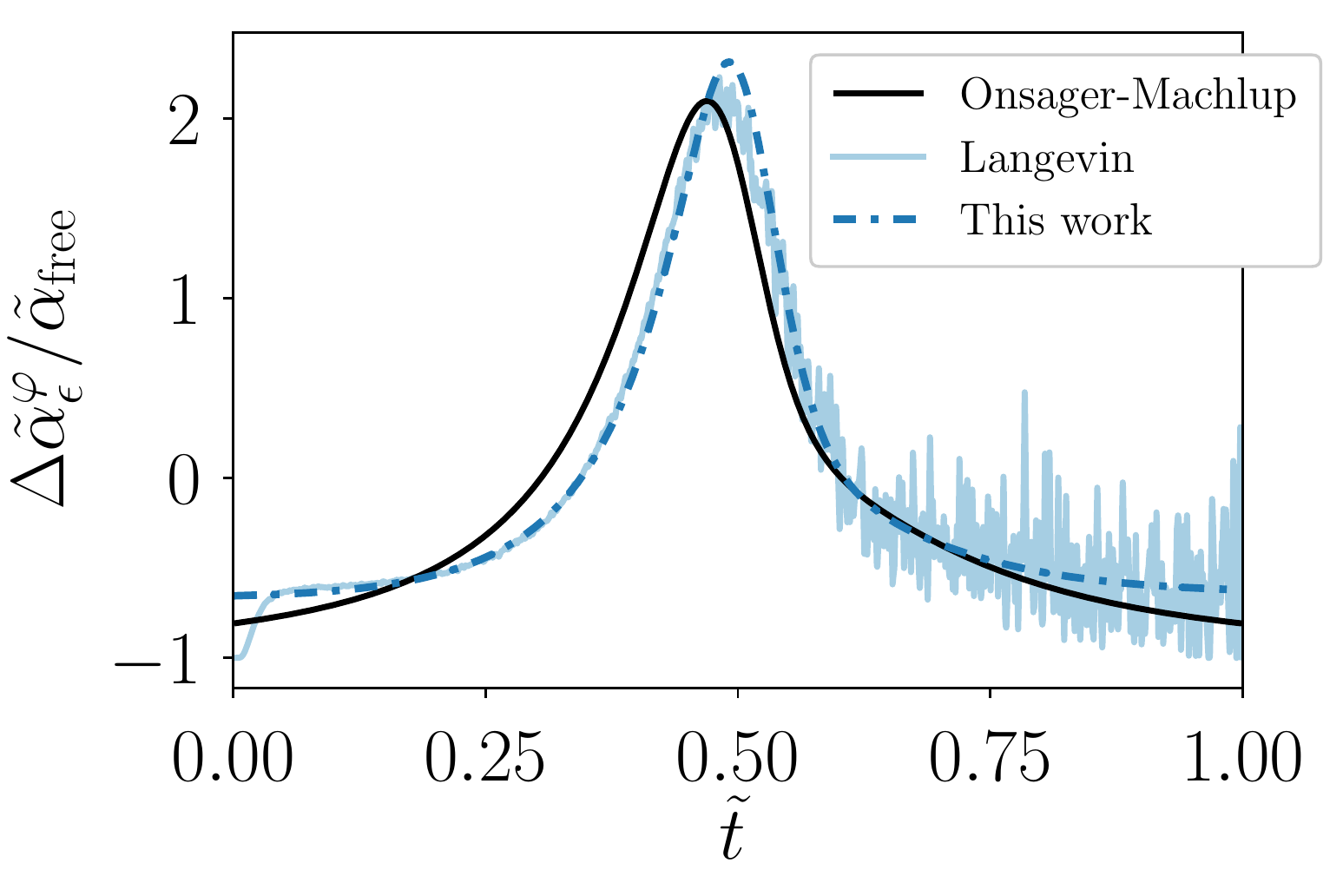} 
\caption{\label{fig:exit_rate_Langevin} {Exit rate as measured directly
from Langevin simulations for tube of radius $\epsilon=R/L=0.5$.}
The black solid and blue dash-dotted lines are replots of the corresponding
lines in Fig.~\ref{fig:densities_and_rates} (e). The blue solid
line represents the exit rate as estimated from simulated trajectories.
For this, $2.4\cdot10^{6}$ independent Langevin simulations in the
quartic double well Eq.~\eqref{eq:U_quartic} with a simulation timestep
$\Delta\tDL=10^{-5}$ are performed, with each trajectory starting
at $x=-L$. From these trajectories, the sojourn probability for a
tube of radius $\epsilon=R/L=0.5$ around a reference path $\traj(t)$,
as defined in Eq.~\eqref{eq:def_traj}, is evaluated directly, by
counting which proportion of trajectories has never left the tube
until any given time. Subsequently, the exit rate is numerically calculated
via Eq.~\eqref{eq:exit_rate_DL}, and the result is smoothed using
a moving average with a Hann-window of width $\Delta\tDL=0.003$.
From this exit rate, finally, the free-diffusion exit rate Eq.~\eqref{eq:aexit_free_DL}
is subtracted, and the result is divided by the free-diffusion exit
rate, c.f.~Eq.~\eqref{eq:def_relative_rate}.}
\end{figure}

To close this section, we illustrate how finite-radius exit rates
obtained directly from measured trajectories compare to our perturbative
analytical results. For this, we consider a tube radius $\epsilon=R/L=0.5$,
as also discussed in Fig.~\ref{fig:densities_and_rates} (e). Figure
\ref{fig:exit_rate_Langevin} depicts the exit rate obtained directly
from a large number of independent simulated time series. As Fig.~\ref{fig:exit_rate_Langevin}
shows, the exit rate obtained directly from Langevin time series agrees
well with our perturbative result Eq.~\eqref{eq:exit_rate_Ndim_expansion_DL},
and shows clear deviations from the OM Lagrangian Eq.~\eqref{eq:aexit_zero_DL}.
This shows that it is possible to measure the exit rate for a finite-radius
tube directly from time series, without fitting any model to the data.
Note that since the FPE, Eq.~\eqref{eq:1D_FP_eq}, with absorbing
boundary conditions is equivalent to the Langevin Eq.~\eqref{eq:Langevin_eq_intro},
with trajectories being discarded once they first cross the absorbing
boundary, it is expected that Fig.~\ref{fig:densities_and_rates}
(e) and Fig.~\ref{fig:exit_rate_Langevin} lead to the same conclusions;
indeed, the agreement between numerical FP solution and
results obtained from Langevin simulations is an important consistency
check for our numerics. Apart from illustrating how our results directly
connect to measured time series, the analysis based on Langevin trajectories
also highlights two features that appear when extracting the exit
rate from recorded data. First, since all Langevin simulations are
initiated at $x=-L$, which can be thought of as a definite experimental
initial condition, the exit rate shows a short transient relaxation
period for times $\tDL\lesssim0.05$, see App.~\ref{app:equilibration}
for further discussion. Second, the number of trajectories inside
the tube decreases over time, so that the statistics for calculating
the exit rate become successively worse; this explains why the exit
rate measured from Langevin trajectories starts to become noisy around
$\tDL\approx0.5$.

\section{Summary and Conclusions}

\label{sec:conclusions}

In this work we establish the finite-radius tubular ensemble, which
consists of all stochastic trajectories that stay close to a smooth
reference path $\trajvec$, as a physically and mathematically useful
concept to regularize and extend the path probabilities of individual
stochastic trajectories. We in particular derive explicit expressions
for the probability to observe any path of the tubular ensemble, thus
generalizing the Onsager-Machlup (OM) stochastic action. Our results
have several important consequences.

From a mathematical perspective, we evaluate and study the probability
$P(\trajset)$, i.e.~the probability that a stochastic trajectory
stays close to a given smooth reference path, for finite radius $R$.
We therefore focus on a \textit{measure}, which is in contrast to
previous work, which aimed to define probability \textit{densities}
on the space of all continuous paths, and therefore always involved
the singular limit $R\rightarrow0$ \citep{stratonovich_probability_1971}.
Compared to the approach to path probabilities via path integrals
\citep{onsager_fluctuations_1953,graham_path_1977,langouche_functional_1979,dekker_path_1980,weber_master_2017,cugliandolo_building_2018},
an advantage of our approach is that at no point we need to discretize
time. Therefore, none of the technical/conceptual difficulties arising
from different time-discretization schemes discussed in the literature
arise \citep{wissel_manifolds_1979,adib_stochastic_2008,cugliandolo_building_2018}.
Furthermore, in our theory smooth and non-differentiable 
stochastic trajectories are
cleanly disentangled. The former are used to parametrize a set (a
moving ball with finite radius), the latter are confined to this set.

In a sense, the approach used in the present paper is opposite to
Freidlin-Wentzell theory \citep{ventsel_small_1970}. While Freidlin
and Wentzell also consider the tubular ensemble Eq.~\eqref{eq:def_trajset},
they investigate the double limit of vanishing radius $R\rightarrow0$
and temperature $1/\beta=k_{\mathrm{B}}T\rightarrow0$. Practically
speaking, in their analysis the deterministic force in the Langevin
Eq.~\eqref{eq:Langevin_eq_intro} is assumed to be the dominant term.
In our perturbative calculation at constant temperature, on the other
hand, we perturb around the free-diffusion solution of the FPE, 
which means that in our analysis the random force term in
the Langevin Eq.~\eqref{eq:Langevin_eq_intro} is assumed to be the
dominant term on short length scales. That random noise dominates
over deterministic forces at short length- and time scales is a basic
feature of the Langevin equation and is in fact the reason why a typical
realization of Eq.~\eqref{eq:Langevin_eq_intro} is nowhere differentiable.

Our theory for the finite-radius tubular ensemble Eq.~\eqref{eq:def_trajset}
is also highly relevant from a physical perspective. By establishing
a direct relation between exit rate and stochastic action Lagrangian,
we put the latter within reach of experiments. Indeed, substituting
Eqs.~\eqref{eq:OM_action_def}, \eqref{eq:aexit_def_intro}, into
Eq.~\eqref{eq:action_difference_intro}, it follows that 
\begin{equation}
\LOM(\vectraj,\vectrajdot)-\LOM(\vectrajTwo,\vectrajdotTwo)=\lim_{R\rightarrow0}\left[\aexit-\alpha_{R}^{\vectrajTwo}\right].\label{eq:LagrangianAndExitRate}
\end{equation}
While directly measuring the probability of an individual given path
is experimentally unfeasible, simply because that probability is zero,
directly measuring experimentally the probability for a trajectory
to stay inside a moving ball with finite radius is possible with present-day
technology \citep{li_measurement_2010,caciagli_optical_2017,gladrow_experimental_2019}.
This means the right-hand side of Eq.~\eqref{eq:LagrangianAndExitRate}
can be measured for finite $R$, as was done in Fig.~\ref{fig:exit_rate_Langevin},
and then extrapolated to the limit $R\rightarrow0$; this can be done
without fitting a stochastic model to the time series. Equation \eqref{eq:LagrangianAndExitRate}
thus allows to compare model-free measurements of exit rates (right-hand
side of the equation) to theoretical predictions for the stochastic
action Lagrangian (left-hand side of the equation). This will allow
to experimentally validate theoretical predictions for the stochastic
action Lagrangian as a measure for relative path likelihoods. Equation
\eqref{eq:LagrangianAndExitRate} can furthermore serve as an operational
and experimentally relevant \textit{definition} for the action Lagrangian
for other models of stochastic dynamics, for example those used to
describe active particles \cite{dabelow_irreversibility_2019}.

While irrelevant in the limit $R\rightarrow0$, for finite tube radius
it will be important to understand in more depth how transient effects
due to the initial distribution $\Xvec_{\tinitial}\sim\Pzero$ affect
the sojourn probability. A basis for investigating such boundary effects
is given by the full perturbative solution considered in App.~\ref{app:perturbative_solution_FP}.

While we assume a smooth path $\trajvec$, 
our derivation in fact only uses 
that it is twice differentiable. The first derivative $\dot{\trajvec}$ emerges
from applying the coordinate transformation Eq.~\eqref{eq:def_tDL}
to the FPE, c.f.~Eq.~\eqref{eq:1D_FPO_DL}.
The second derivative enters because the FPE in
terms of the instantaneous eigenbasis,
Eq.~\eqref{eq:1D_FP_in_eigenbasis},
contains the time-derivative of FP eigenfunctions; 
since these eigenfunctions depend on $\dot{\trajvec}$, 
their derivative depends on $\ddot{\trajvec}$.
It will be interesting to extend our theory to
 reference paths $\trajvec$ that are continuous, but not differentiable, 
such as realizations of the Langevin Eq.~\eqref{eq:Langevin_eq_intro}.
A starting point for this would be to investigate how the 
FPE transforms under a non-differentiable coordinate
transformation \cite{cresson_non_2006}.

Another possible extension of our theory is to include
position-dependent diffusivity, i.e.~to replace the constant diffusion
coefficient $D$ by a function $D(\xvec)$. Assuming that the diffusivity
varies slowly along the tube, a first approximation is to simply replace
$D$ by $D(\xvec)$ in our results. In view of the exit rate Eq.~\eqref{eq:axit_result_intro},
the sojourn probability is then given by 
\begin{align}
\Pabsorbing(\tfinal) & =\exp\left[-\frac{\evn^{(0)}}{R^{2}}\int_{\tinitial}^{\tfinal}\mathrm{d}t\,D(\trajvec(t))\right.\label{eq:position_dependent_diffusivity}\\
 & \qquad\quad\left.-\int_{\tinitial}^{\tfinal}\mathrm{d}t\,\LOM(t,\trajvec(t),\trajdotvec(t))+\mathcal{O}(R^{2})\right],\nonumber 
\end{align}
where the diffusivity in the OM Lagrangian Eq.~\eqref{eq:OM_intro}
is now evaluated at $D(\trajvec(t))$. Equation \eqref{eq:position_dependent_diffusivity}
shows that for position-dependent diffusivity, for small tube radius
$R$ the leading order difference in sojourn probabilities along two
paths $\trajvec$, $\bm{\phi}$, is the mean free-diffusion exit rate
along the paths, and the OM action is now a subleading-order correction.
Thus, in the limit $R\rightarrow0$, instead of Eq.~\eqref{eq:OM_intro}
one would rather want to consider an action 
\begin{equation}
S_{D}[\trajvec]\equiv\int_{\tinitial}^{\tfinal}\mathrm{d}t\,D(\trajvec(t))\label{eq:free_diff_action}
\end{equation}
to quantify physically observed relative path probabilities. Intuitively,
a particle is more likely to diffuse away from a given reference path
in a region with large diffusivity, as compared to a region with low
diffusivity. In the mathematical literature the leading-order effect
due to free diffusion, given by Eq.~\eqref{eq:free_diff_action},
is usually scaled away, essentially by introducing a position-dependent tube radius
$R(\xvec)$ such that $D(\vecx)/R(\xvec)^{2}$ is constant as a function
of $\xvec$ \citep{stratonovich_probability_1971}. Thus, before applying
the OM theory in systems with position-dependent diffusivity, one
should decide whether one wants to quantify relative path probabilities
using a spatially constant threshold $R$, in which case one would
want to use Eq.~\eqref{eq:free_diff_action} as action, or using
a varying threshold $R(\vecx)\sim\sqrt{D(\xvec)}$, in which case
the OM action is the leading order difference in sojourn probabilities
\citep{stratonovich_probability_1971,lau_state-dependent_2007}.

The present work on the tubular ensemble Eq.~\eqref{eq:def_trajset}
offers an intuitive picture on (relative) path probabilities for the
Langevin Eq.~\eqref{eq:Langevin_eq_intro}, providing a physical
approach to this hitherto rather technical subject. Since any question
that can be posed for individual stochastic trajectories is straightforwardly
extended to the tubular ensemble, and through that is made accessible
to simulation or experiment, the theory presented here is expected
to find many applications in the future. The results will be particularly
useful for the field of stochastic thermodynamics, where the concept
of individual trajectories, and ratios of their probabilities, is
employed extensively \citep{seifert_entropy_2005,seifert_stochastic_2012,dabelow_irreversibility_2019}.

\begin{acknowledgments}
We thank Professor Mike Cates and Dr.~Yongjoo Baek, 
Dr.~Jules Guioth, 
Dr.~Rob Jack, and Dr.~Patrick Pietzonka
for stimulating discussions;
furthermore we thank Dr.~Kerstin Burghaus,
Dr.~Heather Partner, and Dr.~Carlos Riofr\'{i}o
for helpful comments on the manuscript.
Work was funded in part by the European Research Council under the EU's Horizon 2020 Program, Grant
No. 740269,
and by an Early Career Grant to RA from the Isaac Newton Trust.
\end{acknowledgments}

\appendix

\section{Perturbative spectrum of N-dimensional FPE}

\label{app:perturbative_calculations_N}

\subsection{Perturbation theory}

\label{app:perturbative_spectrum_N_dimensional}

In the present appendix, we perturbatively solve the eigenvalue Eq.~\eqref{eq:1D_FP_eq_DL_linearized_ev}
up to order $\epsilon^{2}$. For this, we first expand the right-hand
side of the equation as a power series in $\epsilon$.

\subsecthead{Taylor expansion of the force.} The multidimensional
Taylor expansion of the force $\Fvec$ around the tube center $\trajvec(t)$
is given by 
\begin{align}
\Fvec(\xvec,t) & =\sum_{k=0}^{\infty}\frac{1}{k!}\sum_{\alpha_{1},...,\alpha_{k}=1}^{N}\left.\frac{\partial^{k}\Fvec}{\partial x_{\alpha_{1}}...\partial x_{\alpha_{k}}}\right|_{(\trajvec(t),t)}\label{eq:Fvec_multidim_taylor_appendix}\\
 & \qquad\qquad\times(\xvec-\trajvec(t))_{\alpha_{1}}...(\xvec-\trajvec(t))_{\alpha_{k}},\nonumber 
\end{align}
where $(\xvec-\trajvec(t))_{\alpha_{i}}\equiv x_{\alpha_{i}}-\traj_{\alpha_{i}}(t)$
is the ${\alpha_{i}}$-th component of the vector $\xvec-\trajvec(t)$.
Substituting Eq.~\eqref{eq:Fvec_multidim_taylor_appendix} into the
definition of the dimensionless force Eq.~\eqref{eq:def_FDL} and
using Eq.~\eqref{eq:def_tDL}, we obtain that 
\begin{align}
\FDLeffvec(\xDLvec,\tDL) & =-\sum_{k=1}^{\infty}\epsilon^{k-1}k\sum_{\alpha_{1},...,\alpha_{k-1}=1}^{N}\EDLvec_{k,\alpha_{1}...\alpha_{k-1}}(\tDL)\xDL_{\alpha_{1}}...\,\xDL_{\alpha_{k-1}}\\
 & \equiv-\sum_{k=1}^{\infty}\epsilon^{k-1}k\,\EDLvec_{{k},\alphavec}(\tDL)\xDL_{\alphavec}\label{eq:taylor_expansion_multidim_force_appendix}
\end{align}
where we use the Einstein sum convention for the indices $\alphavec\equiv(\alpha_{1},...,\alpha_{k-1})$,
abbreviate $\xDL_{\alphavec}\equiv\xDL_{\alpha_{1}}...\,\xDL_{\alpha_{k-1}}$,
and the vector-valued $(k-1)$-multilinear form $\EDLvec_{k}$ is
defined as 
\begin{align}
\EDLvec_{k,\alpha_{1}...\alpha_{k-1}}(\tDL)\equiv-\frac{1}{k!}L^{k}\beta\left.\frac{\partial^{k-1}\Fvec}{\partial x_{\alpha_{1}}...\partial x_{\alpha_{k-1}}}\right|_{(\trajvec(t),t)}+\delta_{k,1}\trajdotvecDL(\tDL),\label{eq:EDLvec_def_appendix}
\end{align}
where dimensionless quantities (as indicated by a tilde) and quantities
with physical dimensions are related via Eqs.~\eqref{eq:def_tDL},
\eqref{eq:def_trajDL}. Note that if the derivatives of the force
commute, e.g.~if the force is a smooth function of position for a
time $t$, then $\EDLvec_{k}$ is symmetric in the $(\alpha_{1},...,\alpha_{k-1})$.
If the force is locally given by a potential $U$ as $\Fvec=-\gradvec U$,
then the $j$-th vector component of $\EDLvec_{k}$ is given by 
\begin{align}
\EDL_{{k},\alpha_{1}...\alpha_{k-1}}^{j}(\tDL) & \equiv\frac{1}{k!}L^{k}\beta\left.\frac{\partial^{k}U}{\partial x_{\alpha_{1}}...\partial x_{\alpha_{k-1}}\partial x_{j}}\right|_{(\trajvec(t),t)}\label{eq:E_force_from_potential_appendix}\\
 & \qquad\qquad+\delta_{k,1}\trajdotDL_{j}(\tDL),\nonumber 
\end{align}
so that Eq.~\eqref{eq:taylor_expansion_multidim_force_appendix}
a multivariate Taylor expansion of the dimensionless potential $\tilde{U}=\beta U+\xDLvec\cdot\dot{\trajvecDL}$
around the tube center $\xDLvec=0$. Consequently, in that case the unnormalized
instantaneous steady state distribution inside the tube is given by
\begin{align}
\reqDL(\xvecDL,\tDL) & =\exp\left[-\sum_{k=1}^{\infty}\epsilon^{k}\sum_{\alpha_{1},...,\alpha_{k}=1}^{N}\EDLvec_{k,\alpha_{1},...,\alpha_{k-1}}^{\alpha_{k}}\xDL_{\alpha_{1}}...\,\xDL_{\alpha_{k}}\right].\label{eq:req_for_potential_force_appendix}
\end{align}

\subsecthead{Hierarchy of equations for the spectrum.} Inserting
the power series Eq.~\eqref{eq:taylor_expansion_multidim_force_appendix}
into the eigenvalue Eq.~\eqref{eq:1D_FP_eq_DL_linearized_ev}, we
obtain 
\begin{equation}
\laplaceDLvec\efn+\sum_{k=1}^{\infty}k\epsilon^{k}(\EDLvec_{k,\alphavec}\cdot\gradvecDL)\left(\xDL_{\alphavec}\efn\right)=-\evn\efn,\label{eq:FP_eigenvalue_equation_taylor_expanded_appendix}
\end{equation}
where the dot denotes a scalar product and $\gradvecDL$ denotes the
gradient operator with vector components $\tilde{\nabla}_{i}\equiv\partial/\partial\xDL_{i}$.

Expanding both the instantaneous eigenvalues and eigenfunctions as
power series in $\epsilon$, as defined in Eq.~\eqref{eq:evn_power_series},
substituting these into Eq.~\eqref{eq:FP_eigenvalue_equation_taylor_expanded_appendix},
and demanding that the resulting equation hold at each power $\epsilon^{k}$,
we obtain a hierarchy of equations which for the $n$-th eigenvalue/eigenfunction
pair at order $\epsilon^{k}$ read 
\begin{align}
\laplaceDLvec\efn^{(k)}+\evn^{(0)}\efn^{(k)} & =-\sum_{l=1}^{k}\evn^{(l)}\efn^{(k-l)}\label{eq:spectrum_hierarchy}\\
 & \qquad-\sum_{l=1}^{k}l\,(\EDLvec_{l,\alphavec}\cdot\gradvecDL)\left(\xDL_{\alphavec}\efn^{(k-l)}\right),\nonumber 
\end{align}
where we use the convention that for $k=0$, the sums on the right-hand
side are zero. For the absorbing boundary conditions to be fulfilled
independently of $\epsilon$, they need to hold at each order separately,
so that for all $k\in\{0,1,2,...\}$ we have 
\begin{equation}
\efn^{(k)}(\xDLvec,\tDL)=0\quad\forall~\xDL\in\partial\BDL\equiv\left\{ \,\xDLvec~\setbar~||\xDLvec||=1\,\right\} .\label{eq:spectrum_hierarchy_bcs}
\end{equation}
While any solution to Eqs.~\eqref{eq:spectrum_hierarchy}, \eqref{eq:spectrum_hierarchy_bcs}
can be used in practice for the spectrum, the solution to these equations
is not unique. To fix the solution uniquely, we introduce a normalization
condition \label{eq:normalization_condition_app} $\langle\efn,\efn\rangle=1$.
Inserting the power series expansion Eq.~\eqref{eq:evn_power_series}
for the eigenfunction into this normalization condition, and demanding
that the condition hold at each power of $\epsilon$, we obtain for
$k=0$ that 
\begin{equation}
\int_{\BDL}\mathrm{d}^{N}\xvecDL~\efn^{(0)}\efn^{(0)}=1,\label{eq:spectrum_normalization_condition_0_appendix}
\end{equation}
while for $k\geq1$ we have that 
\begin{align}
\int_{\BDL}\mathrm{d}^{N} & \xvecDL~\efn^{(k)}\efn^{(0)}\label{eq:spectrum_normalization_condition_1_appendix}\\
 & =-\frac{1}{2}\sum_{l=0}^{k-1}\sum_{m=0}^{k-\max\left\{ 1,l\right\} }\int_{\BDL}\mathrm{d}^{N}\xvecDL~\efn^{(l)}\efn^{(m)}\left(\reqDL^{-1}\right)^{(k-l-m)},\nonumber 
\end{align}
where we use the convention that for $k=1$ the sum on the right-hand
side is zero and the expansion of $\reqDL^{-1}$ in powers of $\epsilon$
is discussed in App.~\ref{app:steady_state_appendix}. Note that
for any $k$, there only appear perturbation terms $\efn^{(l)}$ with
$l<k$ on the right-hand side of Eq.~\eqref{eq:spectrum_normalization_condition_1_appendix}.

Equations \eqref{eq:spectrum_hierarchy}, \eqref{eq:spectrum_hierarchy_bcs},
\eqref{eq:spectrum_normalization_condition_0_appendix}, \eqref{eq:spectrum_normalization_condition_1_appendix},
constitute a closed system of equations that can be solved recursively
to obtain the spectrum to arbitrary order.

At order $k=0$, the right-hand side of Eq.~\eqref{eq:spectrum_hierarchy}
vanishes, so that the equation is reduced to the eigenvalue equation
of the Laplace operator. Thus, $\evn^{(0)}$, $\efn^{(0)}$ is the
spectrum of the Laplace operator with absorbing boundary conditions
on a unit ball, where we assume that $\efn^{(0)}$ has been normalized
according to Eq.~\eqref{eq:spectrum_normalization_condition_0_appendix}.

Assuming the spectrum has been obtained up to order $k-1$, the contribution
at order $k$ is calculated as follows. An equation for $\evn^{(k)}$
is obtained by multiplying Eq.~\eqref{eq:spectrum_hierarchy} with
$\efn^{(0)}$, and subsequently integrating over $\xDLvec$. Upon
integrating the result by parts and using the absorbing boundary conditions
Eq.~\eqref{eq:spectrum_hierarchy_bcs}, it follows that the equation
is in fact independent of $\efn^{(k)}$ and can be solved directly
for $\evn^{(k)}$, leading to 
\begin{align}
\evn^{(k)} & =-\sum_{l=1}^{k-1}\evn^{(l)}\int_{-1}^{1}\mathrm{d}\xDL~\efn^{(0)}\efn^{(k-l)}\label{eq:eigenvalue_k}\\
 & \qquad-\sum_{l=1}^{k}l\,\int_{\BDL}\mathrm{d}^{N}\xDLvec~\efn^{(0)}\left(\EDLvec_{l,\alphavec}\cdot\gradDL\right)\left(\xDL_{\alphavec}\efn^{(k-l)}\right),\nonumber 
\end{align}
where we used the normalization condition Eq.~\eqref{eq:spectrum_normalization_condition_0_appendix}
for $\efn^{(0)}$. Since the right-hand only depends on $\evn^{(l)}$,
$\efn^{(l)}$ with $l<k$, this equation can be used to calculate
the order $k$ eigenvalue contribution in terms of the lower-order
contributions.

Once $\evn^{(k)}$ has been obtained via Eq.~\eqref{eq:eigenvalue_k},
the right-hand side of Eq.~\eqref{eq:spectrum_hierarchy} is known,
so that to obtain $\efn^{(k)}$ the inhomogeneous Helmholtz Eq.~\eqref{eq:spectrum_hierarchy}
with boundary conditions Eq.~\eqref{eq:spectrum_hierarchy_bcs} has
to be solved. While in general this can be done using the corresponding
Green's function, we calculate the spectrum to order $\epsilon^{2}$
directly using a simple ansatz below. Before that, however, we establish
some general properties of the spectrum which follow from parity symmetry.

\subsecthead{Parity properties of the spectrum.} We introduce the
parity operator $\ParityDL$, defined by its action on a function
$f$ as 
\begin{equation}
(\ParityDL f)(\xvecDL)\equiv f(-\xvecDL).\label{eq:definition_parity_operator_appendix}
\end{equation}
Consequently, for products of functions $f$, $g$, it holds that
$\ParityDL(fg)=(\ParityDL f)(\ParityDL g)$, and for the gradient
we have $\ParityDL\gradDL=-\gradDL$. Therefore the operator $\ParityDL$
commutes with the Laplacian, $\ParityDL\laplaceDLvec=\laplaceDLvec\ParityDL$,
so that we can assume that the eigenfunctions $\efn^{(0)}$ of the
Laplacian diagonalize $\laplaceDLvec$ and $\ParityDL$ simultaneously,
so that 
\begin{equation}
\ParityDL\efn^{(0)}=\ParityZero\,\efn^{(0)},\label{eq:parity_efnzero_appendix}
\end{equation}
with $\ParityZero\in\{-1,1\}$.

Via induction in $k$ it follows from Eqs.~\eqref{eq:spectrum_hierarchy},
\eqref{eq:spectrum_hierarchy_bcs}, \eqref{eq:eigenvalue_k}, \eqref{eq:parity_efnzero_appendix},
that 
\begin{equation}
\evn^{(k)}=0\quad\mathrm{for}~k~\mathrm{odd},\label{eq:lambdak_odd_appendix}
\end{equation}
and furthermore that 
\begin{equation}
\ParityDL\efn^{(k)}=(-1)^{k}\ParityZero\,\efn^{(k)}.\label{eq:efnk_parity_appendix}
\end{equation}
Thus, $\efn^{(k)}$ has the same parity as $\efn^{(0)}$ if $k$ is
even, and the opposite parity as $\efn^{(0)}$ if $k$ is odd.

We now calculate the lowest order contributions to the $N$-dimensional
FP spectrum; higher-order results for one-dimensional systems
are given in App.~\ref{app:perturbative_spectrum}.

\subsecthead{Order $\epsilon^{1}$ contribution to the spectrum.}
For $k=1$, Eq.~\eqref{eq:lambdak_odd_appendix} yields $\evn^{(1)}=0$.
Substituting this into Eq.~\eqref{eq:spectrum_hierarchy} for $k=1$,
we obtain 
\begin{align}
\laplaceDLvec\efn^{(1)}+\evn^{(0)}\efn^{(1)} & =-\left(\EDLvec_{1}\cdot\gradvec\right)\efn^{(0)}.\label{eq:spectrum_hierarchy_k1_appendix}
\end{align}
As can be verified by direct substitution, a solution to this inhomogeneous
Helmholtz equation is given by 
\begin{align}
\efn^{(1)}(\xDLvec,\tDL) & =-\frac{1}{2}\left(\EDLvec_{1}(\tDL)\cdot\xvecDL\right)~\efn^{(0)}(\xDLvec,\tDL),\label{eq:efn1_appendix}
\end{align}
where the dot denotes the standard Euclidean inner product between
the two $N$-dimensional vectors $\EDLvec_{1}(\tDL)$, $\xDLvec$.
Because $\efn^{(0)}$ vanishes on $\partial\BDL$, the result Eq.~\eqref{eq:efn1_appendix}
fulfills the boundary condition Eq.~\eqref{eq:spectrum_hierarchy_bcs}.
For $k=1$, the normalization condition Eq.~\eqref{eq:spectrum_normalization_condition_1_appendix}
is fulfilled by Eq.~\eqref{eq:efn1_appendix}, because upon substitution
of Eq.~\eqref{eq:efn1_appendix} the integrand in Eq.~\eqref{eq:spectrum_normalization_condition_1_appendix}
has odd parity, while the integration domain is symmetric with respect
to a parity transformation. With Eqs.~\eqref{eq:lambdak_odd_appendix},
\eqref{eq:efn1_appendix}, we thus have the order $\epsilon^{1}$
contribution to the $N$-dimensional FP spectrum.

\subsecthead{Order $\epsilon^{2}$ contribution to the spectrum.}
For $k=2$, Eq.~\eqref{eq:eigenvalue_k} becomes 
\begin{align}
\evn^{(2)} & =-\int_{\BDL}\mathrm{d}^{N}\xvecDL~\efn^{(0)}\left(\EDLvec_{1}\cdot\gradDL\right)\left(\efn^{(1)}\right)\label{eq:evn2_1_appendix}\\
 & \qquad-2\int_{\BDL}\mathrm{d}^{N}\xvecDL~\efn^{(0)}\left(\EDLvec_{2,\alpha}\cdot\gradDL\right)\left(\xDL_{\alpha}\efn^{(0)}\right),\nonumber 
\end{align}
where we use $\evn^{(1)}=0$. Substituting Eq.~\eqref{eq:efn1_appendix}
into Eq.~\eqref{eq:evn2_1_appendix}, and performing the integrals,
we obtain the second correction for the eigenvalue as 
\begin{align}
\evn^{(2)} & =\frac{\EDLvec_{1}^{2}}{4}-\mathrm{tr}\left(\EDLvec_{2}\right),\label{eq:evn2_appendix}
\end{align}
where 
\begin{align}
\left(\EDLvec_{2,\alpha}\cdot\gradDL\right)\left(\xDL_{\alpha}\right) & =\sum_{i=1}^{N}\EDLvec_{2,i}^{i}\equiv\mathrm{tr}\left(\EDLvec_{2}\right).
\end{align}
Substituting the definition of $\EDLvec_{k,\alphavec}$, Eq.~\eqref{eq:EDLvec_def_appendix},
and using Eqs.~\eqref{eq:def_FDL}, \eqref{eq:def_trajDL}, it follows
that 
\begin{align}
\evn^{(2)} & =\td\left[\frac{\left(D\beta\left.\Fvec\right|_{\trajvec}-\dot{\trajvec}\right)^{2}}{4D}+\frac{D\beta}{2}\left.\left(\gradvec\cdot\Fvec\right)\right|_{\trajvec}\right],\label{eq:evn2_OM_appendix}
\end{align}
which is the OM stochastic action in units of $1/\td\equiv D/L^{2}$.
To calculate $\efn^{(2)}$, we insert Eqs.~\eqref{eq:lambdak_odd_appendix},
\eqref{eq:efn1_appendix}, \eqref{eq:evn2_appendix} into the right-hand
side of Eq.~\eqref{eq:spectrum_hierarchy} (with $k=2$), resulting
in 
\begin{align}
\laplaceDLvec\efn^{(2)}+ & \,\evn^{(0)}\efn^{(2)}=\left[\frac{\EDLvec_{1}^{2}}{4}-\mathrm{tr}\left(\EDLvec_{2}\right)\right]\efn^{(0)}\label{eq:efn2_equation_appendix}\\
 & ~~\qquad\qquad+\left[\frac{1}{2}\left(\EDLvec_{1}\cdot\xvecDL\right)\EDLvec_{1}-2\EDLvec_{2,\alphavec}\xDL_{\alphavec}\right]\cdot\gradDL\efn^{(0)}.\nonumber 
\end{align}
This equation can be solved directly for the case where the force
inside the tube is given as the gradient of an instantaneous potential,
$\Fvec=-\gradvec U$. According to Eq.~\eqref{eq:E_force_from_potential_appendix},
in that case the 2-tensor (or vector-valued one form) $\EDLvec_{2}$
is symmetric, i.e.~we have $\EDL_{2,i}^{j}=\EDL_{2,j}^{i}$, and
direct substitution shows that Eq.~\eqref{eq:efn2_equation_appendix}
is solved by 
\begin{align}
\efn^{(2)} & =\frac{1}{2}\left[\frac{\left(\EDLvec_{1}\cdot\xDLvec\right)^{2}}{4}-\xvecDL\cdot\EDLvec_{2,\alpha}\xDL_{\alpha}\right]\efn^{(0)},\label{eq:efn2_result_appendix}
\end{align}
which fulfills both the normalization condition Eq.~\eqref{eq:spectrum_normalization_condition_1_appendix}
and the boundary conditions Eq.~\eqref{eq:spectrum_hierarchy_bcs}
(note that $\efn^{(0)}$ vanishes on $\partial\BDL$). The solution
Eq.~\eqref{eq:efn2_result_appendix} is also valid if $\efn^{(0)}$
is radially symmetric, as is the case for $n=1$. In that case $\efn^{(0)}$
depends on $\xDLvec$ only via $||\xDLvec||$, and consequently there
is a scalar function $f$ such that $\gradvecDL\efn^{(0)}=f(||\xDLvec||)~\xDLvec$.
Using this, it is readily verified that Eq.~\eqref{eq:efn2_result_appendix}
is a solution to Eq.~\eqref{eq:efn2_equation_appendix}.

\subsecthead{Order $\epsilon^{3}$ contribution to the eigenvalue.}
According to Eq.~\eqref{eq:lambdak_odd_appendix}, we have $\evn^{(3)}=0$.

\subsection{Parity properties of the reflecting-boundary steady state}

\label{app:steady_state_appendix}

In the present section we discuss the perturbative calculation and
parity properties of both the steady state $\reqDL$ and its multiplicative
inverse $\reqDL^{-1}\equiv1/\reqDL$.

\subsecthead{Perturbative calculation of $\reqDL$.} According to
Eq.~\eqref{eq:1D_FPO_DL}, the instantaneous steady state $\reqDL$
is the solution of the boundary value problem 
\begin{align}
\laplaceDLvec\reqDL-\epsilon\,\gradvecDL\cdot\left[\FDLeffvec\reqDL\right] & =0,\label{eq:req_defining_equation_appendix}
\end{align}
with boundary condition 
\begin{equation}
\hat{{\bf {n}}}\cdot\jveceqDL|_{\partial\BDL}=0,\label{eq:req_boundary_condition_appendix}
\end{equation}
where $\jveceqDL\equiv-\gradDL\reqDL+\epsilon\,\FDLeffvec\reqDL$,
where $\hat{\bf{n}}$ is the outward-pointing unit normal vector on $\BDL$, 
and where $\FDLeffvec=\FDLvec-\trajdotDLvec$, as defined in Eq.~\eqref{eq:effective_total_force}.

If the force $\FDLvec$ originates from a potential, $\FDLvec=-\gradvecDL\tilde{U}$,
then the (unnormalized) instantaneous steady state is a Boltzmann
distribution, c.f.~Eqs.~\eqref{eq:reqDL}, \eqref{eq:req_for_potential_force_appendix}.
Using the Taylor expansion of the exponential function, an expansion
in powers of $\epsilon$ for $\reqDL$ is then obtained from Eq.~\eqref{eq:req_for_potential_force_appendix}.

We now discuss how to perturbatively calculate $\reqDL$ for the general
case, in which the force $\FDLvec$ need not have an instantaneous
potential inside the tube. Substituting into Eq.~\eqref{eq:req_defining_equation_appendix}
the power series expansion Eq.~\eqref{eq:taylor_expansion_multidim_force_appendix}
of $\FDLvec$, we obtain 
\begin{equation}
\laplaceDLvec\reqDL+\sum_{k=1}^{\infty}k\epsilon^{k}(\EDLvec_{k,\alphavec}\cdot\gradvecDL)\left(\xDL_{\alphavec}\reqDL\right)=0,\label{eq:req_defining_equation_taylor_appendix}
\end{equation}
where the dot denotes the standard Euclidean inner product. Expanding
the instantaneous steady state as power series in $\epsilon$, 
\begin{align}
\reqDL & =\sum_{k=0}^{\infty}\epsilon^{k}\reqDL^{(k)},\label{eq:req_power_series}
\end{align}
substituting this expansion into Eq.~\eqref{eq:req_defining_equation_taylor_appendix},
and demanding that the resulting equation hold at each power $\epsilon^{k}$,
we obtain a hierarchy of equations which at order $\epsilon^{k}$
reads 
\begin{align}
\laplaceDLvec\reqDL^{(k)} & =-\sum_{l=1}^{k}l\,(\EDLvec_{l,\alphavec}\cdot\gradvecDL)\left(\xDL_{\alphavec}\reqDL^{(k-l)}\right),\label{eq:req_hierarchy_appendix}
\end{align}
where we use the convention that for $k=0$, the sum on the right-hand
side is zero. Inserting the power series expansions Eq.~\eqref{eq:taylor_expansion_multidim_force_appendix},
\eqref{eq:req_power_series}, into the boundary condition Eq.~\eqref{eq:req_boundary_condition_appendix},
and demanding that the resulting equation be fulfilled at each power
$\epsilon^{k}$, we obtain 
\begin{equation}
(\hat{{\bf {n}}}\cdot\gradvecDL)\reqDL^{(k)}=-\sum_{l=1}^{k}l\,(\hat{{\bf {n}}}\cdot\EDLvec_{l,\alphavec})\xDL_{\alphavec}\reqDL^{(k-l)}\label{eq:req_hierarchy_bc_appendix}
\end{equation}
where $k\geq0$ and we use the convention that for $k=0$, the sum
on the right-hand side is zero.

While at order $\epsilon^{0}$, the (unnormalized) solution to Eqs.~\eqref{eq:req_hierarchy_appendix},
\eqref{eq:req_hierarchy_bc_appendix} is simply given by $\reqDL^{(0)}=1$,
for $k\geq1$ the equations have to be solved recursively, similar
to the spectrum in App.~\ref{app:perturbative_spectrum_N_dimensional}.
The resulting corrections at order one and two are 
\begin{align}
\reqDL^{(1)} & =-\EDLvec_{1}\cdot\xvecDL,\label{eq:req_1_appendix}\\
\reqDL^{(2)} & =\frac{1}{2}\left(\EDLvec_{1}\cdot\xvecDL\right)^{2}-\xvecDL\cdot\EDLvec_{2,\alpha}\xDL_{\alpha}.\label{eq:req_2_appendix}
\end{align}
Note that in Eq.~\eqref{eq:req_2_appendix} only the symmetric part
of the 2-tensor (or vector-valued one form) $\EDLvec_{2}$ contributes.

\subsecthead{Parity properties of the $\reqDL^{(k)}$.} Similar
to the parity properties of the FP spectrum, via induction
in $k$ it can be shown that 
\begin{equation}
\ParityDL\reqDL^{(k)}=(-1)^{k}\reqDL^{(k)},\label{eq:req_parity_properties_appendix}
\end{equation}
where the parity operator $\ParityDL$ is defined in Eq.~\eqref{eq:definition_parity_operator_appendix}.

\subsecthead{Perturbative calculation and parity properties of $\reqDL^{-1}$.}
By definition of the inverse, it holds that 
\begin{equation}
\reqDL\,\reqDL^{-1}=1.\label{eq:req_inv_definition_appendix}
\end{equation}
Substituting the power series expansion Eq.~\eqref{eq:req_power_series}
of $\reqDL$ and the expansion 
\begin{align}
\reqDL^{-1} & =\sum_{k=0}^{\infty}\epsilon^{k}\left(\reqDL^{-1}\right)^{(k)},\label{eq:req_inv_power_series}
\end{align}
into Eq.~\eqref{eq:req_inv_definition_appendix}, and demanding that
the equation hold at any order of $\epsilon$, we obtain a recursive
system of equations for the expansion of $\reqDL^{-1}$ given by 
\begin{align}
\left(\reqDL^{-1}\right)^{(0)} & =1,\label{eq:req_inv_zero_appendix}\\
\left(\reqDL^{-1}\right)^{(k)} & =-\sum_{l=1}^{n}\reqDL^{(l)}\left(\reqDL^{-1}\right)^{(k-l)},\label{eq:req_inv_k_appendix}
\end{align}
where in Eq.~\eqref{eq:req_inv_zero_appendix} we use that $\reqDL^{(0)}=1$.
Using Eqs.~\eqref{eq:req_1_appendix}, \eqref{eq:req_2_appendix},
it follows from Eq.~\eqref{eq:req_inv_k_appendix} that 
\begin{align}
\left(\reqDL^{-1}\right)^{(1)} & =\EDLvec_{1}\cdot\xvecDL,\label{eq:req_inv_1_appendix}\\
\left(\reqDL^{-1}\right)^{(2)} & =\frac{1}{2}\left(\EDLvec_{1}\cdot\xvecDL\right)^{2}+\xvecDL\cdot\EDLvec_{2,\alpha}\xDL_{\alpha}.\label{eq:req_inv_2_appendix}
\end{align}
Note that in Eq.~\eqref{eq:req_inv_2_appendix} only the symmetric
part of the 2-tensor $\EDLvec_{2}$ contributes.

According to Eq.~\eqref{eq:req_inv_zero_appendix}, the parity of
$\left(\reqDL^{-1}\right)^{(0)}$ is $1$. Using induction, and applying
the parity operator to Eq.~\eqref{eq:req_inv_k_appendix}, it furthermore
follows that 
\begin{align}
\ParityDL\left[\left(\reqDL^{-1}\right)^{(k)}\right] & =(-1)^{k}\left(\reqDL^{-1}\right)^{(k)}\label{eq:parity_rhoinv_appendix}
\end{align}
for all $k$.

\subsection{Properties of power series expansions derived from parity}

\label{app:power_series_properties_parity}

We now derive properties of some power series expansions used in the
main text.

\subsecthead{Integral over FP eigenfunction.} We consider
\begin{align}
\Integral_{n} & \equiv\int_{\BDL}\mathrm{d}^{N}\xDLvec~\efn,
\end{align}
which we expand in a power
series 
\begin{align}
\Integral_{n} & \equiv \sum_{k=0}^{\infty} \epsilon^k \Integral_n^{(k)},
\end{align}
with $\Integral_n^{(k)}$
defined by Eq.~\eqref{eq:int_rhon_series}. The integral
on the right-hand side of Eq.~\eqref{eq:int_rhon_series} vanishes
if $\efn^{(k)}$ has odd parity. According to Eq.~\eqref{eq:efnk_parity_appendix},
we thus have 
\begin{equation}
\Integral_{n}^{(k)}=0~\mathrm{if}~\begin{cases}
k~\mathrm{odd~and}~p_{n}=1,\\
k~\mathrm{even~and}~p_{n}=-1.
\end{cases}\label{eq:integral_n_zero_appendix}
\end{equation}
In particular, since the lowest eigenfunction of the Laplace operator
(inside a unit ball and with absorbing boundary conditions) is even,
we have 
\begin{equation}
\Integral_{1}=\Integral_{1}^{(0)}+\epsilon^{2}\Integral_{1}^{(2)}+\epsilon^{4}\Integral_{1}^{(4)}+\mathcal{O}(\epsilon^{6}).
\end{equation}

\subsecthead{Inner product of FP eigenfunctions.} We
consider 
\begin{equation}
\langle\efn,\efm\rangle=\int_{\BDL}\mathrm{d}^{N}\xvecDL~\efn\efm\reqDL^{-1},
\end{equation}
c.f.~Eq.~\eqref{eq:def_inner_product}. The power series expansion
of this inner product is given by 
\begin{align}
\langle\efn,\efm\rangle & =\sum_{l=0}^{\infty}\epsilon^{l}\langle\efn,\efm\rangle^{(l)},
\end{align}
where 
\begin{align}
\langle\efn,\efm\rangle^{(l)} & =\sum_{\substack{i,j,k\geq0\\
i+j+k=l
}
}\int_{\BDL}\mathrm{d}^{N}\xvecDL~\efn^{(i)}\efm^{(j)}\left(\reqDL^{-1}\right)^{(k)}
\end{align}
with the power series expansions Eq.~\eqref{eq:evn_power_series},
\eqref{eq:req_inv_power_series}. If the integrand has odd parity,
the integral on the right-hand side vanishes; applying the parity
operator to the integrand we calculate 
\begin{align}
\ParityDL\left[\efn^{(i)}\efm^{(j)}\left(\reqDL^{-1}\right)^{(k)}\right] & =p_{n}p_{m}(-1)^{l}\left[\efn^{(i)}\efm^{(j)}\left(\reqDL^{-1}\right)^{(k)}\right],
\end{align}
where we use Eqs.~\eqref{eq:efnk_parity_appendix}, \eqref{eq:parity_rhoinv_appendix},
and $i+j+k=l$. Thus, we have 
\begin{equation}
\langle\efn,\efm\rangle^{(l)}=0~\mathrm{if}~\begin{cases}
l~\mathrm{odd~and}~p_{n}p_{m}=1,\\
l~\mathrm{even~and}~p_{n}p_{m}=-1.
\end{cases}\label{eq:efnefm_innerproduct_power_series_zero_appendix}
\end{equation}
In particular, we have 
\begin{equation}
\langle\efn,\efn\rangle=\langle\efn,\efn\rangle^{(0)}+\epsilon^{2}\langle\efn,\efn\rangle^{(2)}+\epsilon^{4}\langle\efn,\efn\rangle^{(4)}+\mathcal{O}(\epsilon^{6})
\end{equation}
for any $n$.

\subsecthead{Inner product of FP eigenfunctions including
time derivative.} Since taking a time derivative does not change
spatial parity we, similar to the previous case, have for the power
series expansion 
\begin{equation}
\langle\efn,\efmdot\rangle=\sum_{l=0}^{\infty}\epsilon^{l}\langle\efn,\efmdot\rangle^{(l)}
\end{equation}
that 
\begin{equation}
\langle\efn,\efmdot\rangle^{(l)}=0~\mathrm{if}~\begin{cases}
l=0,\\
l>0~\mathrm{odd~and}~p_{n}p_{m}=1,\\
l>0~\mathrm{even~and}~p_{n}p_{m}=-1,
\end{cases}\label{eq:efnefmdot_innerproduct_power_series_zero_appendix}
\end{equation}
where we note that $\efmdot^{(0)}=0$ since the spectrum of the Laplace
operator (inside a unit ball and with time-independent absorbing boundary
conditions) is independent of time. In particular we have
\begin{equation}
\langle\efn,\efndot\rangle=\epsilon^{2}\langle\efn,\efndot\rangle^{(2)}+\epsilon^{4}\langle\efn,\efndot\rangle^{(4)}+\mathcal{O}(\epsilon^{6})
\end{equation}
for any $n$.

\subsecthead{Product of inner product of FP eigenfunctions
with time derivative and integral over eigenfunctions.} We now consider
the power series expansion of 
\begin{equation}
\langle\efm,\efonedot\rangle\,\Integral_{m}=\sum_{k=0}^{\infty}\epsilon^{k}\left(\langle\efm,\efonedot\rangle\,\Integral_{m}\right)^{(k)}.
\end{equation}
Since $\efone$ has even parity, $p_{1}=1$, we have according to
Eqs.~\eqref{eq:integral_n_zero_appendix}, \eqref{eq:efnefmdot_innerproduct_power_series_zero_appendix},
that the expansions in powers of $\epsilon$ of both $\langle\efm,\efonedot\rangle$,
$\Integral_{m}$, only have nonzero terms at even powers of $\epsilon$
if $p_{m}=1$, and at odd powers of $\epsilon$ if $p_{m}=-1$; therefore,
regardless of $p_{m}$ the product $\langle\efm,\efonedot\rangle\,\Integral_{m}$
only contains even powers of $\epsilon$, i.e.~ 
\begin{equation}
\left(\langle\efm,\efonedot\rangle\,\Integral_{m}\right)^{(k)}=0~\mathrm{if}~k~\mathrm{odd}.
\end{equation}
The lowest order term of the expansion is therefore 
\begin{equation}
\langle\efm,\efonedot\rangle\,\Integral_{m}=\epsilon^{2}\left(\langle\efm,\efonedot\rangle\,\Integral_{m}\right)^{(2)}+\mathcal{O}(\epsilon^{4})
\end{equation}
with 
\begin{equation}
\left(\langle\efm,\efonedot\rangle\,\Integral_{m}\right)^{(2)}=\begin{cases}
\langle\efm,\efonedot\rangle^{(2)}\,\Integral_{m}^{(0)}~ & \mathrm{if}~p_{m}=1,\\[1.5ex]
\langle\efm,\efonedot\rangle^{(1)}\,\Integral_{m}^{(1)}~ & \mathrm{if}~p_{m}=-1.
\end{cases}
\end{equation}

\section{Perturbative solution of the FPE}

\label{app:perturbative_solution_FP}

\subsecthead{Perturbative solution of the FPE
in terms of the instantaneous spectrum.} We now derive an approximate
solution of Eq.~\eqref{eq:1D_FP_in_eigenbasis}, which incorporations
the coupling between eigenmodes to order $\epsilon^{4}$ (and in the
case of a one-dimensional systen, $N=1$, to order $\epsilon^{5}$).
The following derivation is similar to what in quantum mechanics is
called time-dependent perturbation theory \cite{ballentine_quantum_2010,sakurai_modern_2017}.
To render the following calculation easier to read, we rewrite Eq.~\eqref{eq:1D_FP_in_eigenbasis}
as 
\begin{equation}
-{\acoeffndot}=\frac{1}{\pert^{2}}\Ln\acoeffn+\epsilon\,\sum_{m=1}^{\infty}\Cmat_{nm}\acoeffm,\label{eq:1D_FP_eigenbasis_rewritten}
\end{equation}
where we introduce 
\begin{align}
\Ln(\tDL) & \equiv\evn(\tDL)+\epsilon^{2}\left.\frac{\langle\efn,\efndot\rangle}{\langle\efn,\efn\rangle}\right|_{\tDL},\label{eq:def_Lambda_n}\\
\Cmat_{nm}(\tDL) & \equiv(1-\delta_{n,m})\frac{1}{\epsilon}\left.\frac{\langle\efn,\efmdot\rangle}{\langle\efn,\efn\rangle}\right|_{\tDL},\label{eq:def_Cmat_n}
\end{align}
with $\delta_{n,m}$ the Kronecker delta. From the spectrum
calculated in App.~\ref{app:perturbative_spectrum_N_dimensional}, it follows 
that for all $n$, $m$ we have that $\Ln=\mathcal{O}(\epsilon^{0})$,
$\Cmat_{nm}=\mathcal{O}(\epsilon^{0})$, so that the explicit powers
of $\epsilon$ on the right-hand side of Eq.~\eqref{eq:1D_FP_eigenbasis_rewritten}
represent the leading order scaling of each of the terms.

According to Eq.~\eqref{eq:1D_FP_eigenbasis_rewritten}, the dynamics
of each mode is for small $\epsilon$ dominated by the adiabatic exponential
decay described by the instantaneous decay rate $\Ln(\tDL)/\epsilon^{2}$.
We now derive an approximate solution to Eq.~\eqref{eq:1D_FP_eigenbasis_rewritten}
which incorporates the leading order effects of the mode coupling
described by the coupling matrix $\Cmat_{nm}$.

To separate the adiabatic mode decay and the interaction between modes,
we introduce $\bcoeffn$ via 
\begin{equation}
\acoeffn(\tDL)=\bcoeffn(\tDL)\Padn(\tDL,\tzeroDL),\label{eq:def_bcoeffn}
\end{equation}
with the adiabatic propagator $\Padn$ for mode $n$ given by 
\begin{equation}
\Padn(\tDL,\tzeroDL)\equiv\exp\left[-\dfrac{1}{\epsilon^{2}}\int_{\tzeroDL}^{\tDL}\mathrm{d}\tDLdummyone~\Ln(\tDLdummyone)\right].\label{eq:adiabatic_propagator}
\end{equation}
Substituting Eq.~\eqref{eq:def_bcoeffn} into Eq.~\eqref{eq:1D_FP_eigenbasis_rewritten},
we obtain 
\begin{equation}
{\bcoeffndot}(\tDL)=-\epsilon\,\sum_{\substack{m=1\\
m\neq n
}
}^{\infty}\Cmat_{nm}(\tDL)\,\Padmn(\tDL,\tzeroDL)\,\bcoeffm(\tDL),\label{eq:1D_FP_eigenbasis_new_coordinate_rewritten}
\end{equation}
where in the sum bounds we make explicit the fact that $\Cmat_{nn}=0$,
and where we define 
\begin{align}
\Padmn(\tDL,\tzeroDL) & \equiv\Padm(\tDL,\tzeroDL)/\Padn(\tDL,\tzeroDL)\label{eq:adiabatic_propagator_mn}\\
 & =\exp\left[-\dfrac{1}{\epsilon^{2}}\int_{\tzeroDL}^{\tDL}\mathrm{d}\tDLdummyone~\Lmn(\tDLdummyone)\right],
\end{align}
with $\Lmn(\tDL)\equiv\Lm(\tDL)-\Ln(\tDL)$.

Integrating Eq.~\eqref{eq:1D_FP_eigenbasis_new_coordinate_rewritten},
we obtain 
\begin{align}
\bcoeffn(\tDL) & =\bcoeffn(\tzeroDL)-\epsilon\,\sum_{\substack{m=1\\
m\neq n
}
}^{\infty}\int_{\tzeroDL}^{\tDL}\mathrm{d}\tDLdummyone\,\Cmat_{nm}(\tDLdummyone)\,\Padmn(\tDLdummyone,\tzeroDL)\,\bcoeffm(\tDLdummyone).\label{eq:1D_FP_eigenbasis_new_coordinate_integrated}
\end{align}
To eliminate $\bcoeffm(\tDLdummyone)$ in the second term on the right-hand
side of Eq.~\eqref{eq:1D_FP_eigenbasis_new_coordinate_integrated},
we reinsert the expression Eq.~\eqref{eq:1D_FP_eigenbasis_new_coordinate_integrated},
similar to the construction of the Dyson series in quantum mechanics
\cite{sakurai_modern_2017}. Iterating this procedure, by reinserting
Eq.~\eqref{eq:1D_FP_eigenbasis_new_coordinate_integrated} once more
in the result, we obtain 
\begin{widetext}
\begin{align}
\bcoeffn(\tDL) & =\bcoeffn(\tzeroDL)-\epsilon\sum_{\substack{m=1\\
m\neq n
}
}^{\infty}\MMatrixOne_{nm}(\tDL,\tzeroDL)~\bcoeffm(\tzeroDL)+\epsilon^{2}\sum_{\substack{m=1\\
m\neq n
}
}^{\infty}\sum_{\substack{k=1\\
k\neq m
}
}^{\infty}\MMatrixTwo_{nmk}(\tDL,\tzeroDL)~\bcoeffk(\tzeroDL)\label{eq:dyson_4}\\
 & \qquad\qquad-\epsilon^{3}\sum_{\substack{m=1\\
m\neq n
}
}^{\infty}\sum_{\substack{k=1\\
k\neq m
}
}^{\infty}\sum_{\substack{l=1\\
l\neq k
}
}^{\infty}\int_{\tzeroDL}^{\tDL}\mathrm{d}\tDLdummyone\int_{\tzeroDL}^{\tDLdummyone}\mathrm{d}\tDLdummytwo\int_{\tzeroDL}^{\tDLdummytwo}\mathrm{d}\tDLdummytwo'\Cmat_{nm}(\tDLdummyone)\Padmn(\tDLdummyone,\tzeroDL)\Cmat_{mk}(\tDLdummytwo)\Padkm(\tDLdummytwo,\tzeroDL)\Cmat_{kl}(\tDLdummytwo')\Padlk(\tDLdummytwo',\tzeroDL)\bcoeffl(\tDLdummytwo'),\nonumber 
\end{align}
\end{widetext}

where we define 
\begin{align}
\MMatrixOne_{nm}(\tDL,\tzeroDL) & \equiv\int_{\tzeroDL}^{\tDL}\mathrm{d}\tDLdummyone~\Cmat_{nm}(\tDLdummyone)\Padmn(\tDLdummyone,\tzeroDL),\label{eq:MMatrixOne_def}\\
\MMatrixTwo_{nmk}(\tDL,\tzeroDL) & \equiv\int_{\tzeroDL}^{\tDL}\mathrm{d}\tDLdummyone~\Cmat_{nm}(\tDLdummyone)\Padmn(\tDLdummyone,\tzeroDL)\MMatrixOne_{mk}(\tDLdummyone,\tzeroDL).\label{eq:MMatrixTwo_def}
\end{align}
The first term in Eq.~\eqref{eq:dyson_4} represents the adiabatic
decay of the $n$-th eigenmode, for which according to Eq.~\eqref{eq:def_bcoeffn}
$\bcoeffn$ is constant; the remaining terms describe the mode coupling.
Intuitively, one might interpret $\MMatrixOne_{nm}$ as describing
the direct interaction between two modes $n$, $m$, $\MMatrixTwo_{nmk}$
as describing the second-order interactions between two modes $n$,
$k$, via an intermediate mode $m$. By successively substituting
Eq.~\eqref{eq:1D_FP_eigenbasis_new_coordinate_integrated} into Eq.~\eqref{eq:dyson_4},
interactions mediated by arbitrary many intermediate modes can be
constructed. From the form of Eq.~\eqref{eq:1D_FP_eigenbasis_new_coordinate_integrated}
one might naively expect that to obtain the dynamics of $\bcoeffn$
to order $\epsilon^{4}$, one needs to substitute Eq.~\eqref{eq:1D_FP_eigenbasis_new_coordinate_integrated}
four times (and therefore discuss interactions mediated by up to two
intermediate modes at once). However, since $\Padnm$ itself depends
on $\epsilon$, the interactions $\MMatrixOne_{nm}$, $\MMatrixTwo_{nmk}$,
and their higher-order equivalents, also depend on $\epsilon$; the
naive scaling argument that one substitution of Eq.~\eqref{eq:1D_FP_eigenbasis_new_coordinate_integrated}
corresponds to increasing the order in $\epsilon$ by one therefore
breaks down. As we discuss now, for the steady-state solution of $\bcoeffn$
to order $\epsilon^{4}$ (and $\epsilon^{5}$ for a one-dimensional
system, $N=1$), it is in fact sufficient to discuss the mode-coupling
effects mediated by $\MMatrixOne_{nm}$, $\MMatrixTwo_{nmk}$.

\subsecthead{Direct interaction between two modes.} To lowest order,
the coupling between two modes $n$, $m$, is given by $\MMatrixOne_{nm}(\tDL,\tzeroDL)$
defined in Eq.~\eqref{eq:MMatrixOne_def}. To evaluate this matrix
element, we distinguish three possible scenarios. 
\begin{enumerate}
\item $n<m$, and the eigenvalues $\evn$, $\evm$ are not perturbations
around the same eigenspace of the Laplace operator, denoted by $n\notin\eig(m)$
(i.e.~$\evn^{(0)}\neq\evm^{(0)}$). 
\item $n>m$, and the eigenvalues $\evn$, $\evm$ are not perturbations
around the same eigenspace of the Laplace operator, denoted by $n\notin\eig(m)$
(i.e.~$\evn^{(0)}\neq\evm^{(0)}$). 
\item $n\neq m$, but the eigenvalues $\evn$, $\evm$ are perturbations
around the same eigenspace of the Laplace operator, denoted by $n\in\eig(m)$
(i.e.~$\evn^{(0)}=\evm^{(0)}$). 
\end{enumerate}
Note that scenario 3 can only occur for dimensions $N\geq2$; in one
dimension, $N=1$, the absorbing-boundary spectrum of the Laplace
operator inside a finite interval is non-degenerate.

\textit{Direct interactions between modes, scenario 1.} Since the
eigenvalues of the Laplace operator are ordered, for small $\epsilon$
we have $\Lmn(\tDL)\equiv\Lm(\tDL)-\Ln(\tDL)>0$, and since $n\notin\eig(m)$
it holds that $\Lmn(\tDL)=\mathcal{O}(\epsilon^{0})$. Therefore,
for small $\epsilon$ the exponential in the definition of $\Padmn(\tDL,\tzeroDL)$,
Eq.~\eqref{eq:adiabatic_propagator_mn}, decays on a timescale $\tilde{\tau}_{mn}$
defined by 
\begin{equation}
\int_{\tzeroDL}^{\tzeroDL+\tilde{\tau}_{mn}}\mathrm{d}\tDLdummyone~\Lmn(\tDLdummyone)=\epsilon^{2}
\end{equation}
so that for small $\epsilon$ we have 
\begin{equation}
\tilde{\tau}_{mn}\approx\epsilon^{2}/\Lmn(\tzeroDL)=\mathcal{O}(\epsilon^{2}).\label{eq:def_taumn_appendix}
\end{equation}
Since $\Cmat_{nm}=\mathcal{O}(\epsilon^{0})$, the integral in Eq.~\eqref{eq:MMatrixOne_def}
is in scenario 1 thus dominated by $\tDL\approx\tzeroDL$. Assuming
that $\Cmat_{nm}$ does not vary too rapidly on the time scale $\tilde{\tau}_{mn}$,
we Taylor expand around $\tDL=\tzeroDL$, 
\begin{equation}
\Cmat_{nm}(\tDL)\approx\Cmat_{nm}(\tzeroDL)+(\tDL-\tzeroDL)\cdot\Cmatdot_{nm}(\tzeroDL).\label{eq:scenario1_approximation_1}
\end{equation}
Furthermore assuming that $\Lmn$ does not vary too much on the decay
time scale $\tilde{\tau}_{mn}$, we approximate 
\begin{align}
\Padmn(\tDL,\tzeroDL) & =\exp\left[-\dfrac{1}{\epsilon^{2}}\int_{\tzeroDL}^{\tDL}\mathrm{d}\tDLdummyone~\Lmn(\tDLdummyone)\right]\\
 & \approx\exp\left[-\dfrac{\tDL-\tzeroDL}{\epsilon^{2}}\Lmn(\tzeroDL)\right].\label{eq:scenario1_approximation_2}
\end{align}
Physically speaking, with approximations Eqs.~\eqref{eq:scenario1_approximation_1},
\eqref{eq:scenario1_approximation_2}, we assume that the apparent
force (and hence the FP spectrum) inside the tube varies
slowly as compared to the relaxation times of the individual modes.
Inserting approximations Eqs.~\eqref{eq:scenario1_approximation_1},
\eqref{eq:scenario1_approximation_2}, into Eq.~\eqref{eq:MMatrixOne_def},
the integral is evaluated to 
\begin{align}
\MMatrixOne_{nm}(\tDL,\tzeroDL) & =\frac{\epsilon^{2}}{\Lmn(\tzeroDL)}\left\{ \Cmat_{nm}(\tzeroDL)\vphantom{\frac{1}{2}}\right.\label{eq:scenario1_result}\\
 & \qquad-\left[\Cmat_{nm}(\tzeroDL)+\Cmatdot_{nm}(\tzeroDL)\cdot(\tDL-\tzeroDL)\right]\nonumber \\
 & \left.\qquad\qquad\times\exp\left[-\frac{\tDL-\tzeroDL}{\epsilon^{2}}{\Lmn(\tzeroDL)}\right]\right\} +\mathcal{O}(\epsilon^{5}).\nonumber 
\end{align}
For $\tDL-\tzeroDL\gtrsim\tilde{\tau}_{mn}$, the result Eq.~\eqref{eq:scenario1_result}
simplifies to 
\begin{align}
\MMatrixOne_{nm}(\tDL,\tzeroDL) & =\epsilon^{2}\frac{\Cmat_{nm}(\tzeroDL)}{\Lmn(\tzeroDL)}+\mathcal{O}(\epsilon^{5}).\label{eq:scenario1_result_steady}
\end{align}

\textit{Direct interactions between modes, scenario 2.} 
We first note that 
 $\Padmn(\tDLdummyone,\tzeroDL)=\Padmn(\tDL,\tzeroDL)\Padnm(\tDL,\tDLdummyone)$.
Substituting this into Eq.~\eqref{eq:MMatrixOne_def}, we obtain
\begin{align}
\MMatrixOne_{nm}(\tDL,\tzeroDL) & =\Padmn(\tDL,\tzeroDL)\int_{\tzeroDL}^{\tDL}\mathrm{d}\tDLdummyone~\Cmat_{nm}(\tDLdummyone)\Padnm(\tDL,\tDLdummyone).\label{eq:scenario2_calc1}
\end{align}
Similar to the discussion of scenario 1 the term $\Padnm(\tDL,\tDLdummyone)$
decays exponentially as $\tDLdummyone$ is decreased from $\tDL$,
with a characteristic decay time scale $\tilde{\tau}_{mn}$ defined
by 
\begin{equation}
\int_{\tDL-\tilde{\tau}_{mn}}^{\tDL}\mathrm{d}\tDLdummyone~\Lmn(\tDLdummyone)=\epsilon^{2},
\end{equation}
which for small $\epsilon$ is given by 
\begin{equation}
\tilde{\tau}_{mn}\approx\epsilon^{2}/\Lmn(\tDL)=\mathcal{O}(\epsilon^{2}).\label{eq:def_taumn_2_appendix}
\end{equation}
Thus, in scenario 2 the integral in Eq.~\eqref{eq:MMatrixOne_def}
is dominated by $\tDLdummyone\approx\tDL$. Assuming that $\Cmat_{nm}$
does not vary too rapidly on the time scale $\tilde{\tau}_{mn}$,
we Taylor expand around $\tDL$, 
\begin{equation}
\Cmat_{nm}(\tDLdummyone)\approx\Cmat_{nm}(\tDL)+(\tDLdummyone-\tDL)\cdot\Cmatdot_{nm}(\tDLdummyone),\label{eq:scenario2_approximation_1}
\end{equation}
where a dot here denotes a derivative w.r.t~$\tDL$. Furthermore
assuming that $\Lmn$ does not vary too much on the time scale $\tilde{\tau}_{mn}$,
we approximate 
\begin{align}
\Padnm(\tDL,\tDLdummyone) & =\exp\left[-\dfrac{1}{\epsilon^{2}}\int_{\tDLdummyone}^{\tDL}\mathrm{d}\tDLdummytwo~\Lnm(\tDLdummytwo)\right]\\
 & \approx\exp\left[-\dfrac{\tDL-\tDLdummyone}{\epsilon^{2}}\Lnm(\tDL)\right].\label{eq:scenario2_approximation_2}
\end{align}
Inserting approximations Eqs.~\eqref{eq:scenario2_approximation_1},
\eqref{eq:scenario2_approximation_2}, into Eq.~\eqref{eq:scenario2_calc1},
in scenario 2 the integral is evaluated to 
\begin{align}
\MMatrixOne_{nm}(\tDL,\tzeroDL) & =\Padmn(\tDL,\tzeroDL)\frac{\epsilon^{2}}{\Lnm(\tDL)}\label{eq:scenario2_result}\\
 & \times\left\{ \Cmat_{nm}(\tDL)\vphantom{\frac{1}{2}}\right.-\left[\Cmat_{nm}(\tDL)+\Cmatdot_{nm}(\tDL)\cdot(\tDL-\tzeroDL)\right]\nonumber \\
 & \left.\qquad\qquad\times\exp\left[-\frac{\tDL-\tzeroDL}{\epsilon^{2}}{\Lnm(\tDL)}\right]\right\} +\mathcal{O}(\epsilon^{5}).\nonumber 
\end{align}
After an initial transient decay time, i.e.~for $\tDL-\tzeroDL\gtrsim\tilde{\tau}_{nm}$
with $\tilde{\tau}_{mn}$ defined in Eq.~\eqref{eq:def_taumn_2_appendix},
the result Eq.~\eqref{eq:scenario2_result} simplifies to 
\begin{align}
\MMatrixOne_{nm}(\tDL,\tzeroDL) & =\epsilon^{2}~\Padmn(\tDL,\tzeroDL)\frac{\Cmat_{nm}(\tDL)}{\Lnm(\tDL)}+\mathcal{O}(\epsilon^{5}).\label{eq:scenario2_result_steady}
\end{align}

\textit{Direct interactions between modes, scenario 3.} According
to the perturbative FP spectrum calculated in Appendix
\ref{app:perturbative_spectrum_N_dimensional}, for $n\in\eig(m)$
we have 
\begin{equation}
\Lmn(\tDL)\equiv\Lm(\tDL)-\Ln(\tDL)=\mathcal{O}(\epsilon^{3}),
\end{equation}
c.f.~Eqs.~\eqref{eq:evn_power_series}, \eqref{eq:lambdak_odd_appendix},
\eqref{eq:evn2_appendix}, \eqref{eq:def_Lambda_n}, and note that
$\langle\efn,\efndot\rangle=\mathcal{O}(\epsilon)$. It follows that
\begin{align}
\Padmn(\tDLdummyone,\tzeroDL) & =\exp\left[-\dfrac{1}{\epsilon^{2}}\int_{\tzeroDL}^{\tDL}\mathrm{d}\tDLdummyone~\Lmn(\tDLdummyone)\right]=1+\mathcal{O}(\epsilon),\label{eq:scenario3_padmn_approximation_appendix}
\end{align}
so that to leading order in $\epsilon$ Eq.~\eqref{eq:MMatrixOne_def}
becomes 
\begin{align}
\MMatrixOne_{nm}(\tDL,\tzeroDL)=\int_{\tzeroDL}^{\tDL}\mathrm{d}\tDLdummyone~\Cmat_{nm}(\tDLdummyone)+\mathcal{O}(\epsilon).\label{eq:scenario3_result}
\end{align}

\textit{Direct interactions between modes: summary.} According to
Eqs.~\eqref{eq:dyson_4}, \eqref{eq:scenario1_result_steady}, \eqref{eq:scenario2_result_steady},
\eqref{eq:scenario3_result}, the leading-order contribution to the
coupling between two modes $n$, $m$ scales with $\epsilon$ as 
\begin{align}
\epsilon~\MMatrixOne_{nm}(\tDL,\tzeroDL) & \sim\begin{cases}
\epsilon^{3}, & n<m~\text{and}~n\notin\eig(m),\\
\epsilon^{3}~\Padmn(\tDL,\tzeroDL), & n>m~\text{and}~n\notin\eig(m),\\
\epsilon, & n\in\eig(m),
\end{cases}\label{eq:direct_mode_interactions_results_listing}
\end{align}
where $\Padmn(\tDL,\tzeroDL)$ grows exponentially with an exponent
that scales as $1/\epsilon^{2}$. These scalings are valid after an
initial transient time of the order of 
\begin{align}
\tDL-\tzeroDL\gtrsim\tilde{\tau}_{mn} & \equiv\frac{\epsilon^{2}}{|\Lmn|},\label{eq:intial_decay_time_appendix}
\end{align}
where we assume that the order of magnitude of $\tilde{\tau}_{mn}$
is independent of the time at which $\Lmn(\tDL)$ is evaluated in
Eq.~\eqref{eq:intial_decay_time_appendix}, so that we omit the time-dependence
in $\Lmn$. From the leading-order scalings Eq.~\eqref{eq:direct_mode_interactions_results_listing}
we can infer the largest term in the sum 
\begin{equation}
\epsilon\sum_{\substack{m=1\\
m\neq n
}
}^{\infty}\MMatrixOne_{nm}(\tDL,\tzeroDL)~\bcoeffm(\tzeroDL),\label{eq:first_correction_sum_appendix}
\end{equation}
which appears in Eq.~\eqref{eq:dyson_4}. Assuming that all the $\bcoeffm(\tzeroDL)$
are of comparable order of magnitude, which term dominates in Eq.~\eqref{eq:first_correction_sum_appendix}
depends
on $n$. 
\begin{itemize}
\item For $n=1$ only scenario 1 is relevant (note that the lowest eigenvalue
of the Laplace operator is non-degenerate \cite{grebenkov_geometrical_2013});
the leading-order correction to $\bcoeffn(\tzeroDL)$ is thus at order
$\epsilon^{3}$, and all modes $m>1$ contribute to this correction,
meaning that all terms in Eq.~\eqref{eq:first_correction_sum_appendix}
are relevant. 
\item For $n>1$ the dominant correction is given by scenario 2, $m=1$;
this is because the corresponding $\Padmn(\tDL,\tzeroDL)$ grows fastest,
as 
\begin{equation}
\Delta\Lambda_{1n}=\min_{m<n}\{\Delta\Lambda_{mn}\},
\end{equation}
which follows for small $\epsilon$ from the fact that we perturb
around the ordered eigenvalues of the Laplace operator. In particular,
note that even though in scenario 3, where $n\in\eig(m)$, the coupling
between modes has a lower-order prefactor (order $\epsilon$), the
fact that in scenario 2 the factor $\Padmn(\tDL,\tzeroDL)$ grows
exponentially (with an exponent that scales as $1/\epsilon^{2}$)
makes this the dominant contribution. This means that for $n>1$ the
sum Eq.~\eqref{eq:first_correction_sum_appendix} is dominated by
the term $m=1$, i.e. 
\begin{equation}
\epsilon\sum_{\substack{m=1\\
m\neq n
}
}^{\infty}\MMatrixOne_{nm}(\tDL,\tzeroDL)~\bcoeffm(\tzeroDL)\approx\epsilon~\MMatrixOne_{n1}(\tDL,\tzeroDL)~\bcoeffone(\tzeroDL),
\end{equation}
which is expected to hold after a time $\tilde{\tau}_{1n}$ as defined
in Eq.~\eqref{eq:intial_decay_time_appendix}. 
\end{itemize}
Intuitively, these results tell us that i) the dominant correction
to the adiabatic decay of the lowest mode $n=1$ is due to its interaction
with the modes $m>1$ during the initial relaxation of the initial
conditions (note that $\Cmat_{nm}$, $\Lmn$ in Eq.~\eqref{eq:scenario1_result_steady}
are evaluated at $\tzeroDL$), and ii) the dominant correction to
the adiabatic decay of any mode $n>1$ is due to instantaneous excitation
by the lowest mode $m=1$ (note that $\Cmat_{nm}$, $\Lmn$ in Eq.~\eqref{eq:scenario2_result_steady}
are evaluated at $\tDL$), which after an initial relaxation is expected
to be the dominant mode.

\subsecthead{Higher order coupling between modes.} From Eq.~\eqref{eq:direct_mode_interactions_results_listing}
and the subsequent discussion we see that after the initial relaxation
of the system, the interaction between two modes $n\neq m$ leads
to corrections of order $\epsilon^{3}$ if $n\notin\eig(m)$ (with
an exponentially growing factor if $n>m$), and of order $\epsilon$
if $n\in\eig(m)$. To calculate the leading order corrections to $\bcoeffn$
up to order $\epsilon^{4}$ in the steady-state limit, we therefore
only need to take into account two scenarios for the three-mode coupling
described by Eq.~\eqref{eq:MMatrixTwo_def}, namely 
\begin{enumerate}
\item $k>1$, $m\in\mathrm{eig}(k)$, $n=1$. In this scenario, a mode $k>1$
couples to a mode $m\neq k$ from the same Laplace eigenspace ($\rightarrow$
interaction of order $\epsilon$), which then couples to the lowest
mode $n=1$ ($\rightarrow$ interaction of order $\epsilon^{3}$). 
\item $k=1$, $m>1$, $n\in\mathrm{eig}(m)$. In this scenario, the lowest
mode $k=1$ excites a mode $m>1$ ($\rightarrow$ interaction of order
$\epsilon^{3}$, with an exponentially growing prefactor), which then
couples to a mode $n\neq m$ from the same Laplace eigenspace ($\rightarrow$
interaction of order $\epsilon$). 
\end{enumerate}
Note that these cases are only relevant for dimensions $N\geq2$;
since for $N=1$ the spectrum of the Laplace operator is not degenerate,
higher-order couplings between modes always scale as $\epsilon^{6}$
for a one-dimensional system.

\textit{Higher order coupling between modes, scenario 1.} Since $1=n<m$
the factor $\Padmn(\tDLdummyone,\tzeroDL)$ decays exponentially as
a function of $\tDLdummyone$, so that the $\tDLdummyone$-integral
in Eq.~\eqref{eq:MMatrixTwo_def} is dominated by $\tDLdummyone\approx\tzeroDL$.
We therefore approximate 
\begin{align}
\Padmn(\tDLdummyone,\tzeroDL) & \approx\exp\left[-\dfrac{\tDLdummyone-\tzeroDL}{\epsilon^{2}}\Lmn(\tzeroDL)\right],\label{eq:higher_order_coupling_scenario1_padmn_approximation_appendix}
\end{align}
and furthermore Taylor expand 
\begin{align}
\Cmat_{nm}(\tDLdummyone) & \approx\Cmat_{nm}(\tzeroDL)+(\tDL-\tDLdummyone)\cdot\Cmatdot_{nm}(\tzeroDL),\label{eq:higher_order_coupling_scenario1_Cmat_approximation_appendix}\\
\MMatrixOne_{mk}(\tDLdummyone,\tzeroDL) & \approx(\tDLdummyone-\tzeroDL)\cdot\Cmat_{mk}(\tzeroDL)+\frac{1}{2}(\tDLdummyone-\tzeroDL)^{2}\cdot\Cmatdot_{mk}(\tzeroDL)\nonumber \\
 & \qquad+\mathcal{O}(\epsilon),\label{eq:higher_order_coupling_scenario1_integral_approximation_appendix}
\end{align}
where at the last equality sign we use that for $m\in\eig(k)$ we
have $\Padkm(\tDLdummytwo,\tzeroDL)=1+\mathcal{O}(\epsilon)$, c.f.~Eq.~\eqref{eq:scenario3_padmn_approximation_appendix}.
Substituting Eqs.~\eqref{eq:higher_order_coupling_scenario1_padmn_approximation_appendix},
\eqref{eq:higher_order_coupling_scenario1_Cmat_approximation_appendix},
\eqref{eq:higher_order_coupling_scenario1_integral_approximation_appendix},
into Eq.~\eqref{eq:MMatrixTwo_def}, the $\tDLdummyone$-integral
is evaluated using integration by parts to yield 
\begin{align}
\MMatrixTwo_{nmk}(\tDL,\tzeroDL) & =-\epsilon^{2}\frac{\tDL-\tzeroDL}{\Lmn(\tzeroDL)}\left[\Cmat_{nm}(\tzeroDL)+(\tDL-\tzeroDL)\cdot\Cmatdot_{nm}(\tzeroDL)\right]\label{eq:higher_order_coupling_term_scenario1_result}\\
 & \qquad\times\left[\Cmat_{mk}(\tzeroDL)+\frac{1}{2}(\tDL-\tzeroDL)\cdot\Cmatdot_{mk}(\tzeroDL)\right]\nonumber \\
 & \qquad\times\exp\left[-\frac{\tDL-\tzeroDL}{\epsilon^{2}}\Lmn(\tzeroDL)\right]+\mathcal{O}(\epsilon^{3}),\nonumber 
\end{align}
which vanishes (to order $\epsilon^{2}$) as $\tDL-\tzeroDL\gtrsim\tilde{\tau}_{m1}$
(recall that in the current scenario $n=1$), with $\tilde{\tau}_{m1}$
defined in Eq.~\eqref{eq:intial_decay_time_appendix}.

\textit{Higher order coupling between modes, scenario 2.} 
Exchanging the two integrals that are present in Eq.~\eqref{eq:MMatrixTwo_def} after
substituting Eq.~\eqref{eq:MMatrixOne_def},
we obtain 
\begin{align}
\MMatrixTwo_{nmk}(\tDL,\tzeroDL) & =\Padkm(\tDL,\tzeroDL)\int_{\tzeroDL}^{\tDL}\mathrm{d}\tDLdummytwo~\Cmat_{mk}(\tDLdummytwo)\Padmk(\tDL,\tDLdummytwo)\label{eq:higher_order_coupling_scenario2_1}\\
 & \qquad\qquad\qquad\times\int_{\tDLdummytwo}^{\tDL}\mathrm{d}\tDLdummyone~\Cmat_{nm}(\tDLdummyone)\Padmn(\tDLdummyone,\tzeroDL),\nonumber 
\end{align}
where we use 
$\Padkm(\tDLdummytwo,\tzeroDL)=\Padkm(\tDL,\tzeroDL)\Padmk(\tDL,\tDLdummytwo)$.
Since $1=k<m$, the factor $\Padmk(\tDL,\tDLdummytwo)$ decays exponentially
as $\tDLdummytwo$ is decreased from $\tDL$, so that the $\tDLdummytwo$-integral
is dominated by $\tDLdummytwo\approx\tDL$. Similar to scenario 1,
we therefore approximate 
\begin{align}
\Padmk(\tDL,\tDLdummytwo) & \approx\exp\left[-\dfrac{\tDL-\tDLdummytwo}{\epsilon^{2}}\Lmk(\tDL)\right],\label{eq:higher_order_coupling_scenario2_padmn_approximation_appendix}\\
\Cmat_{mk}(\tDLdummytwo) & \approx\Cmat_{mk}(\tDL)+(\tDLdummytwo-\tDL)\cdot\Cmatdot_{nm}(\tDL),\label{eq:higher_order_coupling_scenario2_Cmat_approximation_appendix}\\
\int_{\tDLdummytwo}^{\tDL}\mathrm{d}\tDLdummyone & ~\Cmat_{nm}(\tDLdummyone)\Padmn(\tDLdummyone,\tzeroDL)\label{eq:higher_order_coupling_scenario2_integral_approximation_appendix}\\
 & \approx(\tDL-\tDLdummytwo)\cdot\Cmat_{nm}(\tDL)
 +\frac{(\tDL-\tDLdummytwo)^{2}}{2}\Cmatdot_{nm}(\tDL)+\mathcal{O}(\epsilon),\nonumber 
\end{align}
where at the last equality sign we use that for $m\in\eig(n)$ we
have $\Padmn(\tDLdummyone,\tzeroDL)=1+\mathcal{O}(\epsilon)$, c.f.~Eq.~\eqref{eq:scenario3_padmn_approximation_appendix}.
Substituting Eqs.~(\ref{eq:higher_order_coupling_scenario2_padmn_approximation_appendix}-\ref{eq:higher_order_coupling_scenario2_integral_approximation_appendix})
into Eq.~\eqref{eq:higher_order_coupling_scenario2_1}, the $\tDLdummytwo$-integral
is then evaluated using integration by parts to yield 
\begin{align}
\MMatrixTwo_{nmk}(\tDL,\tzeroDL) & =-\Padkm(\tDL,\tzeroDL)\epsilon^{2}\frac{\tDL-\tzeroDL}{\Lmk(\tDL)}\label{eq:higher_order_coupling_term_2_result}\\
 & \qquad\times\left[\Cmat_{mk}(\tDL)-(\tDL-\tzeroDL)\cdot\Cmatdot_{mk}(\tDL)\right]\nonumber \\
 & \qquad\times\left[\Cmat_{nm}(\tDL)-\frac{1}{2}(\tDL-\tzeroDL)\cdot\Cmatdot_{nm}(\tDL)\right]\nonumber \\
 & \qquad\times\exp\left[-\frac{\tDL-\tzeroDL}{\epsilon^{2}}\Lmk(\tDL)\right]+\mathcal{O}(\epsilon^{3}).\nonumber 
\end{align}
Comparing this result to Eq.~\eqref{eq:scenario2_result_steady},
we see that after an initial transient time, i.e.~for $\tDL-\tzeroDL\gtrsim\tilde{\tau}_{21}=\max_{m>1}\{\tilde{\tau}_{m1}\}$
(recall that in the current scenario $k=1$), with $\tilde{\tau}_{m1}$
defined in Eq.~\eqref{eq:intial_decay_time_appendix}, the contribution
to the amplitude $\bcoeffn$ from Eq.~\eqref{eq:higher_order_coupling_term_2_result}
is exponentially smaller than the contribution from Eq.~\eqref{eq:scenario2_result_steady};
thus the contribution from Eq.~\eqref{eq:higher_order_coupling_term_2_result}
can be neglected as $\tDL-\tzeroDL\gtrsim\tilde{\tau}_{21}$.

\subsecthead{Final result for approximate FP solution.}
Substituting the results Eqs.~\eqref{eq:scenario1_result_steady},
\eqref{eq:scenario2_result_steady}, \eqref{eq:scenario3_result},
\eqref{eq:higher_order_coupling_term_scenario1_result}, \eqref{eq:higher_order_coupling_term_2_result},
into Eq.~\eqref{eq:dyson_4}, we find that the $\bcoeffn$ are to
exponentially leading order given by 
\begin{align}
\bcoeffone(\tDL) & \approx\bcoeffone(\tzeroDL)-\epsilon^{3}\sum_{m=2}^{\infty}\frac{\Cmat_{1m}(\tzeroDL)}{\Lmone(\tzeroDL)}\bcoeffm(\tzeroDL)+\mathcal{O}(\epsilon^{k}),\label{eq:bcoeffone_final_result_appendix}\\
\bcoeffn(\tDL) & \approx-\Padonen(\tDL,\tzeroDL)~\epsilon^{3}\frac{\Cmat_{n1}(\tDL)}{\Lnone(\tDL)}\bcoeffone(\tzeroDL)+\mathcal{O}(\epsilon^{k}),\label{eq:bcoeffn_final_result_appendix}
\end{align}
where $n>1$ in Eq.~\eqref{eq:bcoeffn_final_result_appendix}, and
$k=6$ for a one-dimensional system, $N=1$, and $k=5$ for $N\geq2$.
These approximate expressions are valid after an initial transient
decay time 
\begin{equation}
\tDL-\tzeroDL\gtrsim\trelDL\equiv\tilde{\tau}_{21}=\max_{m>1}\{\tilde{\tau}_{m1}\},\label{eq:trel_def_appendix}
\end{equation}
with $\tilde{\tau}_{m1}$ defined in Eq.~\eqref{eq:intial_decay_time_appendix}.
Substituting these results for $\bcoeffn$ into Eq.~\eqref{eq:def_bcoeffn},
the coefficients $\acoeffn$ of the eigenfunction-expansion of the
solution of the FPE are finally given by 
\begin{align}
\acoeffone(\tDL) & \approx\Padone(\tDL,\tzeroDL)\left[\acoeffone(\tzeroDL)\vphantom{\frac{1}{2}}\right.\label{eq:acoeffone_final_result_appendix}\\
 & \qquad\left.-\epsilon^{3}\sum_{m=2}^{\infty}\frac{\Cmat_{1m}(\tzeroDL)}{\Lmone(\tzeroDL)}\acoeffm(\tzeroDL)+\mathcal{O}(\epsilon^{k})\right],\nonumber \\
\acoeffn(\tDL) & \approx\Padone(\tDL,\tzeroDL)~\left[-\epsilon^{3}\frac{\Cmat_{n1}(\tDL)}{\Lnone(\tDL)}\bcoeffone(\tzeroDL)+\mathcal{O}(\epsilon^{k})\right]\label{eq:acoeffn_final_result_1_appendix}\\
 & =-\epsilon^{3}\frac{\Cmat_{n1}(\tDL)}{\Lnone(\tDL)}\acoeffone(\tDL)+\mathcal{O}(\epsilon^{k}),\label{eq:acoeffn_final_result_2_appendix}
\end{align}
where $n>1$, and for a one-dimensional system, $N=1$, we have $k=6$,
while for $N\geq2$ we have $k=5$; to obtain Eq.~(\ref{eq:acoeffone_final_result_appendix}-\ref{eq:acoeffn_final_result_2_appendix})
we furthermore use that $\acoeffn(\tzeroDL)=\bcoeffn(\tzeroDL)$,
and at Eq.~\eqref{eq:acoeffn_final_result_2_appendix} we use Eq.~\eqref{eq:acoeffone_final_result_appendix}.
The expressions Eqs.~\eqref{eq:acoeffone_final_result_appendix},
\eqref{eq:acoeffn_final_result_1_appendix} hold after the initial
transient decay time $\trelDL$ defined in Eq.~\eqref{eq:trel_def_appendix},
and neglect both terms of the order $\mathcal{O}(\epsilon^{k})$,
as well as terms exponentially small as compared to the leading-order
contributions.

\section{Explicit results for one-dimensional systems}

\label{app:one_dimensional_system}

In the present section, we consider our theory for a one-dimensional
system, $N=1$.

\subsection{Spectrum of the FPE}

\label{app:perturbative_spectrum}

We now derive explicit expressions for the perturbative spectrum of
the FPE,
 following the strategy from App.~\ref{app:perturbative_spectrum_N_dimensional}.
In particular we show that at order $k$, the perturbative contribution
to the eigenfunction is given by 
\begin{align}
\efn^{(k)}(\xDL,\tDL) & =\Qs^{(k)}(\xDL,\tDL)\cdot\sin\left[n\frac{\pi}{2}(\xDL+1)\right]\label{eq:form_of_perturbation_terms_onedim_app}\\
 & \qquad+\Qc^{(k)}(\xDL,\tDL)\cdot\cos\left[n\frac{\pi}{2}(\xDL+1)\right],\nonumber 
\end{align}
where $\Qs^{(k)}(\xDL,\tDL)$, $\Qc^{(k)}(\xDL,\tDL)$ are polynomials
in $\xDL$ of order $\leq k$.

For $N=1$, the Taylor expansion of the force, Eqs.~\eqref{eq:Fvec_multidim_taylor_appendix},
becomes 
\begin{align}
\FDLeff(\xDL,\tDL) & =-\sum_{k=1}^{\infty}\epsilon^{k-1}k\,\EDL_{k}(\tDL)\xDL^{k-1},\label{eq:taylor_expansion_onedim_force_app}
\end{align}
with 
\begin{align}
\EDL_{k}(\tDL) & \equiv-\frac{1}{k!}L^{k}\beta\left.\frac{\partial^{k-1}F}{\partial x^{k-1}}\right|_{(\trajvec(t),t)}+\delta_{k,1}\trajdotDL(\tDL),\label{eq:EDL_def_onedim_app}
\end{align}
where $(x,t)$ and $(\xDL,\tDL)$ are related via Eq.~\eqref{eq:def_tDL}.
With this, the equation for the $n$-th eigenvalue/eigenfunction pair
at order $\epsilon^{k}$, Eq.~\eqref{eq:spectrum_hierarchy}, becomes
\begin{align}
\partial_{\xDL}^{2}\efn^{(k)}+\evn^{(0)}\efn^{(k)} & =-\sum_{l=1}^{k}\evn^{(l)}\efn^{(k-l)}\label{eq:spectrum_hierarchy_onedim_app}\\
 & \qquad-\sum_{l=1}^{k}l\,\EDL_{l}\partial_{\xDL}\left(\xDL^{l-1}\efn^{(k-l)}\right),\nonumber 
\end{align}
where we use the convention that for $k=0$ the sums on the right-hand
side are zero, and each $\efn^{(k)}$ fulfills the boundary conditions
\begin{equation}
\efn^{(k)}(\xDL=-1,\tDL)=\efn^{(k)}(\xDL=1,\tDL)=0,\label{eq:spectrum_hierarchy_bcs_onedim_app}
\end{equation}
c.f.~Eq.~\eqref{eq:spectrum_hierarchy_bcs}. The normalization condition
at order $k$ is given
by Eqs.~\eqref{eq:spectrum_normalization_condition_0_appendix},
\eqref{eq:spectrum_normalization_condition_0_appendix}, where we
note that 
\begin{align}
\reqDL^{-1}(\xDL,\tDL) & =\exp\left[\sum_{k=1}^{\infty}\epsilon^{k}\EDL_{k}(\tDL)\,\xDL^{k}\right].\label{eq:reqDL_inv_onedim_app}
\end{align}

For $N=1$ the equation for the $\evn^{(k)}$, Eq.~\eqref{eq:eigenvalue_k},
becomes 
\begin{align}
\evn^{(k)} & =-\sum_{l=1}^{k-1}\evn^{(l)}\int_{-1}^{1}\mathrm{d}\xDL~\efn^{(0)}\efn^{(k-l)}\label{eq:eigenvalue_k_onedim_app}\\
 & \qquad-\sum_{l=1}^{k}l\,\int_{\BDL}\mathrm{d}\xDL~\efn^{(0)}\EDL_{l}\partial_{\xDL}\left(\xDL^{l-1}\efn^{(k-l)}\right).\nonumber 
\end{align}
We now show how Eqs.~\eqref{eq:spectrum_hierarchy_onedim_app}, \eqref{eq:spectrum_hierarchy_bcs_onedim_app},
\eqref{eq:eigenvalue_k_onedim_app}, can be solved recursively with
increasing $k$, and that at order $k$ the solution for $\efn^{(k)}$
is of the form Eq.~\eqref{eq:form_of_perturbation_terms_onedim_app}.

At order $k=0$, the right-hand side of Eq.~\eqref{eq:spectrum_hierarchy_onedim_app}
vanishes and we obtain 
\begin{align}
\evn^{(0)} & =\left(\frac{n\pi}{2}\right)^{2},\label{eq:spectrum_evn_zero}\\
\efn^{(0)}(\xDL) & =\sin\left[n\frac{\pi}{2}\left(\xDL+1\right)\right],\label{eq:spectrum_efn_zero}
\end{align}
which is simply the spectrum for free diffusion in a domain $\xDL\in[-1,1]$
with absorbing boundary conditions. Note that Eq.~\eqref{eq:spectrum_efn_zero}
fulfills the normalization condition Eq.~\eqref{eq:spectrum_normalization_condition_0_appendix}.

Assuming the spectrum is known to order $k-1$ and is of the form
Eq.~\eqref{eq:form_of_perturbation_terms_onedim_app}, we now derive
expressions for $\evn^{(k)}$, $\efn^{(k)}$. The correction at order
$k$ to the eigenvalue, $\evn^{(k)}$, is obtained directly from Eq.~\eqref{eq:eigenvalue_k_onedim_app}
by evaluating the right-hand side. According to Eq.~\eqref{eq:form_of_perturbation_terms_onedim_app}
for $n<k$, the integrands on the right-hand side of Eq.~\eqref{eq:eigenvalue_k_onedim_app}
are sums over trigonometric functions multiplied by powers of $\xDL$;
evaluating these integrals in practice is thus straightforward. We
now turn to calculating $\efn^{(k)}$, which according to Eq.~\eqref{eq:spectrum_hierarchy_onedim_app}
obeys an inhomogeneous (undamped) harmonic oscillator equation of
motion. The solution thus has the general form 
\begin{equation}
\efn^{(k)}(\xDL)=\efnhom^{(k)}(\xDL)+\efninhom^{(k)}(\xDL),\label{eq:eigenvector_k_general_solution_onedim_app}
\end{equation}
where 
\begin{equation}
\efnhom^{(k)}(\xDL)=\Xns^{(k)}\sin\left[n\frac{\pi}{2}(\xDL+1)\right]+\Xnc^{(k)}\cos\left[n\frac{\pi}{2}(\xDL+1)\right]\label{eq:eigenvector_k_homogeneous_solution_onedim_app}
\end{equation}
is the homogeneous harmonic oscillator solution (the coefficients
$\Xns^{(k)}$, $\Xnc^{(k)}$ will be determined below), and $\efninhom^{(k)}$
is an inhomogeneous solution of Eq.~\eqref{eq:spectrum_hierarchy_onedim_app}.
To obtain an inhomogeneous solution we note that according to Eq.~\eqref{eq:form_of_perturbation_terms_onedim_app},
the right-hand side of Eq.~\eqref{eq:spectrum_hierarchy_onedim_app}
is a sum over the trigonometric functions $\sin$, $\cos$, multiplied
by powers $\xDL^{l}$, $l\leq k-1$. As we show in App.~\ref{app:inhomogeneous_oscillator},
an inhomogeneous solution $\efninhom^{(k)}$ to Eq.~\eqref{eq:spectrum_hierarchy_onedim_app}
is then again given by a sum over trigonometric functions $\sin$,
$\cos$, multiplied by powers $\xDL^{l}$ with $l\leq k$. Thus, Eq.~\eqref{eq:eigenvector_k_general_solution_onedim_app}
is again of the form Eq.~\eqref{eq:spectrum_hierarchy_onedim_app}.

After an inhomogeneous solution at a given order $k$ has been calculated
using the formulas from App.~\ref{app:inhomogeneous_oscillator},
the coefficient $\Xnc^{(k)}$ in Eq.~\eqref{eq:eigenvector_k_general_solution_onedim_app}
is fixed by the boundary conditions Eq.~\eqref{eq:spectrum_hierarchy_bcs_onedim_app}.
The remaining coefficient $\Xns^{(k)}$ is determined by the normalization
condition Eq.~\eqref{eq:spectrum_normalization_condition_1_appendix}.

Using this algorithm, the spectrum can be calculated to arbitrary
order in $\epsilon^{k}$. While according to Eq.~\eqref{eq:lambdak_odd_appendix}
$\evn^{(k)}=0$ for $k$ odd, for $k\leq5$ the eigenvalues for even
$k$ follow as 
\begin{align}
\evn^{(0)} & =\left(\frac{n\pi}{2}\right)^{2}, & \evn^{(2)} & =\left(\frac{\EDL_{1}}{2}\right)^{2}-\EDL_{2},\label{eq:evn_0}
\end{align}
\begin{align}
\evn^{(4)} & =\frac{1}{6(n\pi)^{2}}\left(3E_{1}E_{3}+2E_{2}^{2}-12E_{4}\right)\left[(n\pi)^{2}-6\right], 
\end{align}
The corresponding eigenfunctions for $l\leq5$ are of the form Eq.~\eqref{eq:spectrum_hierarchy_onedim_app},
with polynomials 
\begin{align}
\Qs^{(0)}(\xDL) & =1, & %
\Qc^{(0)}(\xDL) & =0,\label{eq:Qs_0}\\
\Qs^{(1)}(\xDL) & =-\frac{\EDL_{1}}{2}\xDL, & %
\Qc^{(1)}(\xDL) & =0,\\
\Qs^{(2)}(\xDL) & =\frac{\xDL^{2}}{8}\left(\EDL_{1}^{2}-4\EDL_{2}\right), & %
\Qc^{(2)}(\xDL) & =0,%
\end{align}

\begin{widetext}
\begin{align}
\Qs^{(3)}(\xDL) & =\frac{\xDL}{48(n\pi)^{2}}\left[-\EDL_{1}^{3}(n\pi\xDL)^{2}+12\EDL_{1}\EDL_{2}((n\pi\xDL)^{2}+4)-24\EDL_{3}((n\pi\xDL)^{2}+6)\right],\\
\Qc^{(3)}(\xDL) & =-\frac{1}{2n\pi}(\xDL^{2}-1)\left(\EDL_{1}\EDL_{2}-3\EDL_{3}\right),\\
\Qs^{(4)}(\xDL) & =\frac{1}{384(n\pi)^{4}}\left[\EDL_{1}^{4}(n\pi\xDL)^{4}\ -24\EDL_{1}^{2}\EDL_{2}(n\pi\xDL)^{2}\left((n\pi\xDL)^{2}+8\right)+96\EDL_{3}\EDL_{1}\left((n\pi\xDL)^{4}+2(n\pi)^{2}(6\xDL^{2}-1)+24\right)\right.\nonumber \\
 & \qquad\left.+16\EDL_{2}^{2}\left(3(n\pi\xDL)^{4}+8(n\pi)^{2}(3\xDL^{2}-1)+96\right)\ +192\EDL_{4}\left(-(n\pi\xDL)^{4}+4(n\pi)^{2}(-3\xDL^{2}+1)-48\right)\ \right]\label{eq:Qs_4}\\
\Qc^{(4)}(\xDL) & =\frac{1}{12n\pi}\xDL\cdot(\xDL^{2}-1)\left(3\EDL_{1}^{2}\EDL_{2}-15\EDL_{1}\EDL_{3}-4\EDL_{2}^{2}+24\EDL_{4}\right),\label{eq:Qc_4}\\
\Qs^{(5)}(\xDL) & =\frac{\xDL}{3840(n\pi)^{4}}\left[-\EDL_{1}^{5}(n\pi\xDL)^{4}\ +40\EDL_{1}^{3}\EDL_{2}(n\pi\xDL)^{2}\left((n\pi\xDL)^{2}+12\right)-80\EDL_{1}\EDL_{2}^{2}\left(3(n\pi\xDL)^{4}+8(n\pi)^{2}(6\xDL^{2}-1)+96\right)\right.\nonumber \\
 & \qquad\left.\label{eq:Qs_{5}}+240\EDL_{1}^{2}\EDL_{3}\left(-(n\pi\xDL)^{4}+2(n\pi)^{2}(-9\xDL^{2}+2)-48\right)+960\EDL_{1}\EDL_{4}\left(n\pi)^{2}((n\pi)^{2}\xDL^{4}+20\xDL^{2}-4\right)\right.\\
 & \qquad\left.+960\EDL_{2}\EDL_{3}\left((n\pi\xDL)^{4}+18(n\pi\xDL)^{2}-72\right)+1920\EDL_{5}\left(-(n\pi\xDL)^{4}-20(n\pi\xDL)^{2}+120\right)\right],\nonumber \\
\Qc^{(5)}(\xDL) & =\frac{1}{96(n\pi)^{3}}(\xDL^{2}-1)\left[-6\EDL_{1}^{3}\EDL_{2}(n\pi\xDL)^{2}+40\EDL_{1}\EDL_{2}^{2}(n\pi\xDL)^{2}+42\EDL_{1}^{2}\EDL_{3}(n\pi\xDL)^{2}\right.\label{eq:Qc_5}\\
 & \qquad\left.-48\EDL_{1}\EDL_{4}\left((n\pi)^{2}(3\xDL^{2}+1)-12\right)-24\EDL_{2}\EDL_{3}\left(3(n\pi)^{2}(2\xDL^{2}+1)-36\right)+240\EDL_{5}\left((n\pi)^{2}(\xDL^{2}+1)-12\right)\right],\nonumber 
\end{align}
\end{widetext}

where we suppress the time-dependence via the $\EDL_{l}$, which are
defined in Eq.~\eqref{eq:EDL_def_onedim_app}.

With Eqs.~\eqref{eq:evn_power_series}, \eqref{eq:lambdak_odd_appendix},
\eqref{eq:form_of_perturbation_terms_onedim_app}, \eqref{eq:EDL_def_onedim_app},
(\ref{eq:evn_0}-\ref{eq:Qc_5}), we have explicit expressions for
the spectrum of the FP operator Eq.~\eqref{eq:1D_FPO_DL}
up to order $\epsilon^{6}$ (eigenvalues) and $\epsilon^{5}$ (eigenfunctions)
for the case of a one-dimensional system, $N=1$.

\subsection{Inhomogeneous solution for harmonic oscillator}

\label{app:inhomogeneous_oscillator}

In the present section we explain how to obtain an inhomogeneous solution
to Eq.~\eqref{eq:spectrum_hierarchy_onedim_app}. Since Eq.~\eqref{eq:spectrum_hierarchy_onedim_app}
is linear, we can consider the inhomogeneous solution for each term
on the right-hand side separately, and subsequently add them. According
to Eqs.~\eqref{eq:form_of_perturbation_terms_onedim_app}, \eqref{eq:spectrum_efn_zero},
for each term on the right-hand side of Eq.~\eqref{eq:spectrum_hierarchy_onedim_app}
we have to solve an equation of either of the two forms 
\begin{align}
\partial_{\xDL}^{2}\efninhom+\left(\frac{n\pi}{2}\right)^{2}\efninhom & =T\xDL^{l}\sin\left[n\frac{\pi}{2}\left(\xDL+1\right)\right],\label{eq:inhomogeneous_equation_A}\\
\partial_{\xDL}^{2}\efninhom+\left(\frac{n\pi}{2}\right)^{2}\efninhom & =T\xDL^{l}\cos\left[n\frac{\pi}{2}\left(\xDL+1\right)\right],\label{eq:inhomogeneous_equation_B}
\end{align}
with $T\in\mathbb{R}$, $l\in\{0,1,2,..\}$ and $n\in\{1,2,...\}$.
Direct substitution shows that while a solution to Eq.~\eqref{eq:inhomogeneous_equation_A}
is given by 
\begin{align}
\efninhom(\xDL) & =T\sum_{m=1}^{l+1}\frac{l!}{m!}\frac{\left(n\pi\xDL\right)^{m}}{(n\pi)^{l+2}}\label{eq:inhomogeneous_solution_A}\\
 & \qquad\times\sin\left[n\frac{\pi}{2}\left(\xDL+1\right)+\frac{\pi}{2}(l-m)\right],\nonumber 
\end{align}
an inhomogeneous solution to Eq.~\eqref{eq:inhomogeneous_equation_B}
is given by 
\begin{align}
\efninhom(\xDL) & =T\sum_{m=1}^{l+1}\frac{l!}{m!}\frac{\left(n\pi\xDL\right)^{m}}{(n\pi)^{l+2}}\label{eq:inhomogeneous_solution_B}\\
 & \qquad\times\cos\left[n\frac{\pi}{2}\left(\xDL+1\right)+\frac{\pi}{2}(l-m)\right].\nonumber 
\end{align}
Note that the shifts $(l-m)\pi/2$ with $l-m\in\mathbb{Z}$ in Eqs.~\eqref{eq:inhomogeneous_solution_A},
\eqref{eq:inhomogeneous_solution_B}, simply alternate between $\cos$
and $\sin$ (with possible sign changes), as follows directly from
the trigonometric identities $\sin\left(\theta\pm\pi/2\right)=\pm\cos\theta$,
$\cos\left(\theta\pm{\pi}/{2}\right)=\mp\sin\theta$. Thus, each term
in the solutions Eqs.~\eqref{eq:inhomogeneous_solution_A}, \eqref{eq:inhomogeneous_solution_B}
is again of the form of the right-hand side of Eqs.~\eqref{eq:inhomogeneous_equation_A},
\eqref{eq:inhomogeneous_equation_B}. In particular, if we start with
a power $\xDL^{l}$ in either Eq.~\eqref{eq:inhomogeneous_equation_A},
\eqref{eq:inhomogeneous_equation_B}, the highest power in $\xDL$
that appears in the corresponding inhomogeneous solution Eq.~\eqref{eq:inhomogeneous_solution_A},
\eqref{eq:inhomogeneous_solution_A} is $l+1$. From this and Eq.~\eqref{eq:spectrum_efn_zero}
it follows that the order of the polynomials $\Qs^{(k)}(\xDL)$, $\Qc^{(k)}(\xDL)$
is always less or equal than $k$.

\subsection{Normalized probability density inside the tube to order $\epsilon^{5}$}

\label{app:normalized_density_in_tube}

In the present section we give the explicit expansion of the normalized
probability density Eqs.~\eqref{eq:approximate_normalized_density}
as a power series in $\epsilon$ up to order $\epsilon^{5}$. The
power series is based on the unnormalized density Eq.~\eqref{eq:approximate_propagator_steady}
and has the form Eq.~\eqref{eq:PnormalizedDL_power_series}, reprinted
here for convenience, 
\begin{align}
\PabsorbingNormalizedDL(\xDL,\tDL) & =\sum_{k=0}^{4}\epsilon^{k}\left\{ \PnCoeffADL^{(k)}\sin\left[\frac{\pi}{2}(\xDL+1)\right]\right.\label{eq:PnormalizedDL_power_series_app}\\
 & \qquad\qquad\left.+\PnCoeffBDL^{(k)}\cos\left[\frac{\pi}{2}(\xDL+1)\right]\right\} +\mathcal{O}(\epsilon^{6}).\nonumber 
\end{align}

Substituting the perturbative FP spectrum calculated in
App.~\ref{app:perturbative_spectrum}
 into the propagator Eq.~\eqref{eq:approximate_propagator_steady},
 the infinite sums that appear can be evaluated explicitly.
Using the result to evaluate the normalized probability density
Eq.~\eqref{eq:approximate_normalized_density}, an explicit perturbative
expression for $\PabsorbingNormalizedDL$ of the form Eq.~\eqref{eq:PnormalizedDL_power_series_app}
is obtained. The resulting coefficients $\PnCoeffADL^{(k)}$, $\PnCoeffBDL^{(k)}$
for $k\leq5$ are 
\begin{align}
\PnCoeffADL^{(0)}(\xDL) & =\frac{\pi}{4}, & \PnCoeffBDL^{(0)}(\xDL) & =0,\label{eq:PnCoeffA_0}\\
\PnCoeffADL^{(1)}(\xDL) & =-\frac{\pi}{8}\xDL\EDL_{1}, & \PnCoeffBDL^{(1)}(\xDL) & =0,
\end{align}
\begin{align}
\PnCoeffADL^{(2)}(\xDL) & =\frac{1}{32\pi}\left[\left(\pi^{2}(\xDL^{2}-1)+8\right)\EDL_{1}^{2}\right.\\
 & \qquad\left.\nonumber+\left(-4\pi^{2}(\xDL^{2}-1)-32\right)\EDL_{2}\right],\\
\PnCoeffBDL^{(2)}(\xDL) & =0,
\end{align}
\begin{align}
\PnCoeffADL^{(3)}(\xDL) & =\frac{\xDL}{192\pi}\left[\left(\pi^{2}(-\xDL^{2}+3)-24\right)\EDL_{1}^{3}\right.\nonumber \\
 & \qquad\qquad+12\left(\pi^{2}(\xDL^{2}-1)+12\right)\EDL_{1}\EDL_{2}\\
 & \qquad\qquad-24\left(\pi^{2}\xDL^{2}+6\right)\EDL_{3}\left.-24\dot{\EDL}_{1}\right],\nonumber 
\end{align}
\begin{align}
\PnCoeffBDL^{(3)}(\xDL) & =\frac{1}{16}\left(\xDL^{2}-1\right)\left(-2\EDL_{1}\EDL_{2}+6\EDL_{3}+\dot{\EDL}_{1}\right),
\end{align}
\begin{align}
\PnCoeffADL^{(4)}(\xDL) & =\frac{1}{1536\pi^{3}}\left[\left(\pi^{4}(\xDL^{4}-6\xDL^{2}+5)+48\pi^{2}(\xDL^{2}-1)\right)\EDL_{1}^{4}\right.\nonumber \\
 & \qquad+24\left(-\pi^{4}(\xDL^{2}-1)^{2}-8\pi^{2}(3\xDL^{2}-1)\right)\EDL_{1}^{2}\EDL_{2}\nonumber \\
 & \qquad+96\left(\pi^{4}(\xDL^{4}-1)+4\pi^{2}(3\xDL^{2}+4)-48\right)\EDL_{1}\EDL_{3}\nonumber \\
 & \qquad+16\left(3\pi^{4}(\xDL^{2}-1)^{2}+8\pi^{2}(9\xDL^{2}-1)-192\right)\EDL_{2}^{2}\nonumber \\
 & \qquad+192\left(\pi^{4}(-\xDL^{4}+1)-4\pi^{2}(3\xDL^{2}+5)+96\right)\EDL_{4}\nonumber \\
 & \qquad+96\left(\pi^{2}(\xDL^{2}-3)+32\right)\EDL_{1}\dot{\EDL}_{1}\nonumber \\
 & \qquad\left.+64\left(\pi^{2}(-3\xDL^{2}+7)-72\right)\dot{\EDL}_{2}\right],
\end{align}
\begin{align}
\PnCoeffBDL^{(4)}(\xDL) & =\frac{\xDL}{96}\left(-\xDL^{2}+1\right)\left[-6\EDL_{1}^{2}\EDL_{2}+30\EDL_{1}\EDL_{3}+8\EDL_{2}^{2}\right.\nonumber \\
 & \qquad\qquad\qquad\left.-48\EDL_{4}+3\EDL_{1}\dot{\EDL}_{1}-4\dot{\EDL}_{2}\right],\label{eq:PnCoeffB_4}
\end{align}
\begin{align}
 & \PnCoeffADL^{(5)}(\xDL)=\frac{1}{15360\pi^{3}}\xDL\\
 & ~\times\left[\EDL_{1}^{5}\pi^{2}(-\pi^{2}\xDL^{4}+\xDL^{2}10(\pi^{2}-8)-25\pi^{2}+240)\right.\nonumber \\
 & ~~\left.\nonumber+40\EDL_{1}^{3}\EDL_{2}((\pi\xDL)^{4}-4(\pi\xDL)^{2}(\pi^{2}-11)+3\pi^{4}-36\pi^{2}+96)\right.\\
 & ~~\left.\nonumber+240\EDL_{1}^{2}\EDL_{3}(-(\pi\xDL)^{4}+(\pi\xDL)^{2}(\pi^{2}-26)+2\pi^{4}-26\pi^{2}+48)\right.\\
 & ~~\left.\nonumber+960\EDL_{1}\EDL_{4}((\pi\xDL)^{4}+20(\pi\xDL)^{2}-\pi^{4}+20\pi^{2}-144)\right.\\
 & ~~\left.\nonumber+80\EDL_{1}\EDL_{2}^{2}\pi^{2}(-3\pi^{2}\xDL^{4}+6\xDL^{2}(\pi^{2}-16)-3\pi^{2}+32)\right.\\
 & ~~\left.\nonumber-240\EDL_{1}^{2}\dot{\EDL}_{1}((\pi\xDL)^{2}-7\pi^{2}+72)\right.\\
 & ~~\left.\nonumber+320\EDL_{1}\dot{\EDL}_{2}((\pi\xDL)^{2}-\pi^{2}+96)\right.\\
 & ~~\left.\nonumber+320\dot{\EDL}_{1}\EDL_{2}((\pi\xDL)^{2}+3\pi^{2}+48)\right.\\
 & ~~\left.\nonumber+960\EDL_{2}\EDL_{3}((\pi\xDL)^{4}+(\pi\xDL)^{2}(-\pi^{2}+26)-6\pi^{2}-24)\right.\\
 & ~~\left.\nonumber-5760\dot{\EDL}_{3}(\pi^{2}+2)\right.\\
 & ~~\left.\nonumber+1920\EDL_{5}(-(\pi\xDL)^{4}-20(\pi\xDL)^{2}+120)\right],
\end{align}
\begin{align}
\PnCoeffBDL^{(5)}(\xDL) & =\frac{1}{384\pi^{2}}(\xDL^{2}-1)\left[-6\EDL_{1}^{3}\EDL_{2}((\pi\xDL)^{2}-\pi^{2}+8)\right.\nonumber \\
 & \qquad\qquad\left.+6\EDL_{1}^{2}\EDL_{3}(7(\pi\xDL)^{2}-3\pi^{2}+24)\right.\\
 & \qquad\qquad\left.\nonumber-48\EDL_{1}\EDL_{4}(3(\pi\xDL)^{2}+\pi^{2}-12)\right.\\
 & \qquad\qquad\left.\nonumber+8\EDL_{1}\EDL_{2}^{2}(5(\pi\xDL)^{2}-3\pi^{2}+24)\right.\\
 & \qquad\qquad\left.\nonumber+3\EDL_{1}^{2}\dot{\EDL}_{1}((\pi\xDL)^{2}-\pi^{2}+8)\right.\\
 & \qquad\qquad\left.\nonumber-8\EDL_{1}\dot{\EDL}_{2}((\pi\xDL)^{2}+12)\right.\\
 & \qquad\qquad\left.\nonumber-12\dot{\EDL}_{1}\EDL_{2}((\pi\xDL)^{2}-\pi^{2}+16)\right.\\
 & \qquad\qquad\left.\nonumber+144\EDL_{2}\EDL_{3}(-(\pi\xDL)^{2}+2)\right.\\
 & \qquad\qquad\left.\nonumber+12\dot{\EDL}_{3}((\pi\xDL)^{2}+\pi^{2}+12)\right.\\
 & \qquad\qquad\left.\nonumber+240\EDL_{5}((\pi\xDL)^{2}+\pi^{2}-12)\right],
\end{align}
where the $\EDL_{l}\equiv\EDL_{l}(\tDL)$ are defined in Eq.~\eqref{eq:EDL_def_onedim_app}
and a dot denotes a derivative with respect to $\tDL$.

\subsection{Effect of initial distribution inside tube on exit rate}

\label{app:equilibration}

\begin{figure*}[ht]
\centering 
 \includegraphics[width=1\textwidth]{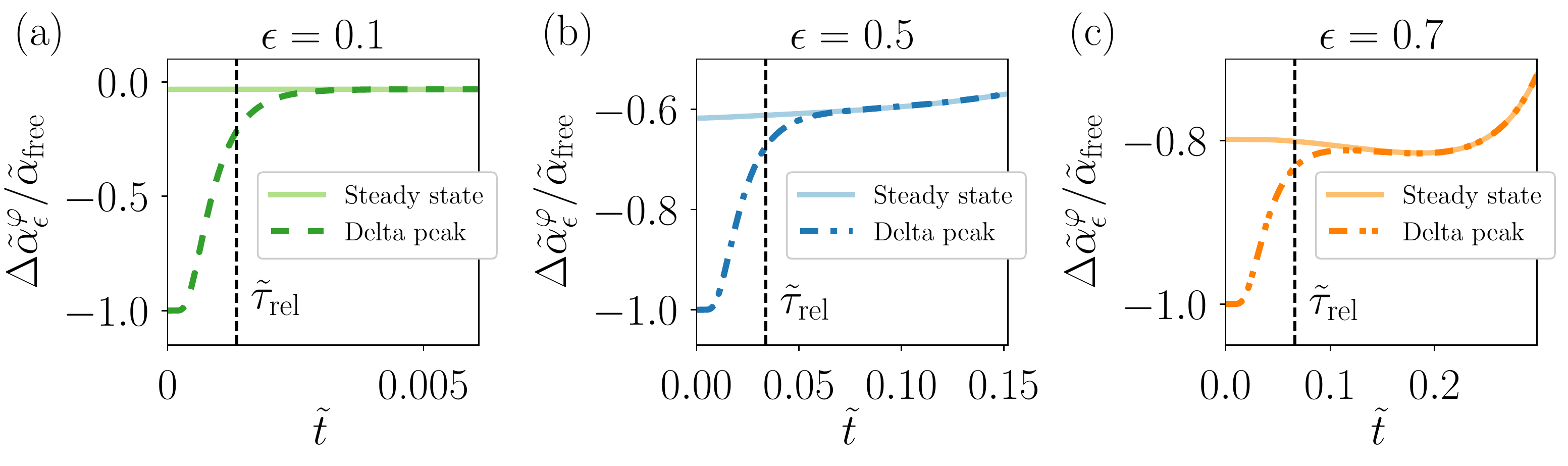} \caption{ \label{fig:delta_peak} 
 {Effect of the initial distribution
$\PDLzero$ inside the tube on the exit rate.} The exit rate $\aexitDL_{\epsilon}^{\traj}$,
defined in Eq.~\eqref{eq:exit_rate_DL}, is shown as a function of
time $\tDL$, for tube radius (a) $\epsilon=0.1$, (b) $\epsilon=0.5$,
and (c) $\epsilon=0.7$. From all rates the free-diffusion exit rate
is subtracted and the result is divided by the free-diffusion exit
rate, as defined in Eq.~\eqref{eq:def_relative_rate}. All data shown
is obtained from numerical simulations of the FPE,
Eq.~\eqref{eq:1D_FP_eq_DL}, from which the exit rate is calculated
using using Eq.~\eqref{eq:exit_rate_DL}. Colored solid lines are
replots of the corresponding curves in Fig.~\ref{fig:densities_and_rates}
(d), (e), (f), and denote exit rates obtained using the instantaneous
steady-state as initial condition for the simulations, as explained
in App.~\ref{app:numerics}. Colored dashed lines show exit rates
obtained using a delta peak at the tube center as initial condition
for the simulations. Vertical dashed lines denote the initial relaxation
time $\trelDL$ given in Eq.~\eqref{eq:app_initial_relaxation_time}. }
\end{figure*}
As described in App.~\ref{app:numerics}, in the numerical examples
in the main text we eliminate transient relaxation effects at the
initial time $\tinitialDL$ by using the instantaneous FP
steady state at $\tinitialDL$ as initial distribution $\PDLzero$.

To illustrate the effect of the initial distribution $\PDLzero$ on
the finite-radius exit rate $\aexitDL_{\epsilon}^{\traj}$ we here numerically consider
the initial condition $\PDLzero(\xDL)=\delta(\xDL)$, which corresponds
to a particle starting out at time $\tinitialDL$ at the center of
the tube.

In Fig.~\ref{fig:delta_peak} we compare numerical exit rates resulting
from this delta-peak initial condition (dashed colored lines) to numerical
exit rate corresponding to the instantaneous steady-state initial
condition (solid colored lines). As in Fig.~\ref{fig:densities_and_rates}
(d), (e), (f), we shift and rescale exit rates according to Eq.~\eqref{eq:def_relative_rate}.
Using the perturbative results from App.~\ref{app:perturbative_spectrum},
the initial relaxation time $\trelDL$, defined in Eq.~\eqref{eq:trel_def_appendix},
is given as power series in $\epsilon$ as 
\begin{equation}
\trelDL=\frac{3\epsilon^{2}}{4\pi^{2}}+\mathcal{O}(\epsilon^{5}).\label{eq:app_initial_relaxation_time}
\end{equation}
This perturbative expression for $\trelDL$ is plotted in Fig.~\ref{fig:delta_peak}
as vertical dashed lines.
Figure \ref{fig:delta_peak} (a) shows data for tube radius $\epsilon=0.1$.
While the data corresponding to the steady-state initial condition
(colored solid line) is practically constant on the time scale depicted,
the exit rate corresponding to the delta-peak initial condition (colored
dashed line) shows relaxation behavior; the curve starts at $\epsilon^{2}\Delta\aexitDL_{\epsilon}^{\traj}/\aexitfreeDL(0)=-1$,
which according to Eq.~\eqref{eq:def_relative_rate} corresponds
to a vanishing exit rate $\aexitDL_{\epsilon}^{\traj}(0)=0$, consistent with the intuition
that a particle starting in the center of a finite-radius ball needs
a finite time to diffusive out of the ball. This exit rate then relaxes
to the steady-state exit rate on a time scale well-approximated by
Eq.~\eqref{eq:app_initial_relaxation_time}; for times larger than
$\tDL\approx2\cdot\trelDL$ all knowledge of the initial condition
has decayed and the two exit rates are indistinguishable.
The data shown for the larger tube radii $\epsilon=0.5$, $0.7$ in
Fig.~\ref{fig:delta_peak} (b), (c) shows the exact same behavior.
As expected from the leading-order scaling $\trelDL\sim\epsilon^{2}$,
the relaxation time increases with tube radius $\epsilon$.

\subsection{Numerical algorithm for one-dimensional FPE}

\label{app:numerics}

To simulate the dimensionless FPE, Eq.~\eqref{eq:1D_FP_eq_DL},
\eqref{eq:1D_FPO_DL}, we discretize space by introducing the grid
\begin{equation}
\xDL_{i}\equiv-1+i\cdot\Delta\xDL\equiv-1+i\cdot\frac{2}{N+1},\qquad i\in\{0,...,N+1\}.
\end{equation}
and discretize time using a timestep $\Delta\tDL$, 
\begin{equation}
\tDL_{j}\equiv j\cdot\Delta\tDL,\qquad j\in\left\{ \left\lfloor \frac{\tzeroDL}{\Delta\tDL}\right\rfloor ,\left\lfloor \frac{\tzeroDL}{\Delta\tDL}\right\rfloor +1,...,\left\lfloor \frac{\tfinalDL}{\Delta\tDL}\right\rfloor \right\} ,
\end{equation}
where $\lfloor~\rfloor$ means we round down to the closest integer.
Introducing the discretized probability 
$\PDLnoeps_{\epsilon,i}^{j}\equiv\PabsorbingDL(\xDL_{i},\tDL_{j})$, 
where $i = 1,...,N$,
we discretize the FPE, Eq.~\eqref{eq:1D_FP_eq_DL}, as 
\begin{align}
\epsilon^{2}\frac{\PDLnoeps_{\epsilon,i}^{j+1}-\PDLnoeps_{\epsilon,i}^{j}}{\Delta\tDL} & =\frac{\PDLnoeps_{\epsilon,i+1}^{j}-2\PDLnoeps_{\epsilon,i}^{j}+\PDLnoeps_{\epsilon,i-1}^{j}}{\Delta\xDL^{2}}\label{eq:discretization_FP}\\
 & \qquad-\epsilon\frac{\tilde{F}_{\mathrm{app},i+1}^{j}\PDLnoeps_{\epsilon,i+1}^{j}-\tilde{F}_{\mathrm{app},i-1}^{j}\PDLnoeps_{\epsilon,i-1}^{j}}{2\Delta\xDL},\nonumber 
\end{align}
where $i\in\{1,...,N\}$, $\tilde{F}_{\mathrm{app},i}^{j}\equiv\tilde{F}_{\mathrm{app}}(\xDL_{i},\tDL_{j})$,
and in accordance with the absorbing boundary conditions we define
$\PDLnoeps_{\epsilon,0}^{j}=\PDLnoeps_{\epsilon,N+1}^{j}=0$ 
for all $j$.
To obtain an explicit formula for the distribution at time $(j+1)\cdot\Delta\tDL$
in terms of the distribution at time $j\cdot\Delta\tDL$, Eq.~\eqref{eq:discretization_FP}
is then solved for $\PDLnoeps_{\epsilon,i}^{j+1}$ (forward Euler
integration scheme).

All numerical results in this work are obtained using $N=100$, $\Delta\tDL=10^{-7}$.

To eliminate boundary effects due to the transient decay of the initial
condition, we pre-equilibrate the system for every $\epsilon$. Starting
from a distribution $\PDLzero(\xDL)=\sin(\pi(\xDL+1)/2)$, we simulate
the FPE, Eq.~\eqref{eq:discretization_FP}, for
a short time of the order of $\trel$, while holding the parameters
for position and velocity of the path $\trajDL$ constant at the initial
values $\trajDL(\tinitialDL)$, $\trajdotDL(\tinitialDL)$. At the
end of this pre-equilibration, the system is in the instantaneous
steady state decay corresponding to $\trajDL(\tinitialDL)$, $\trajdotDL(\tinitialDL)$.
This instantaneous steady state is then normalized and used as initial
condition for the simulation (in which $\trajDL$, $\trajdotDL$ then
vary with time). A brief discussion on the dependence of the exit
rate on the initial condition is given in App.~\ref{app:equilibration}.


%

\end{document}